\documentclass[preprint, showpacs, aps, pra]{revtex4-1}	
\usepackage{latexsym,amssymb,makeidx}

\usepackage{epsfig, color, ulem}
\usepackage{graphicx}
\usepackage{dcolumn}
\usepackage{bm}
\usepackage{amsmath}
\usepackage{setspace}
\usepackage{lineno}
\def\rev#1{\textcolor{black}{#1}}
\def\revv#1{\textcolor{blue}{#1}}

\begin{document}
\bibliographystyle{jasanum}

\title{Wave-field representations with Green's functions, propagator matrices, and Marchenko-type focusing functions}
\author{Kees Wapenaar}
\affiliation{Department of Geoscience and Engineering,\\ Delft University of Technology,\\ Stevinweg 1,\\ 2628 CN Delft,\\ The Netherlands}
\date{\today}
\noindent
\centerline{\revv{Note: equations (\ref{eqBBB35}) and (\ref{eqBBB36}) are corrected in this version. }}
\mbox{}\\
\mbox{}\\

\maketitle

\begin{spacing}{2.0}

\centerline{\Large Abstract}
\noindent
Classical acoustic wave-field representations consist of volume and boundary integrals, of which the integrands contain specific combinations 
of Green's functions, source distributions and wave fields. 
Using a unified matrix-vector wave equation for different wave phenomena, these representations can be reformulated in terms of 
Green's matrices, source vectors and wave\rev{-field} vectors. The matrix-vector formalism also allows the formulation of representations in which propagator matrices
replace the Green's matrices. These propagator matrices, in turn, can be expressed in terms of Marchenko-type focusing functions.
An advantage of the representations with propagator matrices and focusing functions is that the boundary integrals in these representations are limited to a single open boundary.
This makes these representations a \rev{suitable} basis for developing advanced inverse scattering, imaging and monitoring methods 
for wave fields acquired on a single boundary.\\

\newpage

\section{Introduction}

The aim of this paper is to give a systematic treatment of different types of wave-field representation 
(with Green's functions, propagator matrices and Marchenko-type focusing functions),
to discuss their mutual relations, and indicate some new applications.

\begin{itemize}
\item \underline{Representations with Green's functions. }
A Green's function is the response of a medium to an impulsive point source. It is named after George Green, who,
in a privately published essay \cite{Green1828Book}, introduced the use of impulse responses in field representations \cite{Challis2003PhysicsToday}.
Wave-field representations with Green's functions have been formulated, among others, for optics \cite{Born65Book}, acoustics \cite{Rayleigh78Book, Bleistein84Book},
 elastodynamics  \cite{Knopoff56JASA, Hoop58PHD, Gangi70JGR, Pao76JASA} and electromagnetics \cite{Kong86Book, Altman91Book, Hoop95Book}.
They find numerous applications in forward modeling problems \cite{Hilterman70GEO, Frazer85GJRAS, Mansuripur2021NP}, 
inverse source problems \cite{Porter82JOSA, Hoop95Book}, inverse scattering problems
\cite{Devaney82UI, Bojarski83JASA, Bleistein84Book, Oristaglio89IP}, 
imaging \cite{Porter70JOSA, Schneider78GEO, Berkhout82Book,  Maynard85JASA, Esmersoy88GEO, Lindsey2004AJSS}, time-reversal acoustics
\cite{Fink2001IP}, and Green's function retrieval from ambient noise \cite{Derode2003JASA, Wapenaar2003GEO, Weaver2004JASA}.
\item \underline{Representations with propagator matrices. }
In elastodynamic wave theory, a matrix formalism has been introduced to describe the 
propagation of waves in laterally invariant layered media  \cite{Thomson50JAP, Haskell53BSSA} .
This formalism was refined by Gilbert and Backus \cite{Gilbert66GEO}, who coined the name propagator matrix. In essence, a propagator matrix `propagates' a wave field
(represented as a vectorial quantity)  from one plane in space to another. 
Using perturbation theory,  the connection between wave-field representations with Green's functions and the propagator matrix formalism was discussed \cite{Kennett72BS}.
The propagator matrix has been extended for laterally varying isotropic and anisotropic layered media \cite{Kennett72GJRAS, Woodhouse74GJR}.  
Propagation invariants for laterally varying layered media have been introduced
and the propagator matrix concept has been proposed for the modeling of reflection and transmission responses \cite{Haines88GJI, Kennett90GJI, Koketsu91GJI, Takenaka93WM}. 
The propagator matrix has also been used in a seismic imaging method that accounts for multiple scattering in a model-driven way \cite{Wapenaar87GEO}.
\item \underline{Representations with Marchenko-type focusing functions.}
Building on a 1D acoustic autofocusing method, it has been shown that the wave field inside a laterally invariant layered medium can be retrieved with the Marchenko method
from the single-sided reflection response at the surface of the medium \cite{Rose2001PRA, Rose2002IP, Broggini2012EJP}. 
This concept was extended to a 3D Marchenko wave-field retrieval method for laterally varying media \cite{Wapenaar2014JASA}.
Central in the 3D Marchenko method are wave-field representations containing focusing functions. 
These representations have found applications in imaging methods \cite{Ravasi2016GJI, Staring2018GEO, Jia2018GEO} 
and inverse source problems \cite{Neut2017JASA} that account for multiple scattering in a data-driven way.
\end{itemize}

The setup of this paper is as follows. 
In section \ref{sec2} we briefly review the matrix-vector wave equation for laterally varying media 
\cite{Kennett72GJRAS, Woodhouse74GJR}, generalized for 
different wave phenomena, and we briefly discuss the concept of the Green's matrix, the propagator matrix and the Marchenko-type focusing function.
The symmetry properties of the matrix-vector wave equation allow the formulation of unified matrix-vector wave-field reciprocity theorems  \cite{Wapenaar96JASA, Haines96JMP},
which are reviewed in section \ref{sec3}. These reciprocity theorems form the basis for a systematic treatment of the different types of wave-field representation 
mentioned above.
Traditionally, a wave-field representation is obtained by replacing one of the states in a reciprocity theorem by a Green's state. In section \ref{sec4} we follow this approach for the matrix-vector
reciprocity theorems. By replacing one of the wave-field vectors by the Green's matrix (and the source vector by a unit source matrix) we obtain wave-field representations with Green's matrices.
Analogous to this, in section \ref{sec5} we replace one of the wave-field vectors in the reciprocity theorems by the propagator matrix and thus obtain wave-field representations with propagator matrices.
In section \ref{sec6} we discuss a mixed form, obtained by replacing one of the wave-field vectors by the Green's matrix and the other by the propagator matrix. In section \ref{sec7} we
discuss the relation between the propagator matrix and the Marchenko-type focusing functions and use this relation to derive wave-field representations with focusing functions.
We end with conclusions in section \ref{sec8}.

\section{The unified matrix-vector wave equation, the Green's matrix, the propagator matrix and the focusing function}\label{sec2}

\subsection{The matrix-vector wave equation}

We use a unified matrix-vector wave equation as the basis for the derivations in this paper.
In the space-frequency domain it has the following form \cite{Gilbert66GEO, Kennett72BS, Kennett72GJRAS, Woodhouse74GJR, Haines88GJI}
\begin{eqnarray}\label{eq2.1}
\partial_3{\bf q} - {{\mbox{\boldmath ${\cal A}$}}}\,{\bf q} ={\bf d},
\end{eqnarray}
with
\begin{eqnarray}\label{Aeq7mvbbprff}
{\bf q}=\begin{pmatrix} {{\bf q}_1} \\ {{\bf q}_2} \end{pmatrix},\quad
                {\bf d}=\begin{pmatrix} {\bf d}_1 \\ {\bf d}_2 \end{pmatrix},\quad
{{\mbox{\boldmath ${\cal A}$}}}= \begin{pmatrix}{{\mbox{\boldmath ${\cal A}$}}}_{11}      & {{\mbox{\boldmath ${\cal A}$}}}_{12} \\
                {{\mbox{\boldmath ${\cal A}$}}}_{21} & {{\mbox{\boldmath ${\cal A}$}}}_{22}    \end{pmatrix}.
\end{eqnarray}
Here ${\bf q}({\bf x},\omega)$ is a space- and frequency-dependent $N\times1$ wave-field vector, where ${\bf x}$ denotes the
Cartesian coordinate vector $(x_1,x_2,x_3)$ (with the positive $x_3$-axis pointing downward) and $\omega$ the angular frequency.
The $N/2\,\times\,1$ sub-vectors $ {{\bf q}_1}({\bf x},\omega)$ and $ {{\bf q}_2}({\bf x},\omega)$ contain  wave-field quantities, which are specified for different wave phenomena in Table \ref{table1}.
Operator $\partial_3$ stands for the  differential operator $\partial/\partial x_3$. Matrix ${{\mbox{\boldmath ${\cal A}$}}}({\bf x},\omega)$ is an $N\times N$ operator matrix;
it contains the space- and frequency-dependent anisotropic medium parameters and the horizontal differential operators  $\partial_1$ and $\partial_2$.
Definitions of this operator matrix for different wave phenomena can be found in many of the references mentioned in the introduction. 
$N\times1$ vector ${\bf d}({\bf x},\omega)$ contains the space- and frequency-dependent source functions.
A comprehensive overview of the operator matrices and source vectors for the wave phenomena considered in Table \ref{table1} (with some minor modifications)
is given in reference \cite{Wapenaar2019GJI}.

\begin{table}
\caption{Wave-field sub-vectors ${{\bf q}_1}({\bf x},\omega)$ and ${{\bf q}_2}({\bf x},\omega)$ for different wave phenomena. 
For acoustic waves \rev{in stationary fluids}, $p$ and $v_3$ stand for the acoustic pressure and the vertical component of the particle velocity, respectively. 
For electromagnetic waves, $E_\alpha$ and $H_\alpha$ ($\alpha=1,2$) are the \rev{horizontal components of the} electric and magnetic field strength, respectively. 
For elastodynamic waves, $v_k$ and $\tau_{k3}$ ($k=1,2,3$) are the particle velocity and traction components, respectively.
The same quantities appear in the vectors for the other wave phenomena, where,
for poroelastodynamic and seismoelectric waves, the quantities are averaged in the bulk, fluid or solid, as indicated by the superscripts $b$, $f$ and $s$, respectively. Finally, $\phi$ denotes the porosity.
}\label{table1}
\begin{center}
\begin{tabular}{lrcc}
\hline\hline
& $N$ & ${{\bf q}_1}$ & ${{\bf q}_2}$  \\
\hline\hline
Acoustic & 2&$p$ & $v_3$ \\
\hline
Electromagnetic &4&${\bf E}_0=\begin{pmatrix}  E_1 \\  E_2\end{pmatrix}$& ${\bf H}_0=\begin{pmatrix}  H_2 \\ - H_1\end{pmatrix}$ \\
\hline
Elastodynamic  &6& ${\bf v}=\begin{pmatrix}v_1\\v_2\\v_3\end{pmatrix}$ & $-{\mbox{\boldmath $\tau$}}_3=-\begin{pmatrix}\tau_{13}\\ \tau_{23}\\ \tau_{33}\end{pmatrix}$\\
\hline
Poroelastodynamic  &8& $ \begin{pmatrix} {\bf v}^s\\ \phi( v_3^f- v_3^s) \end{pmatrix}$ & $ \begin{pmatrix}- {\mbox{\boldmath $\tau$}}_3^b \\ p^f  \end{pmatrix}$\\
\hline
Piezoelectric  &10& $\begin{pmatrix} {\bf v}\\  {\bf H}_0 \end{pmatrix}$ & $\begin{pmatrix}- {\mbox{\boldmath $\tau$}}_3 \\ {\bf E}_0 \end{pmatrix}$ \\
\hline
Seismoelectric  &12& $\begin{pmatrix} {\bf v}^s\\ \phi( v_3^f- v_3^s) \\ {\bf H}_0\end{pmatrix}$ & $\begin{pmatrix}- {\mbox{\boldmath $\tau$}}_3^b \\ p^f \\ {\bf E}_0 \end{pmatrix}$\\
\hline
\hline
\end{tabular}
\end{center}\end{table}

For all wave phenomena considered in Table \ref{table1}, operator matrix ${{\mbox{\boldmath ${\cal A}$}}}$ obeys the following symmetry properties 
\begin{eqnarray}
{{\mbox{\boldmath ${\cal A}$}}}^t{\bf N}&=&-{\bf N}{{\mbox{\boldmath ${\cal A}$}}},\label{eqsym}\\
{{\mbox{\boldmath ${\cal A}$}}}^\dagger{\bf K}&=&-{\bf K}{{\,\,\,\bar{\!\!\!{{\mbox{\boldmath ${\cal A}$}}}}}},\label{eqsymad}\\
{{\mbox{\boldmath ${\cal A}$}}}^*{\bf J}&=&{\bf J}{{\,\,\,\bar{\!\!\!{{\mbox{\boldmath ${\cal A}$}}}}}},\label{eqsymcon}
\end{eqnarray}
with
\begin{eqnarray}\label{eq4.3}
{{\bf N}}=\begin{pmatrix} {\bf O} & {\bf I} \\ -{\bf I} & {\bf O} \end{pmatrix},
\quad {{\bf K}}=\begin{pmatrix} {\bf O} & {\bf I} \\ {\bf I} & {\bf O} \end{pmatrix},
\quad {{\bf J}}=\begin{pmatrix} {\bf I} & {\bf O} \\ {\bf O} & -{\bf I} \end{pmatrix},
\end{eqnarray}
where ${\bf O}$ and ${\bf I}$ are $N/2\,\times\, N/2$ zero and identity matrices. 
Superscript $t$ denotes transposition (meaning that the matrix is transposed and the operators in the matrix are also transposed, 
with $\partial_1^t=-\partial_1$ and $\partial_2^t=-\partial_2$), $*$ denotes complex conjugation,
and $\dagger$  transposition and complex conjugation.
In general, the medium parameters in ${{\mbox{\boldmath ${\cal A}$}}}$ are complex-valued and frequency-dependent, accounting for losses. 
 The bar above a quantity means that this quantity is defined in the adjoint medium. Hence, \rev{if} ${{\mbox{\boldmath ${\cal A}$}}}$ is defined in a lossy medium, 
then ${{\,\,\,\bar{\!\!\!{{\mbox{\boldmath ${\cal A}$}}}}}}$ is defined in an effectual medium and vice versa \cite{Hoop88JASA} (a wave propagating through an effectual medium gains energy). 
For lossless media the bar can be dropped.
For all wave phenomena considered in Table \ref{table1}, 
the power-flux density $j$ in the $x_3$-direction is related to the sub-vectors ${{\bf q}_1}$ and ${{\bf q}_2}$ according to
\begin{eqnarray}\label{eq50ab}
j={\frac{1}{4}}({{\bf q}_1}^\dagger{{\bf q}_2}+{{\bf q}_2}^\dagger{{\bf q}_1}).
\end{eqnarray}

As a special case we consider acoustic waves in an inhomogeneous \rev{stationary} medium, with complex-valued and frequency-dependent compressibility 
$\kappa({\bf x},\omega)$ and mass density  $\rho_{kl}({\bf x},\omega)$. 
The latter is defined as a tensor, to account for effective anisotropy, for example due to fine layering at the micro scale \cite{Schoenberg83JASA}.
The mass density tensor is symmetric, that is, $\rho_{kl}({\bf x},\omega)=\rho_{lk}({\bf x},\omega)$.
We introduce the inverse of the mass density tensor, the specific volume tensor $\vartheta_{kl}({\bf x},\omega)$, via $\vartheta_{kl}\rho_{lm}=\delta_{km}$. 
Einstein's summation convention holds for repeated subscripts (unless otherwise noted); Latin subscripts run from 1 to 3 and Greek subscripts from 1 to 2.
For the acoustic situation the vectors and matrix in Eq. (\ref{eq2.1}) are given by
\begin{eqnarray}
&&\hspace{-0.7cm}{\bf q}=\begin{pmatrix} p \\ v_3 \end{pmatrix},\quad
{\bf d}=\begin{pmatrix} \vartheta_{33}^{-1}\vartheta_{3l}f_l \\ \frac{1}{i\omega}\partial_\alpha({ b_{\alpha \beta} f_\beta}) + q \end{pmatrix},\label{eq9992}\\
&&\hspace{-0.7cm}{{\mbox{\boldmath ${\cal A}$}}}= \begin{pmatrix}-\vartheta_{33}^{-1}\vartheta_{3 \beta}\partial_\beta      & i\omega \vartheta_{33}^{-1} \\
                 i\omega\kappa-\frac{1}{i\omega}\partial_\alpha{b_{\alpha\beta}}\partial_\beta & -\partial_\alpha \vartheta_{\alpha 3}\vartheta_{33}^{-1}   \end{pmatrix},\label{eq9993}
\end{eqnarray}
with
\begin{eqnarray}\label{eq9994}
b_{\alpha \beta}=\vartheta_{\alpha \beta}-  \vartheta_{\alpha 3}\vartheta_{33}^{-1} \vartheta_{3\beta},
\end{eqnarray}
where $i$ is the imaginary unit and $q({\bf x},\omega)$ and $f_l({\bf x},\omega)$ are sources in terms of volume-injection rate density and external force density, respectively.

Finally, for an isotropic medium, using $\vartheta_{kl}=\rho^{-1}\delta_{kl}$ \rev{(with $\rho$ the scalar mass density)}, 
we obtain for the source vector and operator matrix \cite{Corones75JMAA, Ursin83GEO, Fishman84JMP, Wapenaar89Book, Hoop96JMP}
\begin{eqnarray}
{\bf d}=\begin{pmatrix}f_3 \\ \frac{1}{i\omega}\partial_\alpha({ \frac{1}{\rho} f_\alpha}) +q \end{pmatrix},\quad
{{\mbox{\boldmath ${\cal A}$}}}= \begin{pmatrix}0      &i\omega \rho \\
               i\omega\kappa-\frac{1}{i\omega}\partial_\alpha\frac{1}{\rho}\partial_\alpha &0    \end{pmatrix}.\label{eq9996ge}
\end{eqnarray}

\subsection{The Green's matrix}

In the space-time domain, a Green's function is the response to an impulsive point source, with the impulse defined as $\delta(t)$. The Fourier transform of $\delta(t)$ equals $1$, 
hence, in the space-frequency domain the Green's function is the response to a point source with unit amplitude for all frequencies. 
We introduce the $N\times N$ Green's matrix ${\bf G}({\bf x},{\bf x}_A,\omega)$ for an \rev{unbounded}, arbitrary inhomogeneous, anisotropic medium   as the solution of
\begin{eqnarray}\label{eq2.1g}
\partial_3{\bf G} -{{\mbox{\boldmath ${\cal A}$}}}{\bf G} ={\bf I}\delta({\bf x}-{\bf x}_A),
\end{eqnarray}
where  ${\bf x}_A=(x_{1,A},x_{2,A},x_{3,A})$ defines the position of the point source and ${\bf I}$ is an $N\times N$ identity matrix.
Here ${\bf I}$ has a size different from that in Eq. (\ref{eq4.3}).
For simplicity we use one notation for differently sized identity matrices (the size always follows  from the context). Also for the zero matrix ${\bf O}$ we use a single notation for 
differently sized matrices.
Similar to operator matrix ${{\mbox{\boldmath ${\cal A}$}}}$, the Green's matrix is partitioned as
\begin{eqnarray}
{\bf G}({\bf x},{\bf x}_A,\omega)= \begin{pmatrix}{\bf G}_{11}      & {\bf G}_{12} \\
                {\bf G}_{21} & {\bf G}_{22}    \end{pmatrix}({\bf x},{\bf x}_A,\omega).\label{eq1213}
\end{eqnarray}
Equation (\ref{eq2.1g}) does not have a unique solution. \rev{To specify a unique solution}, we demand that the time-domain Green's function ${\bf G}({\bf x},{\bf x}_A,t)$ is causal, hence
\begin{eqnarray}
{\bf G}({\bf x},{\bf x}_A,t<0)={\bf O}.\label{eq1014}
\end{eqnarray}
This condition implies that ${\bf G}$ is outward propagating for $|{\bf x}-{\bf x}_A|\to\infty$.

The simplest representation involving the Green's matrix is obtained when ${\bf q}$ and ${\bf G}$ reside in the same medium throughout space and both are outward propagating for $|{\bf x}-{\bf x}_A|\to\infty$.
Whereas ${\bf q}({\bf x},\omega)$ is the response to a source distribution ${\bf d}({\bf x},\omega)$ (Eq. (\ref{eq2.1})), 
${\bf G}({\bf x},{\bf x}_A,\omega)$ is the response to a point source ${\bf I}\delta({\bf x}-{\bf x}_A)$ for an arbitray source position ${\bf x}_A$ (Eq. (\ref{eq2.1g})). 
Since both equations are linear, \rev{a representation for} ${\bf q}({\bf x},\omega)$ follows by applying the superposition principle, according to 
\begin{eqnarray}
{\bf q}({\bf x},\omega)=\int_{{\mathbb{R}}^3} {\bf G}({\bf x},{\bf x}_A,\omega){\bf d}({\bf x}_A,\omega){\rm d}^3{\bf x}_A,\label{eq1115}
\end{eqnarray}
where ${\mathbb{R}}$ is the set of real numbers. This representation is a special case of  more general representations with Green's matrices, derived in a more formal way in section \ref{sec4}.

Next, we discuss the $2\times 2$ acoustic Green's matrix as a special case of the $N\times N$ Green's matrix. 
For this situation ${\bf G}$ is partitioned as
\begin{eqnarray}\label{eq423}
{\bf G}({\bf x},{\bf x}_A,\omega)= \begin{pmatrix}G^{p,f}      & G^{p,q} \\
                G^{v,f} & G^{v,q}    \end{pmatrix}({\bf x},{\bf x}_A,\omega).
\end{eqnarray}
Here the first superscript ($p$ or $v$) refers to the observed wave-field quantity at ${\bf x}$ (acoustic pressure or vertical component of particle velocity), whereas the second superscript
($f$ or $q$) refers to the source type at ${{\bf x}_A}$ (vertical component of force or volume injection rate). 
\rev{The unit of a specific element of the Green's matrix is the ratio of the units of the observed wave-field quantity at ${\bf x}$ and the source quantity at ${{\bf x}_A}$. For example,
$[G^{p,f}({\bf x},{\bf x}_A,\omega)]=[p]/[f]={\rm m}^{-2}$.}
For an isotropic medium, all elements can be expressed in terms of the upper-right element, as follows
\begin{eqnarray}
&&\hspace{-0.7cm}G^{v,q}({\bf x},{\bf x}_A,\omega) = \frac{1}{i\omega\rho({\bf x},\omega)}\partial_3G^{p,q}({\bf x},{\bf x}_A,\omega),\label{eq1124}\\
&&\hspace{-0.7cm}G^{p,f}({\bf x},{\bf x}_A,\omega) = -\frac{1}{i\omega\rho({\bf x}_A,\omega)}\partial_{3,A} G^{p,q}({\bf x},{\bf x}_A,\omega),\label{eq1125}\\
&&\hspace{-0.7cm}G^{v,f}({\bf x},{\bf x}_A,\omega) = \frac{1}{i\omega\rho({\bf x},\omega)}\Bigl(\partial_3G^{p,f}({\bf x},{\bf x}_A,\omega)-\delta({\bf x}-{{\bf x}_A})\Bigr).
\label{eq1126}
\end{eqnarray}
Here $\partial_{3,A}$ stands for differentiation with respect to the source coordinate $x_{3,A}$. 
Equations (\ref{eq1124}) and (\ref{eq1126}) follow directly from Eqs. (\ref{eq9996ge}), (\ref{eq2.1g}) and (\ref{eq423}). 
Equation (\ref{eq1125}) follows from Eq. (\ref{eq1124}) and a source-receiver reciprocity relation, which is derived in section \ref{sec4.1}. 

We illustrate $G^{p,q}$, decomposed into plane waves, for a horizontally layered lossless isotropic medium. 
To this end, we first define the spatial Fourier transform of a space- and frequency-dependent function $u({\bf x},\omega)$ along the horizontal coordinates ${{\bf x}_{\rm H}}=(x_1,x_2)$,
according to
\begin{eqnarray}
&&\hspace{-0.9cm}\tilde u({{\bf s}},x_3,\omega)=\int_{{\mathbb{R}}^2}\exp\{-i\omega{{\bf s}}\cdot{{\bf x}_{\rm H}}\}u({{\bf x}_{\rm H}},x_3,\omega){\rm d}^2{{\bf x}_{\rm H}},\label{eq99950b}
\end{eqnarray}
with ${{\bf s}}=(s_1,s_2)$, where $s_1$ and $s_2$ are horizontal slownesses.
This transform decomposes the function $u({\bf x},\omega)$ \rev{at a given depth level $x_3$} into monochromatic plane-wave components. 
Next, we define the inverse temporal Fourier transform, per slowness value ${{\bf s}}$, as
\begin{eqnarray}
&&\hspace{-.9cm}u({{\bf s}},x_3,\tau)=
\frac{1}{\pi}\Re\int_0^\infty \tilde u({{\bf s}},x_3,\omega)\exp\{-i\omega\tau\rev{\}}{\rm d}\omega,\label{eq500a}
\end{eqnarray}
where $\Re$ denotes the real part and $\tau$ is the intercept time \cite{Stoffa89Book}.
We apply these transforms to the Green's function $G^{p,q}({\bf x},{\bf x}_A,\omega)$ and the source function $\delta({\bf x}-{\bf x}_A)$, choosing
${\bf x}_A=(0,0,{x_{3,0}})$ and setting $s_2=0$  (the field is cylindrically symmetric in the considered horizontally layered isotropic medium). We thus obtain 
$G^{p,q}(s_1,x_3,{x_{3,0}},\tau)$, which is the plane-wave response (as a function of $x_3$ and $\tau$) to a source function
$\delta(x_3-{x_{3,0}})\delta(\tau)$.

\begin{figure}
\vspace{0cm}
\centerline{\epsfysize=5.8 cm \epsfbox{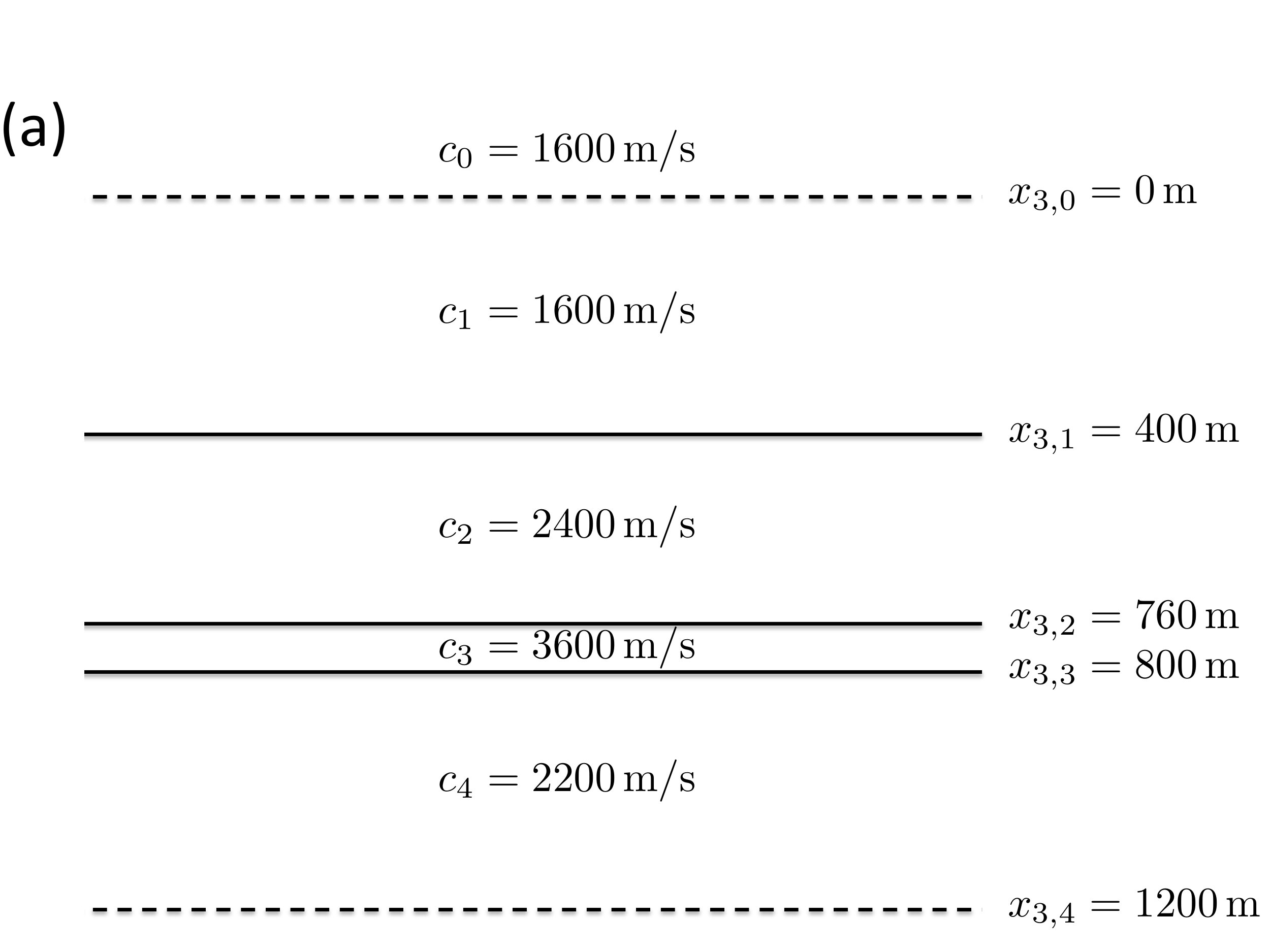}}
\vspace{.5cm}\centerline{\epsfysize=5.8 cm \epsfbox{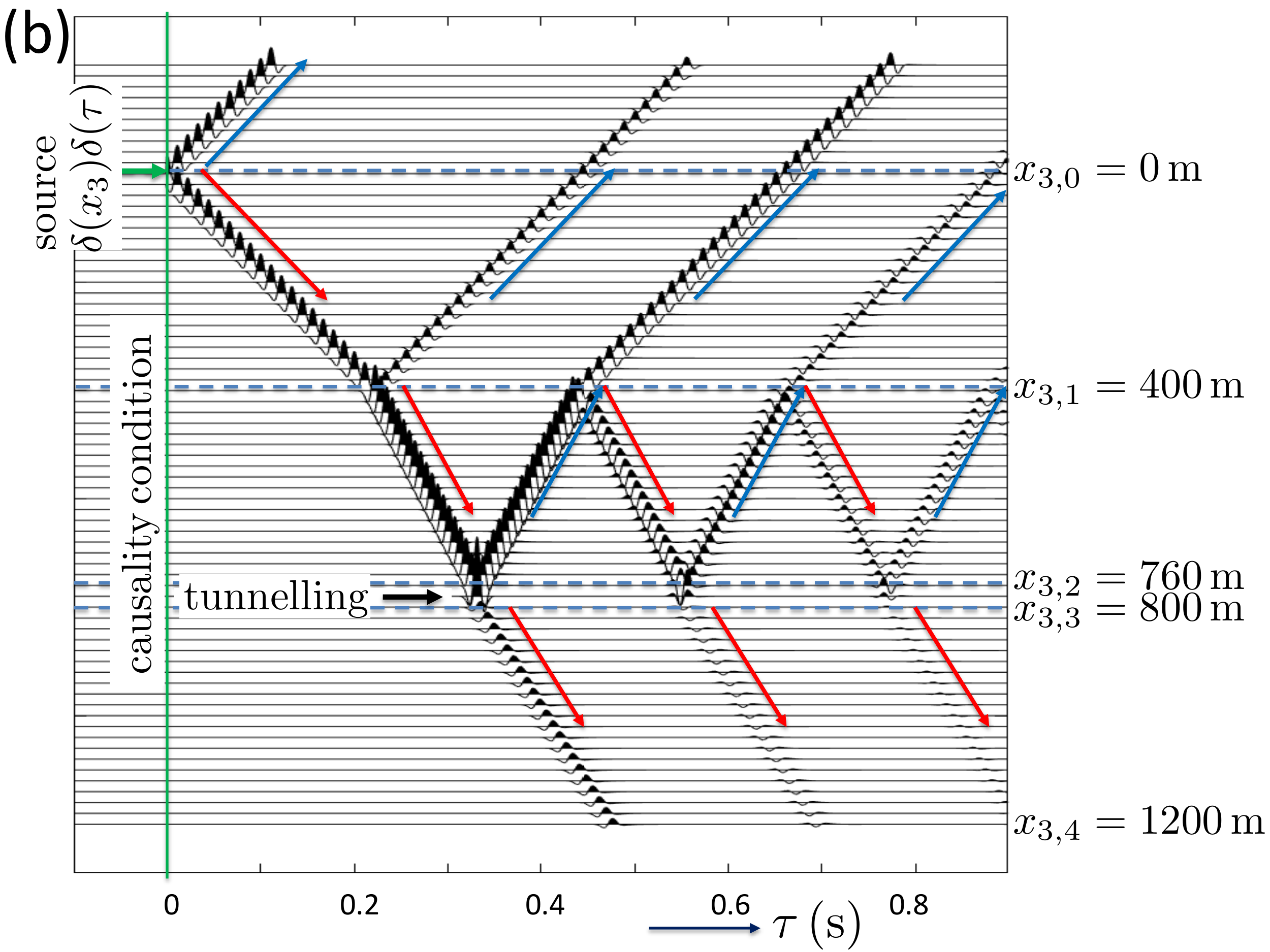}}
\vspace{-0cm}
\caption{ (a) Horizontally layered medium. (b) Green's function $G^{p,q}(s_1,x_3,{x_{3,0}},\tau)$ (fixed $s_1$), convolved with a wavelet. 
}\label{Figure1}
\end{figure}

An example horizontally layered medium is shown in Fig. \ref{Figure1}(a). The propagation velocities in the layers are indicated by $c_n$ (with $c=1/\sqrt{\kappa\rho}$). 
The depth level of the source is chosen as ${x_{3,0}}=0$ m, see Fig. \ref{Figure1}(b). The vertical green line in this figure 
is the line $\tau=0$, left of which the field is zero due to the causality condition (similar as in Eq. (\ref{eq1014})).
Figure \ref{Figure1}(b) further shows the numerically modelled Green's function $G^{p,q}(s_1,x_3,{x_{3,0}},\tau)$ 
as a function of $x_3$ and $\tau$, for a single horizontal slowness $s_1=1/3500$ s/m (red and blue arrows indicate downgoing and upgoing waves). 
This is the wave field that would be measured by a series of  acoustic pressure receivers, vertically above and below the 
volume-injection rate source at ${x_{3,0}}=0$ m. 
Each trace shows $G^{p,q}(s_1,x_3,{x_{3,0}},\tau)$  for a specific depth $x_3$ as a function 
of $\tau$ (actually, at each depth the Green's function has  been convolved with a time-symmetric wavelet, with a central frequency of 50 Hz, to get a nicer display). 

The horizontal slowness $s_1$ is related to the propagation angle $\alpha_n$ \rev{in layer $n$} via $\sin\alpha_n=s_1c_n$, hence, in layer 1 (with $c_1=1600$ m/s) the propagation angle is $\alpha_1=27.2^o$. 
In the thin layer, with $c_3=3600$ m/s, we obtain $\sin\alpha_3=1.03$,
meaning that $\alpha_3$ is complex-valued. This implies that the wave is evanescent in this layer. Since the layer is thin, the wave tunnels through the layer and continues with a lower amplitude
as a downgoing wave in the lower half-space. 

\subsection{The homogeneous Green's matrix}

For a lossless medium, a homogeneous Green's function is the superposition of a Green's function and its complex conjugate (or, in the time domain, its time-reversed version).
The superposition is chosen such that the source terms of the two functions cancel each other, hence, 
a homogeneous Green's function obeys a wave equation without a source term  \cite{Porter70JOSA, Oristaglio89IP}.
Here we extend this concept for the matrix-vector wave equation for a medium with losses.

Let ${\bf G}({\bf x},{\bf x}_A,\omega)$ be again the outward propagating solution of Eq. (\ref{eq2.1g}) for an \rev{unbounded}, arbitrary inhomogeneous, anisotropic medium.
We introduce the Green's matrix of the adjoint medium, $\bar{\bf G}({\bf x},{\bf x}_A,\omega)$, 
as the outward propagating solution of
\begin{eqnarray}\label{eq2.1gadj}
\partial_3\bar{\bf G} -{{\,\,\,\bar{\!\!\!{{\mbox{\boldmath ${\cal A}$}}}}}}\bar{\bf G} ={\bf I}\delta({\bf x}-{\bf x}_A).
\end{eqnarray}
Pre- and post multiplying all terms by ${\bf J}$ and subsequently using Eq. (\ref{eqsymcon}) and ${\bf J}{\bf J}={\bf I}$ gives 
\begin{eqnarray}\label{eq2.1gadjb}
\partial_3{\bf J}\bar{\bf G}{\bf J} -{{\mbox{\boldmath ${\cal A}$}}}^*{\bf J}\bar{\bf G}{\bf J} ={\bf I}\delta({\bf x}-{\bf x}_A).
\end{eqnarray}
Subtracting the complex conjugate of all terms in this equation from the corresponding terms in Eq. (\ref{eq2.1g}) yields
\begin{eqnarray}\label{eq2.1gadjd}
\partial_3{\bf G}_{\rm h} -{{\mbox{\boldmath ${\cal A}$}}}{\bf G}_{\rm h}={\bf O},
\end{eqnarray}
with
\begin{eqnarray}\label{eq750}
{\bf G}_{\rm h}({\bf x},{\bf x}_A,\omega)={\bf G}({\bf x},{\bf x}_A,\omega)-{\bf J}\bar{\bf G}^*({\bf x},{\bf x}_A,\omega){\bf J}.
\end{eqnarray}
Since ${\bf G}_{\rm h}({\bf x},{\bf x}_A,\omega)$ obeys a  wave equation without a source term, we call it the homogeneous Green's matrix.
According to Eqs. (\ref{eq4.3}), (\ref{eq1213}) and (\ref{eq750}), it is partitioned as
\begin{eqnarray}
&&\hspace{-0.7cm}{\bf G}_{\rm h}({\bf x},{\bf x}_A,\omega)= \begin{pmatrix}\{{\bf G}_{11}-\bar{\bf G}_{11}^*\}      & \{{\bf G}_{12}+\bar{\bf G}_{12}^*\}\\
                \{{\bf G}_{21}+\bar{\bf G}_{21}^*\} & \{{\bf G}_{22}-\bar{\bf G}_{22}^*\}   \end{pmatrix}({\bf x},{\bf x}_A,\omega).
\label{eq1213h}
\end{eqnarray}

\subsection{The propagator matrix}
We introduce the $N\times N$ propagator matrix ${\bf W}({\bf x},{\bf x}_A,\omega)$ for an arbitrary inhomogeneous anisotropic medium
as the solution of the unified matrix-vector wave equation (\ref{eq2.1}), but without the source vector 
${\bf d}$. Hence,
\begin{eqnarray}\label{eq2.1gw}
\partial_3{\bf W} -{{\mbox{\boldmath ${\cal A}$}}}{\bf W}={\bf O}.
\end{eqnarray}
Similar to operator matrix ${{\mbox{\boldmath ${\cal A}$}}}$ and Green's matrix ${\bf G}$, the propagator matrix is partitioned as
\begin{eqnarray}
{\bf W}({\bf x},{\bf x}_A,\omega)= \begin{pmatrix}{\bf W}_{11}      & {\bf W}_{12} \\
                {\bf W}_{21} & {\bf W}_{22}    \end{pmatrix}({\bf x},{\bf x}_A,\omega).\label{eq10W}
\end{eqnarray}
Equation (\ref{eq2.1gw}) does not have a unique solution. \rev{To specify a unique solution}, we impose the  boundary condition
\begin{eqnarray}
{\bf W}({\bf x},{{\bf x}_A},\omega)|_{x_3=x_{3,A}} = {\bf I}\delta({{\bf x}_{\rm H}}-{{\bf x}_{{\rm H},A}}),\label{eq9998d}
\end{eqnarray}
where ${{\bf x}_{{\rm H},A}}$ denotes the horizontal coordinates of ${{\bf x}_A}$, hence, ${{\bf x}_{{\rm H},A}}=(x_{1,A},x_{2,A})$.  
Since Eq. (\ref{eq2.1gw}) is  first order in $\partial_3$, a single boundary condition suffices. 
Note that ${\bf W}({\bf x},{{\bf x}_A},\omega)$ only depends on the medium parameters between depth levels $x_{3,A}$ and $x_3$. 
This is different from the Green's matrix ${\bf G}({\bf x},{\bf x}_A,\omega)$, which, for an arbitrary inhomogeneous medium, depends on the entire medium (this is easily understood from the illustration in Fig. \ref{Figure1}(b)). 

The simplest representation involving the propagator matrix is obtained when ${\bf q}$ and ${\bf W}$ reside in the same medium in the region between $x_{3,A}$ and $x_3$ and both have no sources in this region.
Whereas ${\bf q}({\bf x},\omega)$ obeys no boundary conditions in this region,  ${\bf W}({\bf x},{{\bf x}_A},\omega)$ collapses to $ {\bf I}\delta({{\bf x}_{\rm H}}-{{\bf x}_{{\rm H},A}})$  at depth level $x_{3,A}$
(Eq. (\ref{eq9998d})). Applying the superposition principle again yields
\begin{eqnarray}
{\bf q}({\bf x},\omega)=\int_{{{\partial\mathbb{D}}}_A} {\bf W}({\bf x},{{\bf x}_A},\omega){\bf q}({{\bf x}_A},\omega){\rm d}^2{{\bf x}_A},\label{eq1330}
\end{eqnarray}
where 
${{\partial\mathbb{D}}}_A$ is the horizontal boundary defined as $x_3=x_{3,A}$ \cite{Gilbert66GEO, Kennett72BS, Kennett72GJRAS, Woodhouse74GJR, Haines88GJI}. 
Note that ${\bf W}({\bf x},{{\bf x}_A},\omega)$ `propagates' the field vector ${\bf q}$ from depth level $x_{3,A}$ to $x_3$, hence the name `propagator matrix'.
This representation is a special case of more general representations with propagator matrices, derived in a more formal way in section \ref{sec5}.

When we replace ${\bf q}({{\bf x}_A},\omega)$ by a Green's matrix ${\bf G}({{\bf x}_A},{{\bf x}_B},\omega)$, with ${{\bf x}_B}$ outside the region between $x_{3,A}$ and $x_3$, we obtain
\begin{eqnarray}
{\bf G}({\bf x},{{\bf x}_B},\omega)=\int_{{{\partial\mathbb{D}}}_A} {\bf W}({\bf x},{{\bf x}_A},\omega){\bf G}({{\bf x}_A},{{\bf x}_B},\omega){\rm d}^2{{\bf x}_A}.\label{eq1331}
\end{eqnarray}
This is the simplest relation between the Green's matrix and the propagator matrix. It is a special case of more general representations with Green's matrices and propagator matrices, 
derived in a more formal way in section \ref{sec6.1}.
Alternatively, we may replace ${\bf G}({\bf x},{{\bf x}_B},\omega)$ by the homogeneous Green's matrix ${\bf G}_{\rm h}({\bf x},{{\bf x}_B},\omega)$, where ${{\bf x}_B}$ may be located anywhere, since the 
homogeneous Green's function has no source at ${{\bf x}_B}$, hence
\begin{eqnarray}
{\bf G}_{\rm h}({\bf x},{{\bf x}_B},\omega)=\int_{{{\partial\mathbb{D}}}_A} {\bf W}({\bf x},{{\bf x}_A},\omega){\bf G}_{\rm h}({{\bf x}_A},{{\bf x}_B},\omega){\rm d}^2{{\bf x}_A}.\label{eq1332}
\end{eqnarray}
This relation will be derived  in a more formal way in section \ref{sec6.2}.

Next, we discuss the $2\times 2$ acoustic propagator matrix as a special case of the $N\times N$ propagator matrix. 
For this situation ${\bf W}$ is partitioned as
\begin{eqnarray}\label{eq424}
{\bf W}({\bf x},{\bf x}_A,\omega)= \begin{pmatrix}W^{p,p}      & W^{p,v} \\
                W^{v,p} & W^{v,v}    \end{pmatrix}({\bf x},{\bf x}_A,\omega).
\end{eqnarray}
The first and second superscripts refer to the wave-field quantities at ${\bf x}$ and ${\bf x}_A$, respectively (with superscript $p$ again standing for acoustic pressure and $v$ for the
 vertical component of particle velocity).
 \rev{The unit of a specific element of the propagator matrix is the ratio of the units of the wave-field quantities at ${\bf x}$ and ${{\bf x}_A}$. For example,
$[W^{p,v}({\bf x},{\bf x}_A,\omega)]=[p]/[v]={\rm kg\,s}^{-1}{\rm m}^{-2}$.}
For an isotropic medium, the elements can be expressed in terms of the upper-right element, as follows
\begin{eqnarray}
&&\hspace{-0.9cm}W^{v,v}({\bf x},{\bf x}_A,\omega) = \frac{1}{i\omega\rho({\bf x},\omega)}\partial_3W^{p,v}({\bf x},{\bf x}_A,\omega),\label{eq1124W}\\
&&\hspace{-0.9cm}W^{p,p}({\bf x},{\bf x}_A,\omega) = -\frac{1}{i\omega\rho({\bf x}_A,\omega)}\partial_{3,A} W^{p,v}({\bf x},{\bf x}_A,\omega),\label{eq1125W}\\
&&\hspace{-0.9cm}W^{v,p}({\bf x},{\bf x}_A,\omega) = \frac{1}{i\omega\rho({\bf x},\omega)}\partial_3W^{p,p}({\bf x},{\bf x}_A,\omega).\label{eq1126W}
\end{eqnarray}
Equations (\ref{eq1124W}) and (\ref{eq1126W}) follow directly from Eqs. (\ref{eq9996ge}), (\ref{eq2.1gw}) and (\ref{eq424}). Equation (\ref{eq1125W}) follows from Eqs. (\ref{eq1124W}) and 
a source-receiver reciprocity relation, which is derived in section \ref{sec5.1}.

\begin{figure}
\vspace{0cm}
\centerline{\epsfysize=5.8 cm \epsfbox{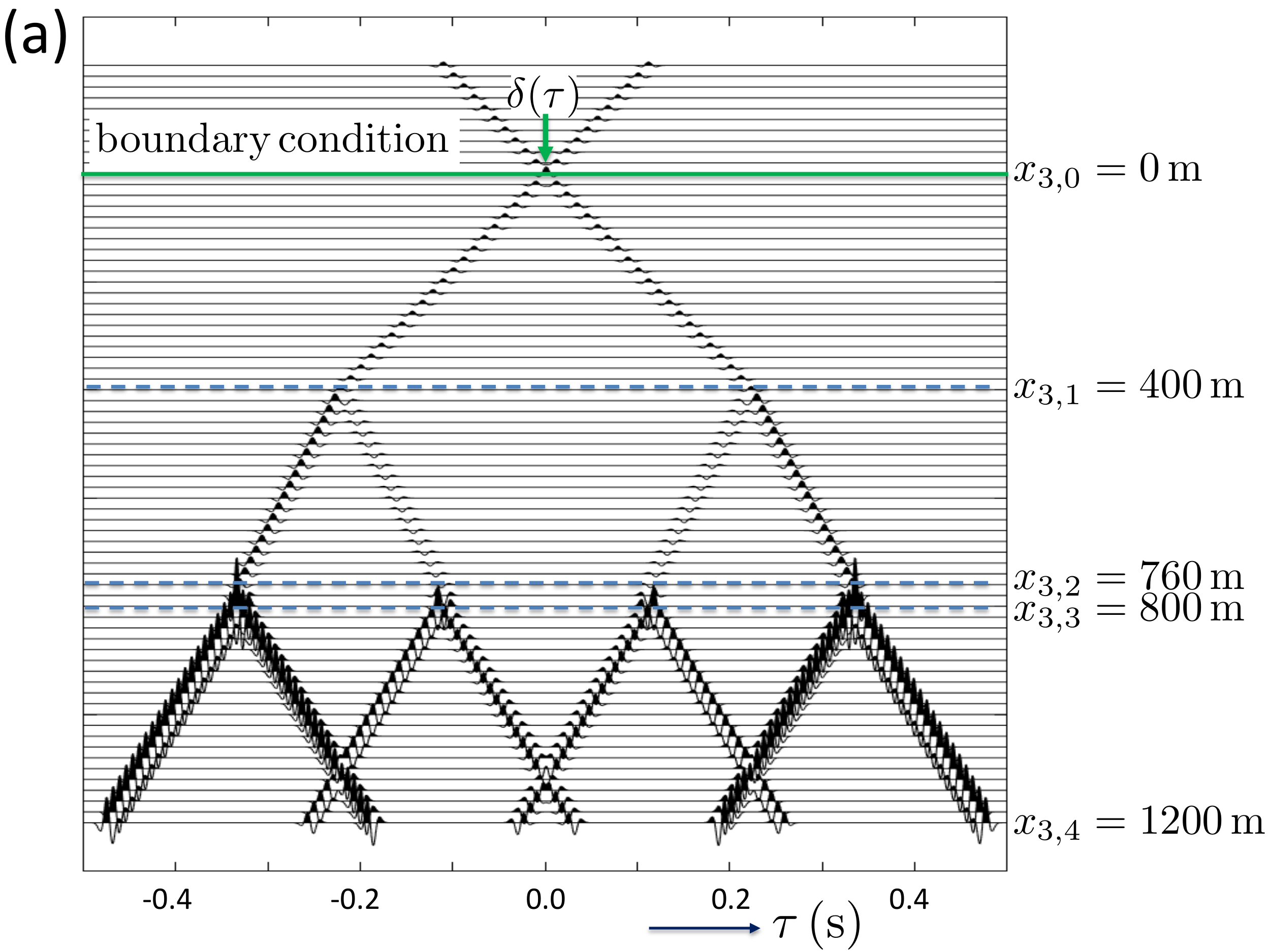}}
\centerline{\epsfysize=5.8 cm \epsfbox{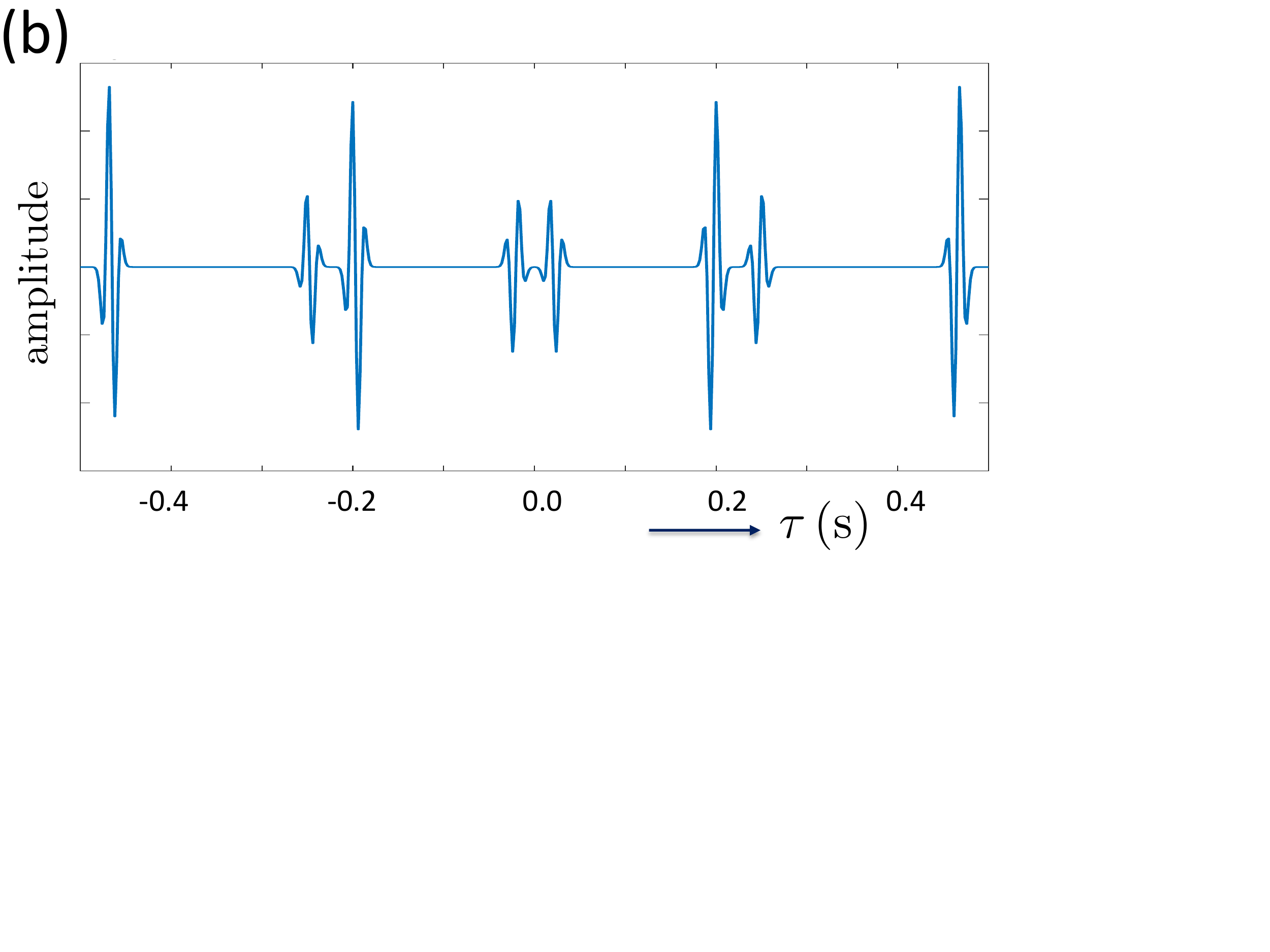}}\vspace{-2.4cm}
\caption{ (a) Propagator element $W^{p,p}(s_1,x_3,{x_{3,0}},\tau)$ (fixed $s_1$), convolved with a wavelet. (b) Last trace of (a).
}\label{Figure2}
\end{figure}

\begin{figure}
\vspace{0cm}
\centerline{\epsfysize=5.8 cm \epsfbox{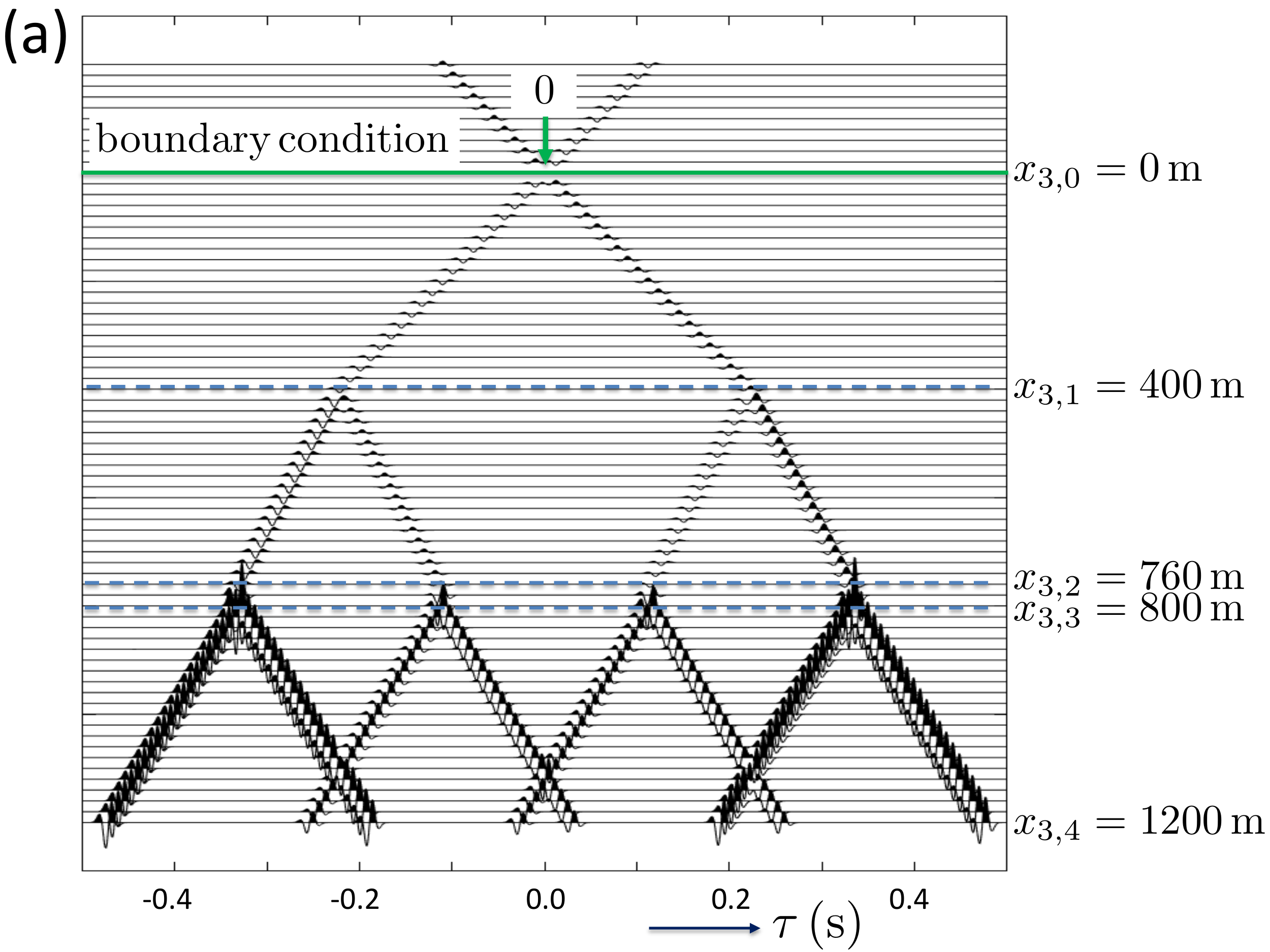}}
\centerline{\epsfysize=5.8 cm \epsfbox{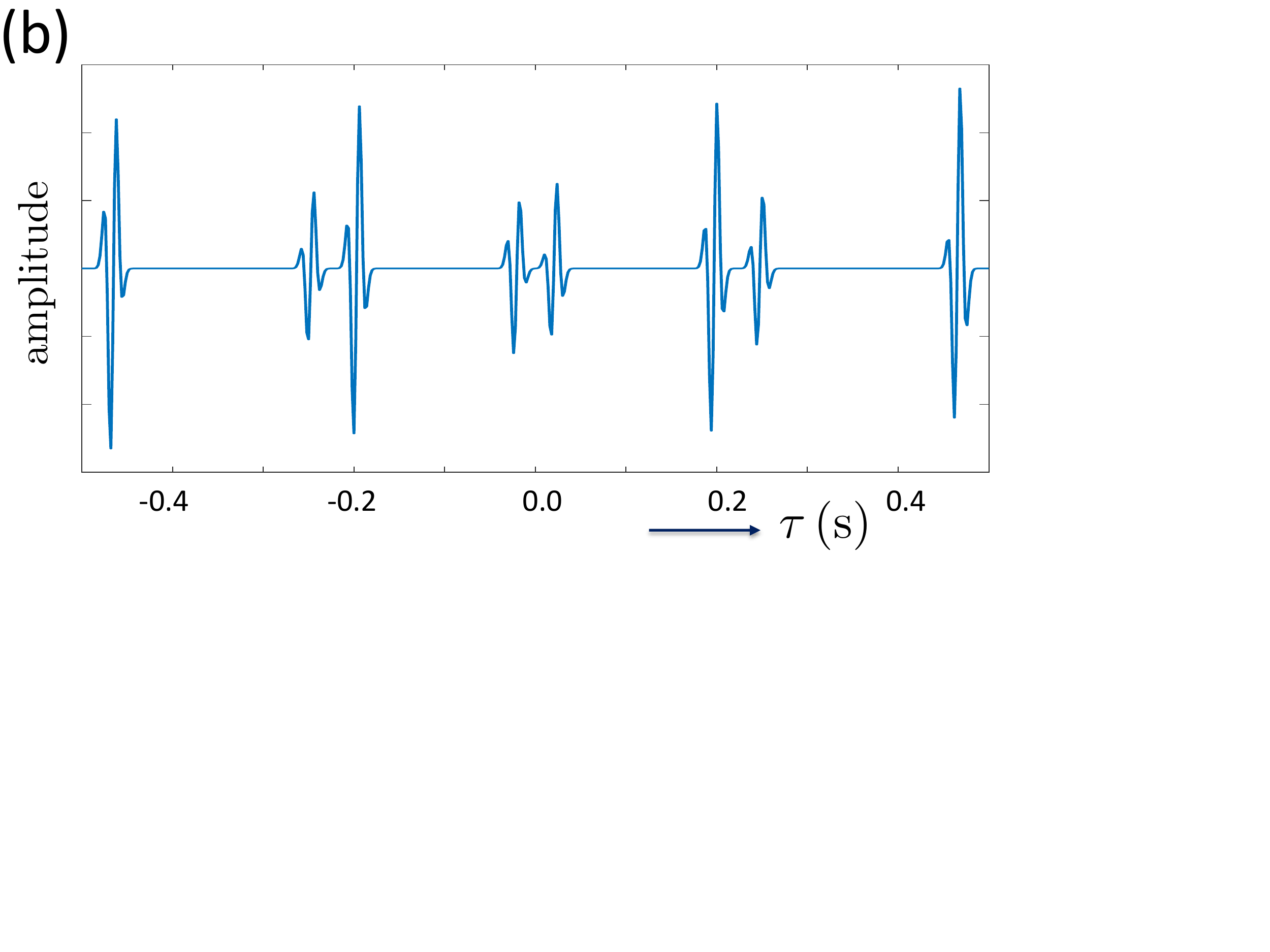}}\vspace{-2.4cm}
\caption{ (a) Propagator element $W^{p,v}(s_1,x_3,{x_{3,0}},\tau)$ (fixed $s_1$), convolved with a wavelet. (b) Last trace of (a).
}\label{Figure3}
\end{figure}

We illustrate the elements $W^{p,p}$ and $W^{p,v}$, decomposed into plane waves, for the horizontally layered medium of Fig. \ref{Figure1}(a).
We choose again ${\bf x}_A=(0,0,{x_{3,0}})$ and set $s_2=0$.
Usually the propagator matrix is considered in the frequency domain, but to facilitate the comparison with the
Green's function in Fig. \ref{Figure1}(b), we consider the time-domain functions $W^{p,p}(s_1,x_3,{x_{3,0}},\tau)$ and $W^{p,v}(s_1,x_3,{x_{3,0}},\tau)$
(obtained via the transforms of Eqs. (\ref{eq99950b}) and (\ref{eq500a})), again for
a single horizontal slowness $s_1=1/3500$ s/m, see Figs. \ref{Figure2}(a) and  \ref{Figure3}(a). At $x_3={x_{3,0}}$ (the horizontal green lines in these figures) 
the boundary conditions for these functions are $W^{p,p}(s_1,{x_{3,0}},{x_{3,0}},\tau)=\delta(\tau)$
and $W^{p,v}(s_1,{x_{3,0}},{x_{3,0}},\tau)=0$, respectively (this follows from applying the transforms of Eqs. (\ref{eq99950b}) and (\ref{eq500a}) to Eq. (\ref{eq9998d}), with ${{\bf x}_{{\rm H},A}}=(0,0)$).
Following these functions along the depth coordinate, starting at $x_3={x_{3,0}}$, we observe an acausal upgoing wave and 
a causal downgoing wave until we reach the first interface. Here both events split into upgoing and downgoing waves below the interface.
The waves tunnel through the high-velocity layer, split again, and continue with higher amplitudes in the next layer. 
This illustrates that evanescent waves may lead to unstable behaviour of the propagator matrix and should be handled with care.
Note that $W^{p,p}(s_1,x_3,{x_{3,0}},\tau)$ and $W^{p,v}(s_1,x_3,{x_{3,0}},\tau)$ are symmetric and asymmetric in time, 
respectively. This is best seen in Figs. \ref{Figure2}(b) and \ref{Figure3}(b), which show the last trace of both elements. The same holds for elements $W^{v,v}$ and $W^{v,p}$, which are not shown.

\subsection{The Marchenko-type focusing function}\label{sec2.5}

\begin{figure}
\vspace{0cm}
\centerline{\epsfysize=5.8 cm \epsfbox{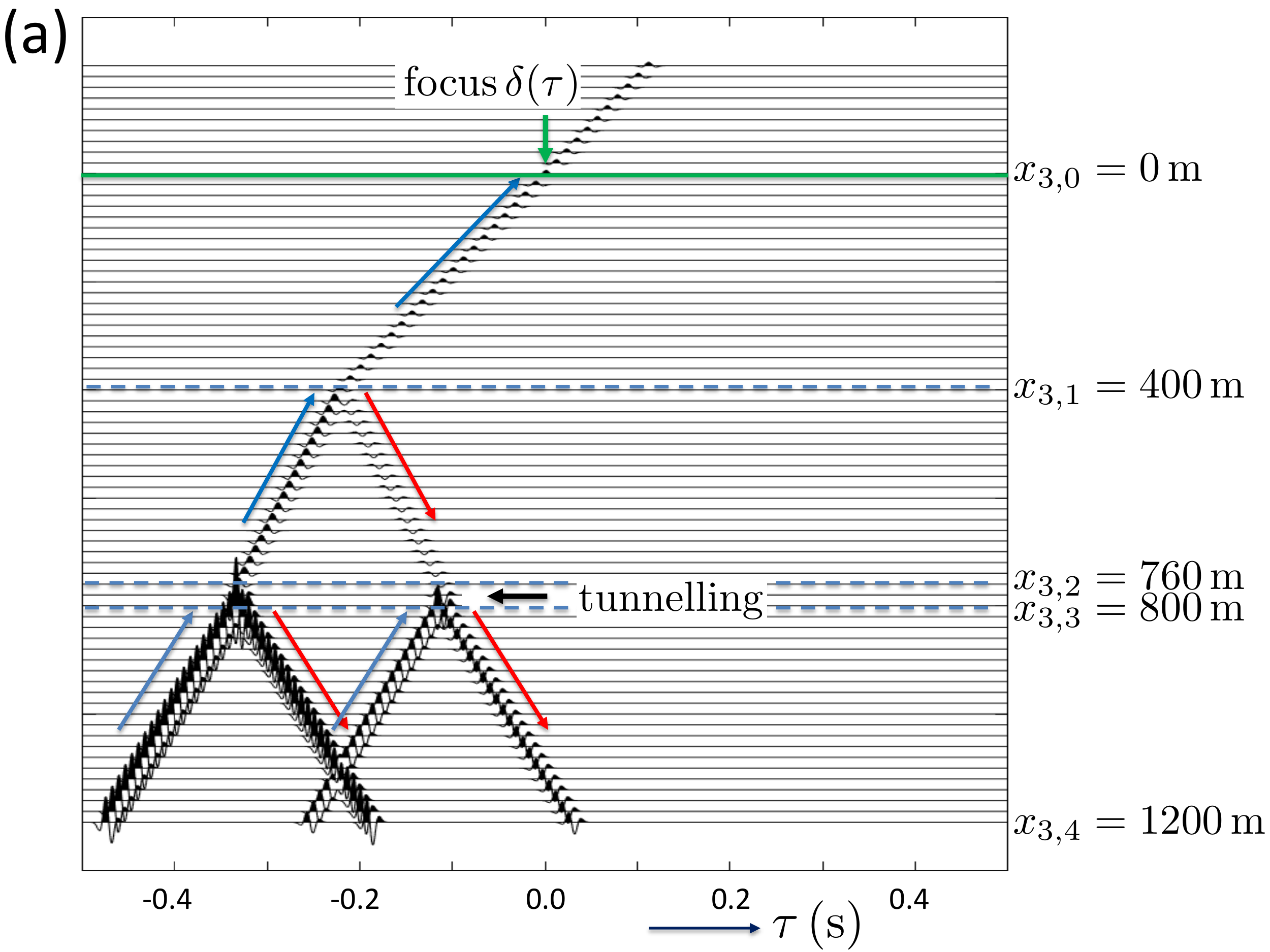}}
\centerline{\epsfysize=5.8 cm \epsfbox{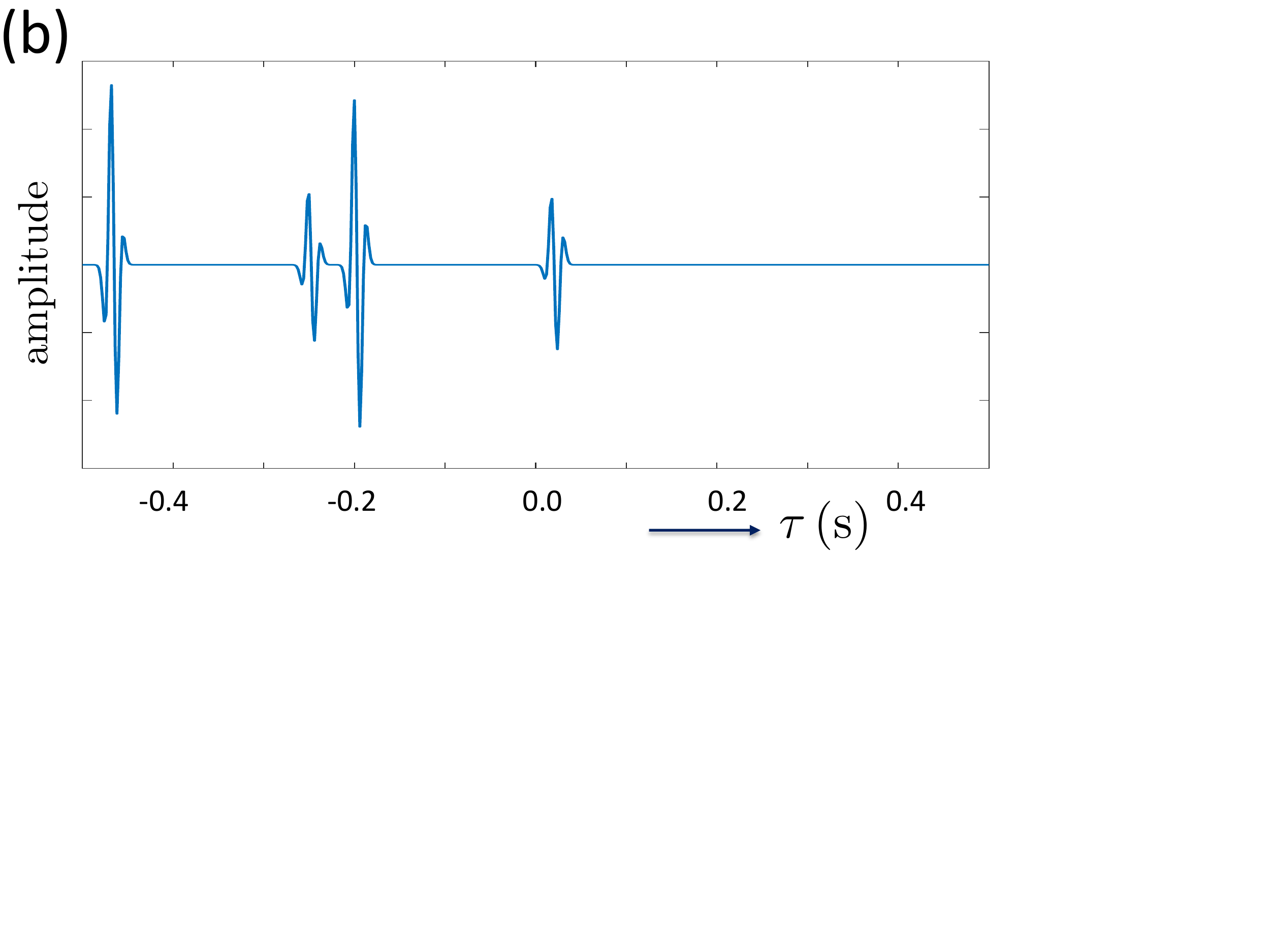}}\vspace{-2.4cm}
\caption{ (a) Focusing function $F^p(s_1,x_3,{x_{3,0}},\tau)$ (fixed $s_1$), convolved with a wavelet.  (b) Last trace of (a).
}\label{Figure7}
\end{figure}

For an arbitrary inhomogeneous anisotropic medium, the time-domain version of boundary condition  (\ref{eq9998d})  reads
\begin{eqnarray}
{\bf W}({\bf x},{{\bf x}_A},t)|_{x_3=x_{3,A}} = {\bf I}\delta({{\bf x}_{\rm H}}-{{\bf x}_{{\rm H},A}})\delta(t).\label{eq9998dtd}
\end{eqnarray}
This boundary condition has a similar form as the focusing condition for the focusing functions appearing in the multidimensional Marchenko method \cite{Wapenaar2014JASA}.
In section \ref{sec7.2} we discuss the general relations between the propagator matrix and the Marchenko-type focusing functions.
Here we present a short preview of these relations by considering the acoustic propagator matrix in the horizontally layered lossless isotropic medium of Fig. \ref{Figure1}(a).
We combine the elements $W^{p,p}$ and $W^{p,v}$, decomposed into plane waves (Figs. \ref{Figure2} and \ref{Figure3}), as follows \cite{Wapenaar2021GEO}
\begin{eqnarray}
&&\hspace{-0.99cm}F^p(s_1,x_3,{x_{3,0}},\tau)=
W^{p,p}(s_1,x_3,{x_{3,0}},\tau) - \frac{s_{3,0}}{\rho_0}W^{p,v}(s_1,x_3,{x_{3,0}},\tau),\label{eq1135}
\end{eqnarray}
where the vertical slowness $s_{3,0}$ is defined as
\begin{eqnarray}
s_{3,0}=\sqrt{1/c_0^2-s_1^2},
\end{eqnarray}
with $c_0$ and $\rho_0$ being the propagation velocity and mass density of the upper half-space \rev{$x_3\le x_{3,0}$}. Due to the symmetric form of $W^{p,p}$ and the asymmetric form of $W^{p,v}$,
half of the events double in amplitude  and the other half of the events cancel. The result is shown in Fig. \ref{Figure7}.
For the interpretation of this focusing function we start at the bottom of Fig. \ref{Figure7}(a). The blue arrows indicate upgoing waves, which are tuned such that, 
after interaction with the tunneling waves in the thin layer and the downgoing wave just above the thin layer (indicated by a red arrow), 
they continue as a single upgoing wave, which finally focuses at depth level ${x_{3,0}}$ as a temporal delta function $\delta(\tau)$ and continues as an upgoing
wave into the homogeneous upper half-space. 

Note that, although we interpret the focusing function here in terms of upgoing and  downgoing waves, it is derived from the propagator matrix 
(which does not rely on up/down decomposition) and the vertical slowness
$s_{3,0}$ and mass density $\rho_0$ of the upper half-space. Moreover, the focusing function is defined in the actual medium rather than in a truncated version of the actual medium, 
as is usually the case for Marchenko-type focusing functions \cite{Wapenaar2014JASA}. 
Hence, the only assumption is that a real-valued vertical slowness $s_{3,0}$ exists in the upper half-space. Other than that, 
the focusing function $F^p(s_1,x_3,{x_{3,0}},\tau)$, as defined in Eq. (\ref{eq1135}),
does not require a truncated medium, does not rely on up/down decomposition inside the medium, and does (at least in principle) 
not break down when waves become evanescent inside the medium. 
These properties also hold for the more general version of the Marchenko-type focusing functions defined in section \ref{sec7}.

\section{Unified matrix-vector wave-field reciprocity theorems}\label{sec3}

\begin{figure}
\vspace{0cm}
\centerline{\epsfysize=6.8 cm \epsfbox{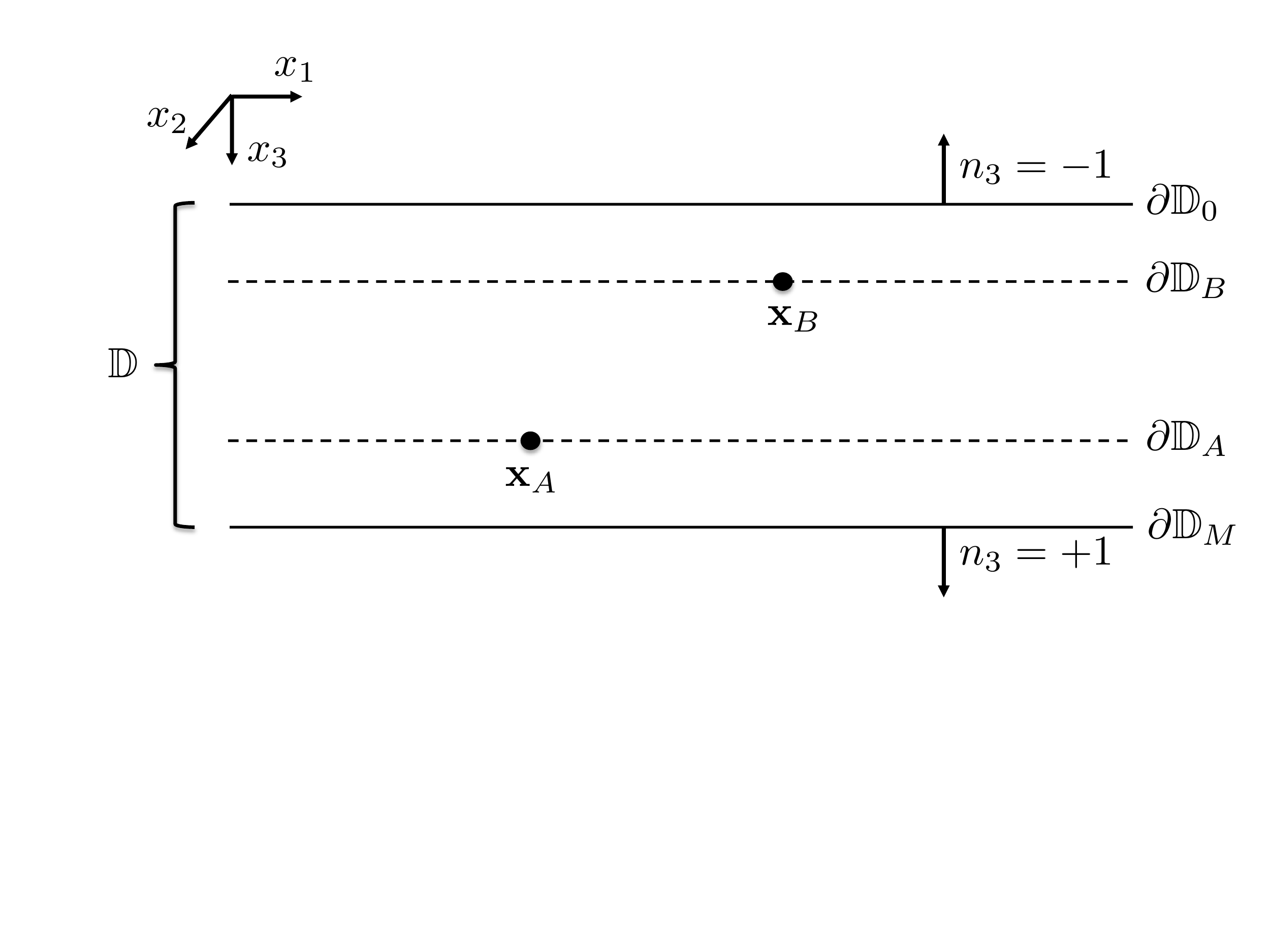}}\vspace{-2cm}
\caption{Configuration for the matrix-vector reciprocity theorems, Eqs. (\ref{eq4.1}) and (\ref{eq4.2}) and for the representations with Green's matrices.
}\label{Fig1}
\end{figure}

As the basis for the derivation of the general representations with Green's matrices (section \ref{sec4}), propagator matrices (section \ref{sec5}), 
or a combination thereof (section \ref{sec6}),
we introduce unified matrix-vector wave-field reciprocity theorems. In general, a wave-field reciprocity theorem interrelates two wave states 
(sources, wave fields and medium parameters) in the same spatial domain
\cite{Hoop95Book}. Reciprocity theorems have been formulated for acoustic  \cite{Rayleigh78Book}, 
electromagnetic  \cite{Lorentz1895KNAW}, elastodynamic  \cite{Knopoff59GEO, Hoop66ASR}, 
poroelastodynamic  \cite{Flekkoy99PRE}, piezoelectric  \cite{Auld79WM, Achenbach2003Book} and seismoelectric waves \cite{Pride96JASA}.
The matrix-vector equation discussed in section \ref{sec2} allows a unified formulation of the reciprocity theorems for
 these different wave phenomena \cite{Wapenaar96JASA, Haines96JMP},
which extends the theory of propagation invariants \cite{Haines88GJI, Kennett90GJI, Koketsu91GJI, Takenaka93WM}. 
Using the symmetry properties of operator matrix ${{\mbox{\boldmath ${\cal A}$}}}$, formulated in Eqs. (\ref{eqsym}) and (\ref{eqsymad}), 
the following matrix-vector reciprocity theorems can be derived  \cite{Wapenaar96JASA, Haines96JMP}
\begin{eqnarray}\label{eq4.1}
&&\hspace{-0.7cm}\int_{\mathbb{D}}\bigl({\bf d}_A^t{\bf N}{\bf q}_B +{\bf q}_A^t{\bf N}{\bf d}_B \bigr){\rm d}^3{\bf x}=
\int_{{{\partial\mathbb{D}}}_0\cup{{\partial\mathbb{D}}_M}}{\bf q}_A^t{\bf N}{\bf q}_B n_3{\rm d}^2{\bf x}+\int_{\mathbb{D}}{\bf q}_A^t{\bf N}({{\mbox{\boldmath ${\cal A}$}}}_A-{{\mbox{\boldmath ${\cal A}$}}}_B){\bf q}_B {\rm d}^3{\bf x}
\end{eqnarray}
and
\begin{eqnarray}\label{eq4.2}
&&\hspace{-0.7cm}\int_{\mathbb{D}}\bigl({\bf d}_A^\dagger{\bf K}{\bf q}_B + {\bf q}_A^\dagger{\bf K}{\bf d}_B\bigr){\rm d}^3{\bf x}=
\int_{{{\partial\mathbb{D}}}_0\cup{{\partial\mathbb{D}}_M}}{\bf q}_A^\dagger{\bf K}{\bf q}_B n_3{\rm d}^2{\bf x}+\int_{\mathbb{D}}{\bf q}_A^\dagger{\bf K}({{\,\,\,\bar{\!\!\!{{\mbox{\boldmath ${\cal A}$}}}}}}_A-{{\mbox{\boldmath ${\cal A}$}}}_B){\bf q}_B {\rm d}^3{\bf x}.
\end{eqnarray}
Here ${\mathbb{D}}$ denotes a domain enclosed by two infinite horizontal boundaries ${{\partial\mathbb{D}}}_0$ and ${{\partial\mathbb{D}}_M}$ at depth levels ${x_{3,0}}$ and ${x_{3,M}}$ with outward pointing
normals $n_3=-1$ and $n_3=1$, respectively, see Fig. \ref{Fig1}. Subscripts $A$ and $B$ refer to  two independent states. 
These theorems hold for lossless media and for media with losses \cite{Wapenaar2019GJI}.
Equation (\ref{eq4.1}) is a convolution-type reciprocity theorem, since products like ${\bf q}_A^t{\bf N}{\bf q}_B$ in the frequency domain correspond to convolutions in the time domain (like in reference \cite{Hoop66ASR}).
Equation (\ref{eq4.2}) is a correlation-type reciprocity theorem, since products like ${\bf q}_A^\dagger{\bf K}{\bf q}_B$ in the frequency domain correspond to correlations in the time domain (like in reference \cite{Bojarski83JASA}).

A special case is obtained when the sources, wave fields and medium parameters are identical in both states. We may then drop the subscripts $A$ and $B$ and Eq. (\ref{eq4.2}) simplifies to
\begin{eqnarray}\label{eq4.3pp}
&&\hspace{-0.99cm}\int_{\mathbb{D}}{\frac{1}{4}}\bigl({\bf d}^\dagger{\bf K}{\bf q} + {\bf q}^\dagger{\bf K}{\bf d}\bigr){\rm d}^3{\bf x}=
\int_{{{\partial\mathbb{D}}}_0\cup{{\partial\mathbb{D}}_M}}{\frac{1}{4}}{\bf q}^\dagger{\bf K}{\bf q} n_3{\rm d}^2{\bf x}+\int_{\mathbb{D}}{\frac{1}{4}}{\bf q}^\dagger{\bf K}({{\,\,\,\bar{\!\!\!{{\mbox{\boldmath ${\cal A}$}}}}}}-{{\mbox{\boldmath ${\cal A}$}}}){\bf q} {\rm d}^3{\bf x}.
\end{eqnarray}
Since ${\frac{1}{4}}{\bf q}^\dagger{\bf K}{\bf q}={\frac{1}{4}}({\bf q}_1^\dagger{{\bf q}_2}+{\bf q}_2^\dagger{{\bf q}_1})=j$ (see Eq. (\ref{eq50ab})), Eq. (\ref{eq4.3pp}) formulates the unified power  balance.
The term on the left-hand side is the power generated by the sources in ${\mathbb{D}}$.
The first term on the right-hand side is the power flux  through the  boundary ${{\partial\mathbb{D}}}_0\cup{{\partial\mathbb{D}}_M}$ (i.e., the power leaving the domain ${\mathbb{D}}$)
 and the second term on the right-hand side is the dissipated power in ${\mathbb{D}}$.

\section{Representations with Green's matrices}\label{sec4}

A wave-field representation is obtained by replacing one of the states in a reciprocity theorem by a Green's state 
\cite{Knopoff56JASA, Hoop58PHD, Gangi70JGR, Pao76JASA}. In this section we follow this approach
 to derive wave-field representations with Green's matrices  
from the matrix-vector reciprocity theorems discussed in section \ref{sec3}.

\subsection{Symmetry property of the Green's matrix}\label{sec4.1}

Before we derive wave-field representations, we first derive a symmetry property of the Green's matrix.
To this end, we replace both wave-field vectors  ${\bf q}_A$ and ${\bf q}_B$ in reciprocity theorem (\ref{eq4.1}) by Green's matrices ${\bf G}({\bf x},{\bf x}_A,\omega)$ and
${\bf G}({\bf x},{\bf x}_B,\omega)$, respectively. Accordingly, we replace the source vectors ${\bf d}_A$ and ${\bf d}_B$ by ${\bf I}\delta({\bf x}-{\bf x}_A)$ and ${\bf I}\delta({\bf x}-{\bf x}_B)$, respectively,
\rev{with ${\bf x}_A$ and ${\bf x}_B$ denoting the source positions, see Figure \ref{Fig1}}.
Furthermore, we replace ${\mathbb{D}}$ by ${\mathbb{R}}^3$, so that the boundary integral in Eq. (\ref{eq4.1}) vanishes (Sommerfeld radiation condition).
Both Green's matrices are defined in the same medium, hence, ${{\mbox{\boldmath ${\cal A}$}}}_A={{\mbox{\boldmath ${\cal A}$}}}_B$. This implies that
the second integral on the right-hand side of Eq. (\ref{eq4.1}) also vanishes.
From the remaining integral we thus obtain the following symmetry property of the Green's matrix
\begin{eqnarray}\label{eq65a}
{\bf G}^t({\bf x}_B,{\bf x}_A,\omega){\bf N}=-{\bf N}{\bf G}({\bf x}_A,{\bf x}_B,\omega).
\end{eqnarray}
The Green's matrix on the left-hand side is the response to a source at ${\bf x}_A$, observed by a receiver at ${\bf x}_B$.
Similarly, the Green's matrix  on the right-hand side is the response to a source at ${\bf x}_B$, observed by a receiver at ${\bf x}_A$.
Hence, Eq. (\ref{eq65a}) is a unified source-receiver reciprocity relation. 

Using this relation and ${\bf J}{\bf N}=-{\bf N}{\bf J}$, we find for the homogeneous Green's matrix defined in Eq. (\ref{eq750})
\begin{eqnarray}\label{eq65ah}
{\bf G}_{\rm h}^t({\bf x}_B,{\bf x}_A,\omega){\bf N}=-{\bf N}{\bf G}_{\rm h}({\bf x}_A,{\bf x}_B,\omega).
\end{eqnarray}

\subsection{Representations of the convolution type with the Green's matrix}\label{sec4.2}

We derive a representation of the convolution type for the actual wave-field vector ${\bf q}({\bf x},\omega)$, emitted by the actual source distribution ${\bf d}({\bf x},\omega)$ in the actual medium;
the operator matrix in the actual medium is defined as ${{\mbox{\boldmath ${\cal A}$}}}({\bf x},\omega)$. 
We let state $B$ in reciprocity theorem (\ref{eq4.1}) be this actual state, hence, we drop
subscript $B$ from ${\bf q}_B$, ${\bf d}_B$ and ${{\mbox{\boldmath ${\cal A}$}}}_B$. For state $A$ we choose the Green's state.
Hence, we replace  ${\bf q}_A({\bf x},\omega)$ in reciprocity theorem (\ref{eq4.1}) by ${\bf G}({\bf x},{\bf x}_A,\omega)$ and ${\bf d}_A({\bf x},\omega)$ by ${\bf I}\delta({\bf x}-{\bf x}_A)$. 
We keep the subscript $A$ in ${{\mbox{\boldmath ${\cal A}$}}}_A({\bf x},\omega)$, to account for the fact that in general this operator matrix is defined in a medium that may be different from the actual medium.
We thus obtain
\begin{eqnarray}
\chi_{\mathbb{D}}({\bf x}_A){\bf N}{\bf q}({\bf x}_A,\omega)&=&-\int_{\mathbb{D}} {\bf G}^t({\bf x},{\bf x}_A,\omega){\bf N}{\bf d}({\bf x},\omega){\rm d}^3{\bf x}
+\int_{{{\partial\mathbb{D}}}_0\cup{{\partial\mathbb{D}}_M}} {\bf G}^t({\bf x},{\bf x}_A,\omega){\bf N}{\bf q}({\bf x},\omega)n_3{\rm d}^2{\bf x}
\nonumber\\
&+&\int_{\mathbb{D}} {\bf G}^t({\bf x},{\bf x}_A,\omega){\bf N}\{{{\mbox{\boldmath ${\cal A}$}}}_A-{{\mbox{\boldmath ${\cal A}$}}}\}{\bf q}({\bf x},\omega){\rm d}^3{\bf x},
\label{eqrepgenbef}
\end{eqnarray}
where $\chi_{\mathbb{D}}({\bf x})$ is the characteristic function for domain ${\mathbb{D}}$, defined as
\begin{eqnarray}\label{eqC3.2}
\chi_{\mathbb{D}}({\bf x})=
\begin{cases}
1,   &\text{for } {\bf x}\text{ inside }{{\mathbb{D}}}, \\
{\frac{1}{2}}, &\text{for } {\bf x}\text{ on }{{\partial\mathbb{D}}}_0\cup{{\partial\mathbb{D}}_M},\\
0,  &\text{for } {\bf x}\text{ outside }{{\mathbb{D}}}.
\end{cases}
\end{eqnarray}
Using the symmetry property of the Green's matrix, formulated by Eq. (\ref{eq65a}), we obtain
\begin{eqnarray}
\chi_{\mathbb{D}}({\bf x}_A){\bf q}({\bf x}_A,\omega)&=&\int_{\mathbb{D}} {\bf G}({\bf x}_A,{\bf x},\omega){\bf d}({\bf x},\omega){\rm d}^3{\bf x}
-\int_{{{\partial\mathbb{D}}}_0\cup{{\partial\mathbb{D}}_M}} {\bf G}({\bf x}_A,{\bf x},\omega){\bf q}({\bf x},\omega)n_3{\rm d}^2{\bf x}\nonumber\\
&-&\int_{\mathbb{D}} {\bf G}({\bf x}_A,{\bf x},\omega)\{{{\mbox{\boldmath ${\cal A}$}}}_A-{{\mbox{\boldmath ${\cal A}$}}}\}{\bf q}({\bf x},\omega){\rm d}^3{\bf x}.
\label{eqrepgen}
\end{eqnarray}
This is the unified wave-field 
representation of the convolution type with the Green's matrix. The left-hand side is the wave-field vector ${\bf q}$ at a specific point ${\bf x}_A$, multiplied with the
characteristic function. According to the right-hand side, this field  consists of a contribution from the source distribution in ${\mathbb{D}}$ (the first integral), a contribution 
from the wave field on the boundary of ${\mathbb{D}}$ (the second integral), and a contribution caused by the contrasts between the operator matrices in the Green's and the actual state 
(the third integral).

Note that Eq. (\ref{eq1115}) follows as a special case of Eq. (\ref{eqrepgen}) if we choose the same medium parameters in both states and replace ${\mathbb{D}}$ by ${\mathbb{R}}^3$  
(except that the roles of ${\bf x}$ and ${{\bf x}_A}$ are interchanged since we used the symmetry property of Eq. (\ref{eq65a}) in the derivation of Eq. (\ref{eqrepgen})).

Next we consider another special case. We choose again the same medium parameters in both states,
but this time we replace ${\mathbb{D}}$ by the entire half-space below ${{\partial\mathbb{D}}}_0$, which we choose to be  source-free in the actual state.
Assuming ${\bf x}_A$ lies in the lower half-space, we thus obtain
\begin{eqnarray}
&&\hspace{-.7cm}{\bf q}({\bf x}_A,\omega)=\int_{{{\partial\mathbb{D}}}_0} {\bf G}({\bf x}_A,{\bf x},\omega){\bf q}({\bf x},\omega){\rm d}^2{\bf x},\quad\mbox{for}\,x_{3,A}>{x_{3,0}}.
\label{eqrepgens}
\end{eqnarray}
Using Eqs. (\ref{eq9992}), (\ref{eq423}) and (\ref{eq1125}) for the isotropic acoustic  situation, Eq. (\ref{eqrepgens}) yields for the upper element of ${\bf q}({\bf x}_A,\omega)$
\begin{eqnarray}
&&\hspace{-0.7cm}p({\bf x}_A,\omega)=\int_{{{\partial\mathbb{D}}}_0} \Bigl(-\frac{1}{i\omega\rho({\bf x},\omega)}\{\partial_3G^{p,q}({\bf x}_A,{\bf x},\omega)\}p({\bf x},\omega)
+G^{p,q}({\bf x}_A,{\bf x},\omega)v_3({\bf x},\omega)\Bigr){\rm d}^2{\bf x},\label{eqrepgenwghgtraburtsrepag}
\end{eqnarray}
which is the well-known acoustic Kirchhoff-Helmholtz integral.
Hence, the representation of Eq. (\ref{eqrepgens}) is the unified Kirchhoff-Helmholtz integral. It can be used for 
forward wave-field extrapolation of the wave\rev{-field} vector ${\bf q}({\bf x},\omega)$ from ${{\partial\mathbb{D}}}_0$ to any point ${\bf x}_A$ below ${{\partial\mathbb{D}}}_0$. 
Note that the Green's matrix ${\bf G}({\bf x}_A,{\bf x},\omega)$ depends on the medium parameters of the entire half-space below ${{\partial\mathbb{D}}}_0$.
In section \ref{sec5.2} we derive a relation similar to Eq. (\ref{eqrepgens}), but with the Green's matrix replaced by the propagator matrix,
which depends only on the medium parameters between ${x_{3,0}}$ and $x_{3,A}$.

Finally, we replace ${\bf q}({\bf x},\omega)$ by ${\bf G}({\bf x},{\bf x}_B,\omega)$ in Eq. (\ref{eqrepgens}). 
Since we assumed that the lower half-space is source-free for ${\bf q}$, we choose ${\bf x}_B$ in the upper half-space. We thus obtain
\begin{eqnarray}
&&\hspace{-0.7cm}{\bf G}({\bf x}_A,{\bf x}_B,\omega)=\int_{{{\partial\mathbb{D}}}_0}{\bf G}({\bf x}_A,{\bf x},\omega){\bf G}({\bf x},{\bf x}_B,\omega){\rm d}^2{\bf x},
\quad\mbox{for}\,x_{3,A}>{x_{3,0}}>x_{3,B}.\label{eq4charasp}
\end{eqnarray}
This expression shows that the Green's matrix can be composed from a cascade of Green's matrices, assuming a specific order of the depth levels at which the 
Green's sources and receivers are situated. In section \ref{sec5.2} we derive a similar relation for  propagator matrices, for an arbitrary  order of depth levels.

Equation (\ref{eq4charasp}) can be seen as a generalization of the representation that underlies Green's function retrieval by cross-convolution 
(where ${\bf G}({\bf x}_A,{\bf x}_B,\omega)$ is the unknown \cite{Slob2007GRL})
or by multidimensional deconvolution (where ${\bf G}({\bf x}_A,{\bf x},\omega)$ is the unknown \cite{Wapenaar2010JASA}).

\subsection{Representations of the correlation type with the Green's matrix}\label{sec4.3}

We derive a representation of the correlation type for the actual wave-field vector ${\bf q}({\bf x},\omega)$, emitted by the actual source distribution ${\bf d}({\bf x},\omega)$ in the actual medium.
Similar as in section \ref{sec4.2}, we let state $B$  be this actual state, and state $A$ the Green's state. 
Hence, making the same substitutions as in section \ref{sec4.2}, this time in reciprocity theorem (\ref{eq4.2}), using Eq. (\ref{eq65a}) and ${\bf N}^{-1}{\bf K}=-{\bf J}$, yields
\begin{eqnarray}
\chi_{\mathbb{D}}({\bf x}_A){\bf q}({\bf x}_A,\omega)&=&\int_{\mathbb{D}} {\bf J}{\bf G}^*({\bf x}_A,{\bf x},\omega){\bf J}{\bf d}({\bf x},\omega){\rm d}^3{\bf x}
-\int_{{{\partial\mathbb{D}}}_0\cup{{\partial\mathbb{D}}_M}} {\bf J}{\bf G}^*({\bf x}_A,{\bf x},\omega){\bf J}{\bf q}({\bf x},\omega)n_3{\rm d}^2{\bf x}\nonumber\\
&-&\int_{\mathbb{D}} {\bf J}{\bf G}^*({\bf x}_A,{\bf x},\omega){\bf J}\{{{\,\,\,\bar{\!\!\!{{\mbox{\boldmath ${\cal A}$}}}}}}_A-{{\mbox{\boldmath ${\cal A}$}}}\}{\bf q}({\bf x},\omega){\rm d}^3{\bf x}.
\label{eqrepgencor}
\end{eqnarray}
This is the unified wave-field representation of the correlation type with the Green's matrix.  

As a special case we derive a representation for the homogeneous Green's matrix ${\bf G}_{\rm h}$.
To this end, for state $B$ we choose the Green's matrix in the actual medium, hence, we replace ${\bf q}({\bf x},\omega)$ by ${\bf G}({\bf x},{\bf x}_B,\omega)$, 
and ${\bf d}({\bf x},\omega)$ by ${\bf I}\delta({\bf x}-{\bf x}_B)$.
For state $A$ we replace the Green's matrix by that  in the adjoint of the actual medium, hence, we replace ${\bf G}({\bf x}_A,{\bf x},\omega)$ by $\bar{\bf G}({\bf x}_A,{\bf x},\omega)$, 
and ${{\mbox{\boldmath ${\cal A}$}}}_A$ by ${{\,\,\,\bar{\!\!\!{{\mbox{\boldmath ${\cal A}$}}}}}}$. With this choice the contrast operator ${{\,\,\,\bar{\!\!\!{{\mbox{\boldmath ${\cal A}$}}}}}}_A-{{\mbox{\boldmath ${\cal A}$}}}=\,\,\,\,\bar{\bar{\!\!\!\!{{\mbox{\boldmath ${\cal A}$}}}}}-{{\mbox{\boldmath ${\cal A}$}}}$ vanishes.
Making these substitutions in Eq. (\ref{eqrepgencor}), taking ${\bf x}_A$ and ${\bf x}_B$ both inside ${\mathbb{D}}$,
using Eq. (\ref{eq750}), 
we obtain
\begin{eqnarray}
&&\hspace{-0.7cm}{\bf G}_{\rm h}({\bf x}_A,{\bf x}_B,\omega)=
-\int_{{{\partial\mathbb{D}}}_0\cup{{\partial\mathbb{D}}_M}} {\bf J}\bar{\bf G}^*({\bf x}_A,{\bf x},\omega){\bf J}{\bf G}({\bf x},{\bf x}_B,\omega)n_3{\rm d}^2{\bf x},\nonumber\\
&&\hspace{6.cm}\mbox{for}\,\,{x_{3,M}}>\rev{x_{3,\{A,B\}}}>{x_{3,0}}.\label{eqhomgts}
\end{eqnarray}
This is the unified homogeneous Green's matrix representation.
It finds applications in optical, acoustic and seismic holography \cite{Porter70JOSA, Lindsey2004AJSS}, inverse source problems \cite{Porter82JOSA}, inverse
scattering methods \cite{Oristaglio89IP} and Green's function retrieval by cross-correlation 
\cite{Derode2003JASA, Wapenaar2003GEO, Weaver2004JASA, Wapenaar2006PRL, Snieder2007PRE}.
A disadvantage is that the integral is taken along two boundaries ${{\partial\mathbb{D}}}_0$ and ${{\partial\mathbb{D}}_M}$,
whereas in many practical situations, measurements are only available on a single boundary. Using the propagator matrix, in section \ref{sec6.2} we present a single-sided
unified homogeneous Green's function representation.

Finally, using Eqs. (\ref{eq423}), (\ref{eq1124}) and (\ref{eq1125}) for the isotropic acoustic situation, Eq. (\ref{eqhomgts}) yields for the upper-right element of 
${\bf G}_{\rm h}({\bf x}_A,{\bf x}_B,\omega)$
\begin{eqnarray}
G_{\rm h}^{p,q}({\bf x}_A,{\bf x}_B,\omega)&=&
-\frac{1}{i\omega}\int_{{{\partial\mathbb{D}}}_0\cup{{\partial\mathbb{D}}_M}} \Biggl(\frac{1}{\bar\rho^*({\bf x},\omega)}\{\partial_3\bar G^{p,q}({\bf x}_A,{\bf x},\omega)\}^*G^{p,q}({\bf x},{\bf x}_B,\omega)
\nonumber\\
&-&\frac{1}{\rho({\bf x},\omega)}\{\bar G^{p,q}({\bf x}_A,{\bf x},\omega)\}^*\partial_3G^{p,q}({\bf x},{\bf x}_B,\omega)\Biggr)n_3{\rm d}^2{\bf x},\label{eqrepgenwghgtraburtsag}
\end{eqnarray}
with
\begin{eqnarray}
&&\hspace{-0.7cm}G_{\rm h}^{p,q}({\bf x}_A,{\bf x}_B,\omega)=G^{p,q}({\bf x}_A,{\bf x}_B,\omega)+\{\bar G^{p,q}({\bf x}_A,{\bf x}_B,\omega)\}^*,
\label{eq1254}
\end{eqnarray}
\rev{according to Eqs. (\ref{eq423}) and (\ref{eq1213h}).}
For a lossless constant density medium, using $G^{p,q}({\bf x},{\bf x}_B,\omega)=-i\omega\rho {\cal G}({\bf x},{\bf x}_B,\omega)$, 
Eq. (\ref{eqrepgenwghgtraburtsag}) is the well-known scalar homogeneous  Green's function representation  \cite{Porter70JOSA, Oristaglio89IP}, applied to the 
configuration of Fig. \ref{Fig1}.

\section{Representations with propagator matrices}\label{sec5}

We follow a similar approach as in section \ref{sec4} to derive wave-field representations. However, 
this time we replace one of the states in the reciprocity theorems by a propagator state (instead of a Green's state). 
 Hence, this leads to wave-field representations with propagator matrices.

\subsection{Symmetry properties of the propagator matrix}\label{sec5.1}

We start with deriving symmetry properties of the propagator matrix.
To this end, we replace both wave-field vectors  ${\bf q}_A$ and ${\bf q}_B$ in reciprocity theorem (\ref{eq4.1}) by 
propagator matrices ${\bf W}({\bf x},{\bf x}_A,\omega)$ and ${\bf W}({\bf x},{\bf x}_B,\omega)$, respectively.
\rev{We define horizontal boundaries ${{\partial\mathbb{D}}}_A$ and ${{\partial\mathbb{D}}}_B$, containing the points ${\bf x}_A$ and ${\bf x}_B$, respectively, see Fig. \ref{Fig1}.} 
We replace ${{\partial\mathbb{D}}}_0$ and ${{\partial\mathbb{D}}_M}$ in Eq. (\ref{eq4.1}) by these boundaries (and ${\mathbb{D}}$ by the region enclosed by these boundaries).
Note that boundary condition (\ref{eq9998d})  implies ${\bf W}({\bf x},{\bf x}_A,\omega)=  {\bf I}\delta({{\bf x}_{\rm H}}-{{\bf x}_{{\rm H},A}})$ for ${\bf x}$ at ${{\partial\mathbb{D}}}_A$ and 
${\bf W}({\bf x},{\bf x}_B,\omega)=  {\bf I}\delta({{\bf x}_{\rm H}}-{{\bf x}_{{\rm H},B}})$ for ${\bf x}$ at ${{\partial\mathbb{D}}}_B$.
The propagators obey  wave equation (\ref{eq2.1gw}) without sources, hence, we set  ${\bf d}_A$ and ${\bf d}_B$ to ${\bf O}$. 
This implies that the integral on the left-hand side of Eq. (\ref{eq4.1}) vanishes.
Both propagator matrices are defined in the same medium, hence, ${{\mbox{\boldmath ${\cal A}$}}}_A={{\mbox{\boldmath ${\cal A}$}}}_B$. This implies that
the second integral on the right-hand side vanishes.
Evaluating the remaining integral along the boundary ${{\partial\mathbb{D}}}_A\cup{{\partial\mathbb{D}}}_B$ yields 
\rev{(irrespective of the arrangement of ${{\partial\mathbb{D}}}_A$ and ${{\partial\mathbb{D}}}_B$)}
\begin{eqnarray}\label{eq65aw}
{\bf W}^t({\bf x}_B,{\bf x}_A,\omega){\bf N}={\bf N}{\bf W}({\bf x}_A,{\bf x}_B,\omega).
\end{eqnarray}
This is the first unified  symmetry relation for the propagator matrix.

A second symmetry relation can be derived from reciprocity theorem  (\ref{eq4.2}).
To this end we replace   ${\bf q}_A$ and ${\bf q}_B$ by $\bar{\bf W}({\bf x},{\bf x}_A,\omega)$ and ${\bf W}({\bf x},{\bf x}_B,\omega)$, respectively.  
We replace ${{\partial\mathbb{D}}}_0\cup{{\partial\mathbb{D}}_M}$ in Eq. (\ref{eq4.2})  by ${{\partial\mathbb{D}}}_A\cup{{\partial\mathbb{D}}}_B$ (and ${\mathbb{D}}$ by the enclosed region).
We set  ${\bf d}_A$ and ${\bf d}_B$ to ${\bf O}$, hence, the integral on the left-hand side vanishes.
The propagator matrices are defined in mutually adjoint media, hence, ${{\mbox{\boldmath ${\cal A}$}}}_A={{\,\,\,\bar{\!\!\!{{\mbox{\boldmath ${\cal A}$}}}}}}_B$. This implies that
the second integral on the right-hand side of Eq. (\ref{eq4.2}) vanishes.
From the remaining integral we thus obtain
 \begin{eqnarray}\label{eq65awk}
\bar{\bf W}^\dagger({\bf x}_B,{\bf x}_A,\omega){\bf K}={\bf K}{\bf W}({\bf x}_A,{\bf x}_B,\omega).
\end{eqnarray}
From Eqs. (\ref{eq65aw}) and (\ref{eq65awk}), using ${\bf K}{\bf N}^{-1}={\bf J}$, we find
\begin{eqnarray}\label{eq65aws}
\bar{\bf W}^*({\bf x}_B,{\bf x}_A,\omega){\bf J}={\bf J}{\bf W}({\bf x}_B,{\bf x}_A,\omega).
\end{eqnarray}
Note that in this last equation ${\bf x}_B$ and ${\bf x}_A$ appear in the same order on the left- and right-hand sides.

\subsection{Representations of the convolution type with the propagator matrix}\label{sec5.2}

\begin{figure}
\vspace{0cm}
\centerline{\epsfysize=6.8 cm \epsfbox{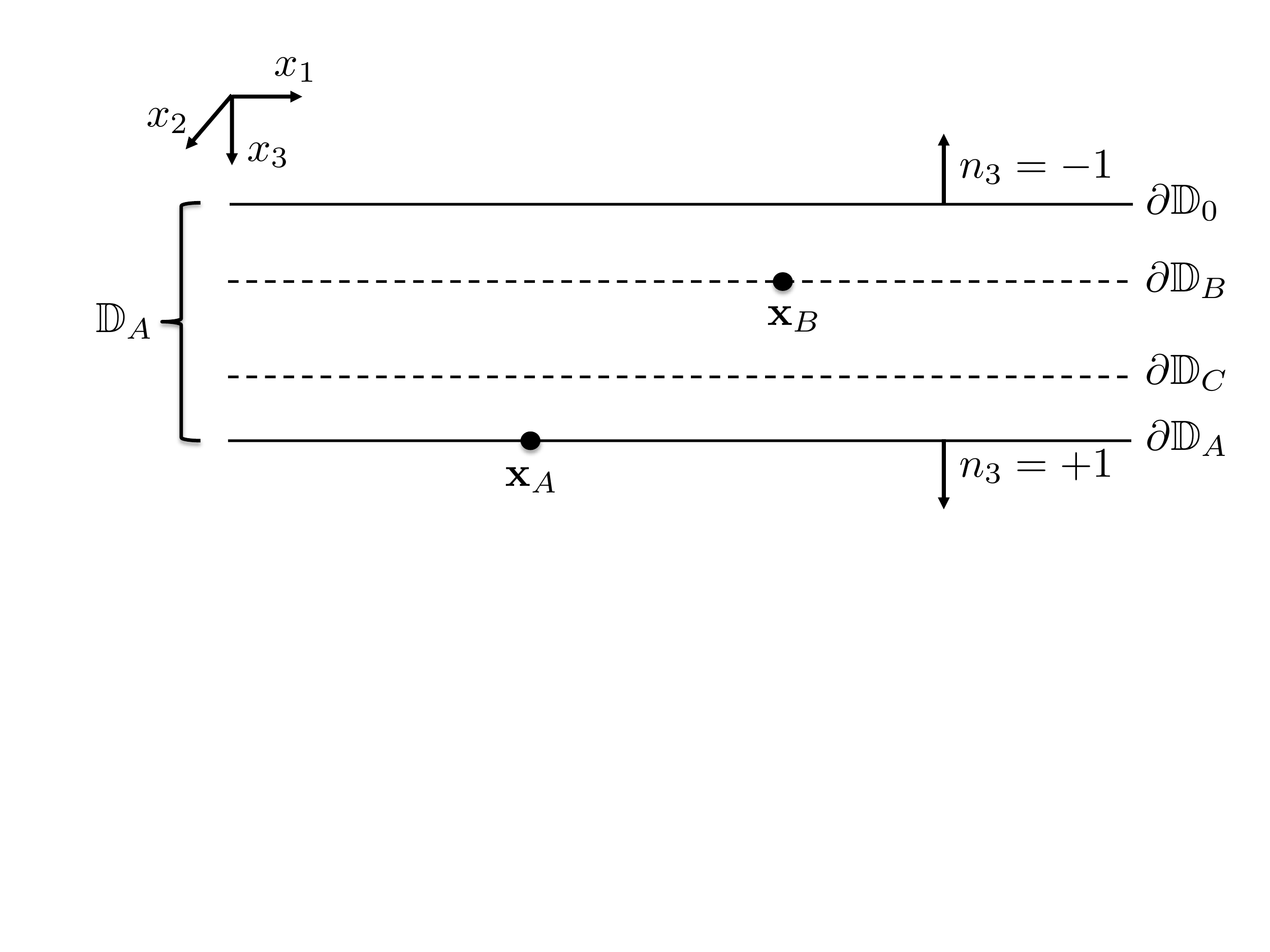}}\vspace{-3cm}
\caption{Configuration for the representations with propagator matrices. 
}\label{Fig2}
\end{figure}

We derive a representation of the convolution type for the actual wave-field vector ${\bf q}({\bf x},\omega)$, 
emitted by the actual source distribution ${\bf d}({\bf x},\omega)$ in the actual medium;
the operator matrix in the actual medium is defined as ${{\mbox{\boldmath ${\cal A}$}}}({\bf x},\omega)$. 
We let state $B$ in reciprocity theorem (\ref{eq4.1}) be this actual state, hence, we drop
subscript $B$ from ${\bf q}_B$, ${\bf d}_B$ and ${{\mbox{\boldmath ${\cal A}$}}}_B$. 
For state $A$ we choose the propagator state.
Hence, we replace  ${\bf q}_A({\bf x},\omega)$ in reciprocity theorem (\ref{eq4.1}) by ${\bf W}({\bf x},{\bf x}_A,\omega)$ and ${\bf d}_A({\bf x},\omega)$ by ${\bf O}$. 
We keep the subscript $A$ in ${{\mbox{\boldmath ${\cal A}$}}}_A({\bf x},\omega)$, to account for the fact that in general this operator matrix is defined in a medium that may be different from the actual medium.
We replace ${{\partial\mathbb{D}}}_0\cup{{\partial\mathbb{D}}_M}$ in reciprocity theorem (\ref{eq4.1}) by ${{\partial\mathbb{D}}}_0\cup{{\partial\mathbb{D}}}_A$, where ${{\partial\mathbb{D}}}_A$ is the boundary containing ${\bf x}_A$.
Here and in the following, ${{\partial\mathbb{D}}}_A$ is below ${{\partial\mathbb{D}}}_0$, hence $x_{3,A}>{x_{3,0}}$. 
The domain enclosed by this boundary is called ${\mathbb{D}_A}$,  see Fig. \ref{Fig2}.
Unlike in the classical, decomposition-based derivations of the Marchenko method \cite{Wapenaar2014JASA, Slob2014GEO}, the domain ${\mathbb{D}_A}$ does not define a truncated medium; 
in general the medium below ${{\partial\mathbb{D}}}_A$ is inhomogeneous.
Applying the mentioned substitutions in Eq. (\ref{eq4.1}), using boundary condition (\ref{eq9998d}) and symmetry relation (\ref{eq65aw}) (with ${\bf x}_B$ replaced by ${\bf x}$) yields
\begin{eqnarray}
\hspace{-0.7cm}{\bf q}({\bf x}_A,\omega)&=&\int_{{\mathbb{D}_A}} {\bf W}({\bf x}_A,{\bf x},\omega){\bf d}({\bf x},\omega){\rm d}^3{\bf x}
+\int_{{{\partial\mathbb{D}}}_0} {\bf W}({\bf x}_A,{\bf x},\omega){\bf q}({\bf x},\omega){\rm d}^2{\bf x}\nonumber\\
&-&\int_{{\mathbb{D}_A}} {\bf W}({\bf x}_A,{\bf x},\omega)\{{{\mbox{\boldmath ${\cal A}$}}}_A-{{\mbox{\boldmath ${\cal A}$}}}\}{\bf q}({\bf x},\omega){\rm d}^3{\bf x}.\label{eqrepgenw}
\end{eqnarray}
This is the unified wave-field 
representation of the convolution type with the propagator matrix. Note the analogy with the representation of the convolution type with the Green's matrix, Eq. (\ref{eqrepgen}).
An important difference is that the boundary integral in Eq. (\ref{eqrepgenw}) is single-sided.

We consider a special case by choosing  the same medium parameters in ${\mathbb{D}_A}$ in both states,
and choosing ${\mathbb{D}_A}$ to be  source-free. We thus obtain
\begin{eqnarray}
{\bf q}({\bf x}_A,\omega)&=&\int_{{{\partial\mathbb{D}}}_0} {\bf W}({\bf x}_A,{\bf x},\omega){\bf q}({\bf x},\omega){\rm d}^2{\bf x}.\label{eqrepgenws}
\end{eqnarray}
This is the special case that was already presented in Eq. (\ref{eq1330}) 
(except that the roles of ${\bf x}$ and ${{\bf x}_A}$ are interchanged since we used the symmetry property of Eq. (\ref{eq65aw}) in the derivation of Eq. (\ref{eqrepgenw})).

Note the analogy of Eq. (\ref{eqrepgenws}) with Eq. (\ref{eqrepgens}), which contains a Green's matrix instead of the propagator matrix.
The propagator matrix in Eq. (\ref{eqrepgenws}) depends only on the medium parameters between ${x_{3,0}}$ and $x_{3,A}$.
Equation (\ref{eqrepgenws}) is used in section \ref{sec7} as the basis for deriving Marchenko-type representations.

Next, we consider again Eq. (\ref{eqrepgenw}), in which we replace  ${\bf q}({\bf x},\omega)$ by ${\bf W}({\bf x},{\bf x}_B,\omega)$, 
and hence ${\bf d}({\bf x},\omega)$ by ${\bf O}$ and ${{\mbox{\boldmath ${\cal A}$}}}$ by ${{\mbox{\boldmath ${\cal A}$}}}_A$. This implies that the first and the third integral on the right-hand side vanish.
Moreover, we replace ${{\partial\mathbb{D}}}_0$  by ${{\partial\mathbb{D}}}_C$ at a newly chosen depth level $x_{3,C}$, \rev{see Fig. \ref{Fig2}.}
This yields
\begin{eqnarray}\label{eq65awc}
{\bf W}({\bf x}_A,{\bf x}_B,\omega)=\int_{{{\partial\mathbb{D}}}_C}{\bf W}({\bf x}_A,{\bf x},\omega){\bf W}({\bf x},{\bf x}_B,\omega){\rm d}^2{\bf x}.
\end{eqnarray}
This expression shows that the propagator matrix can be composed from a cascade of propagator matrices \cite{Gilbert66GEO, Kennett72BS, Kennett72GJRAS, Woodhouse74GJR, Haines88GJI}.
It is similar to Eq. (\ref{eq4charasp}) with the cascade of 
Green's matrices, but unlike in Eq. (\ref{eq4charasp}), where $x_{3,A}>{x_{3,0}}>x_{3,B}$, 
the arrangement of $x_{3,A}$, $x_{3,B}$ and $x_{3,C}$ in Eq. (\ref{eq65awc}) is arbitrary (since there are no sources in both states).

Finally, we replace ${\bf x}_B$ in Eq. (\ref{eq65awc})  by ${\bf x}_A'=({{\bf x}'_{{\rm H},A}},x_{3,A})$. Applying boundary condition (\ref{eq9998d}) to the left-hand side, we obtain
\begin{eqnarray}\label{eq65awcin}
{\bf I}\delta({{\bf x}_{{\rm H},A}}-{{\bf x}'_{{\rm H},A}})=\int_{{{\partial\mathbb{D}}}_C}{\bf W}({\bf x}_A,{\bf x},\omega){\bf W}({\bf x},{\bf x}_A',\omega){\rm d}^2{\bf x}.
\end{eqnarray}
This equation defines ${\bf W}({\bf x}_A,{\bf x},\omega)$ as the inverse of ${\bf W}({\bf x},{\bf x}_A,\omega)$ \cite{Gilbert66GEO, Kennett72BS, Kennett72GJRAS, Woodhouse74GJR, Haines88GJI}.

\subsection{Representations of the correlation type with the propagator matrix}

We aim to derive a representation of the correlation type for the actual wave-field vector ${\bf q}({\bf x},\omega)$, 
emitted by the actual source distribution ${\bf d}({\bf x},\omega)$ in the actual medium.
Similar as in section \ref{sec5.2}, we let state $B$  be this actual state.
For state $A$ we choose the adjoint propagator state.
Hence, we replace  ${\bf q}_A({\bf x},\omega)$, ${{\mbox{\boldmath ${\cal A}$}}}_A({\bf x},\omega)$ and ${\bf d}_A({\bf x},\omega)$ by
 $\bar{\bf W}({\bf x},{\bf x}_A,\omega)$, ${{\,\,\,\bar{\!\!\!{{\mbox{\boldmath ${\cal A}$}}}}}}_A({\bf x},\omega)$  and ${\bf O}$, respectively. 
Furthermore, we replace ${{\partial\mathbb{D}}}_0\cup{{\partial\mathbb{D}}_M}$ again by ${{\partial\mathbb{D}}}_0\cup{{\partial\mathbb{D}}}_A$. 
Applying these substitutions  in reciprocity theorem (\ref{eq4.2}), using boundary condition (\ref{eq9998d}) 
and symmetry relation (\ref{eq65awk}) (with ${\bf x}_B$ replaced by ${\bf x}$) yields again Eq. (\ref{eqrepgenw}).
Hence, due to the additional symmetry property of the propagator matrix (Eq. (\ref{eq65awk})), 
the correlation-type reciprocity theorem does not lead to a new representation.

\section{Representations with Green's matrices and propagator matrices}\label{sec6}

We follow a similar approach as in sections \ref{sec4} and \ref{sec5} to derive wave-field representations. However, 
this time we replace one of the states in the reciprocity theorem of the convolution type by a Green's state and the other by a propagator state. 
This leads to wave-field representations with Green's matrices and propagator matrices.  

\subsection{Single-sided Green's matrix representation}\label{sec6.1}

In reciprocity theorem (\ref{eq4.1}) we choose for state $A$ the propagator state and for state $B$ the Green's state.
Hence, in state $A$ we replace  ${\bf q}_A({\bf x},\omega)$ and ${\bf d}_A({\bf x},\omega)$ by
 ${\bf W}({\bf x},{\bf x}_A,\omega)$ and ${\bf O}$, respectively. In state $B$ we replace
 ${\bf q}_B({\bf x},\omega)$ and ${\bf d}_B({\bf x},\omega)$ by
 ${\bf G}({\bf x},{\bf x}_B,\omega)$  and ${\bf I}\delta({\bf x}-{\bf x}_B)$, respectively.
 The medium parameters in states $A$ and $B$ may be different, hence, we keep the subscripts in ${{\mbox{\boldmath ${\cal A}$}}}_A({\bf x},\omega)$ and ${{\mbox{\boldmath ${\cal A}$}}}_B({\bf x},\omega)$.
Last but not least, we replace ${{\partial\mathbb{D}}}_0\cup{{\partial\mathbb{D}}_M}$  by ${{\partial\mathbb{D}}}_0\cup{{\partial\mathbb{D}}}_A$ and ${\mathbb{D}}$ by the enclosed region ${\mathbb{D}_A}$, see Fig. \ref{Fig2}.
With these substitutions, using boundary condition (\ref{eq9998d}) and symmetry relation (\ref{eq65aw}), reciprocity theorem (\ref{eq4.1}) yields
\begin{eqnarray}
&&\hspace{-1.1cm}{\bf G}({\bf x}_A,{\bf x}_B,\omega)-\chi_{{\mathbb{D}_A}}({\bf x}_B){\bf W}({\bf x}_A,{\bf x}_B,\omega)=\nonumber\\
&&\hspace{-0.4cm}\int_{{{\partial\mathbb{D}}}_0} {\bf W}({\bf x}_A,{\bf x},\omega){\bf G}({\bf x},{\bf x}_B,\omega){\rm d}^2{\bf x}
-\int_{{\mathbb{D}_A}} {\bf W}({\bf x}_A,{\bf x},\omega)\{{{\mbox{\boldmath ${\cal A}$}}}_A-{{\mbox{\boldmath ${\cal A}$}}}_B\}{\bf G}({\bf x},{\bf x}_B,\omega){\rm d}^3{\bf x},\label{eqrepgenwg}
\end{eqnarray}
where $\chi_{{\mathbb{D}_A}}({\bf x})$ is the characteristic function for domain ${\mathbb{D}_A}$, defined \rev{similarly as $\chi_{{\mathbb{D}}}({\bf x})$ in equation (\ref{eqC3.2}).}
Equation (\ref{eqrepgenwg}) is a representation for ${\bf G}-{\bf W}$ (when ${{\bf x}_B}$ is inside ${\mathbb{D}_A}$) in terms of integrals containing ${\bf G}$ and ${\bf W}$.
When ${\bf G}$ and ${\bf W}$ are defined in the same medium, this representation simplifies to
\begin{eqnarray}
&&\hspace{-0.7cm}{\bf G}({\bf x}_A,{\bf x}_B,\omega)-\chi_{{\mathbb{D}_A}}({\bf x}_B){\bf W}({\bf x}_A,{\bf x}_B,\omega)=
\int_{{{\partial\mathbb{D}}}_0} {\bf W}({\bf x}_A,{\bf x},\omega){\bf G}({\bf x},{\bf x}_B,\omega){\rm d}^2{\bf x}.\label{eqrepgenwgs}
\end{eqnarray}
Finally, when ${{\bf x}_B}$ (the position of the source of the Green's function) lies outside ${\mathbb{D}_A}$ (i.e., above ${{\partial\mathbb{D}}}_0$ or below ${{\partial\mathbb{D}}}_A$), 
then the latter  expression simplifies to
\begin{eqnarray}
&&\hspace{-0.7cm}{\bf G}({\bf x}_A,{\bf x}_B,\omega)=
\int_{{{\partial\mathbb{D}}}_0} {\bf W}({\bf x}_A,{\bf x},\omega){\bf G}({\bf x},{\bf x}_B,\omega){\rm d}^2{\bf x}, 
\quad\mbox{for}\,x_{3,B}<{x_{3,0}}\,\mbox{or}\,x_{3,B}>x_{3,A}.\label{eqrepgenwgssim}
\end{eqnarray}
This is a single-sided representation of the Green's matrix that uses the propagator matrix.
This  special case  was already presented in Eq. (\ref{eq1331}) 
(except that the roles of ${\bf x}$ and ${{\bf x}_A}$ are interchanged since we used the symmetry property of Eq. (\ref{eq65aw}) in the derivation of Eq. (\ref{eqrepgenwgssim})).

Note the analogy of Eq. (\ref{eqrepgenwgssim}) with Eq. (\ref{eq4charasp}), which contains a Green's matrix instead of the propagator matrix.
The propagator matrix in Eq. (\ref{eqrepgenwgssim}) depends only on the medium parameters between ${x_{3,0}}$ and $x_{3,A}$.
Moreover, unlike the representation of Eq. (\ref{eq4charasp}), which only holds for $\,x_{3,B}<{x_{3,0}}$, Eq. (\ref{eqrepgenwgssim}) also holds for $x_{3,B}>x_{3,A}$.

\subsection{Single-sided homogeneous Green's matrix representation}\label{sec6.2}

We follow the same procedure as in section \ref{sec6.1}, except in state $B$ we choose the homogeneous Green's function, hence, we replace
${\bf q}_B({\bf x},\omega)$ and ${\bf d}_B({\bf x},\omega)$  \rev{in reciprocity theorem (\ref{eq4.1})} by ${\bf G}_{\rm h}({\bf x},{\bf x}_B,\omega)$  and ${\bf O}$, respectively. 
We thus obtain
\begin{eqnarray}
{\bf G}_{\rm h}({\bf x}_A,{\bf x}_B,\omega)&=&\int_{{{\partial\mathbb{D}}}_0} {\bf W}({\bf x}_A,{\bf x},\omega){\bf G}_{\rm h}({\bf x},{\bf x}_B,\omega){\rm d}^2{\bf x}
\nonumber\\
&-&\int_{{\mathbb{D}_A}} {\bf W}({\bf x}_A,{\bf x},\omega)\{{{\mbox{\boldmath ${\cal A}$}}}_A-{{\mbox{\boldmath ${\cal A}$}}}_B\}{\bf G}_{\rm h}({\bf x},{\bf x}_B,\omega){\rm d}^3{\bf x}.\label{eqrepgenwghom}
\end{eqnarray}
When ${\bf G}_{\rm h}$ and ${\bf W}$ are defined in the same medium, this representation simplifies to
\begin{eqnarray}
&&\hspace{-1.2cm}{\bf G}_{\rm h}({\bf x}_A,{\bf x}_B,\omega)=\int_{{{\partial\mathbb{D}}}_0} {\bf W}({\bf x}_A,{\bf x},\omega){\bf G}_{\rm h}({\bf x},{\bf x}_B,\omega){\rm d}^2{\bf x}.\label{eqrepgenwghomag}
\end{eqnarray}
This is a single-sided representation of the homogeneous Green's matrix that uses the propagator matrix.
This  special case  was already presented in Eq. (\ref{eq1332}) 
(except that the roles of ${\bf x}$ and ${{\bf x}_A}$ are interchanged since we used the symmetry property of Eq. (\ref{eq65aw}) in the derivation of Eq. (\ref{eqrepgenwghomag})).
Note that, unlike the  Green's matrix representation of Eq. (\ref{eqrepgenwgssim}), the homogeneous Green's matrix representation of Eq. (\ref{eqrepgenwghomag})
has no restrictions for the position of ${\bf x}_B$ (since there is no source at ${\bf x}_B$).

Equation  (\ref{eqrepgenwghomag}) is the single-sided counterpart of the unified homogeneous Green's matrix representation of Eq. (\ref{eqhomgts}). 
Whereas in Eq. (\ref{eqhomgts}) the integral is taken along two boundaries ${{\partial\mathbb{D}}}_0$ and ${{\partial\mathbb{D}}_M}$, 
in Eq. (\ref{eqrepgenwghomag}) the integral is taken only along ${{\partial\mathbb{D}}}_0$. This is an important advantage for practical situations,
where measurements are often available only on a single boundary. In section \ref{sec6.3} we indicate how Eq. (\ref{eqrepgenwghomag}) can be used
in a process called source and receiver redatuming.

Finally, using Eqs. (\ref{eq423}), (\ref{eq1124}), (\ref{eq1213h}), (\ref{eq424}) and (\ref{eq1125W}) 
for the isotropic acoustic  situation, assuming real-valued $\rho({\bf x})$, Eq. (\ref{eqrepgenwghomag}) yields for the upper-right element of 
${\bf G}_{\rm h}({\bf x}_A,{\bf x}_B,\omega)$
\begin{eqnarray}
G_{\rm h}^{p,q}({\bf x}_A,{\bf x}_B,\omega)&=&
-\frac{1}{i\omega}\int_{{{\partial\mathbb{D}}}_0}\frac{1}{\rho({\bf x})} \Bigl(\{\partial_3W^{p,v}({\bf x}_A,{\bf x},\omega)\}G_{\rm h}^{p,q}({\bf x},{\bf x}_B,\omega)\nonumber\\
&-&W^{p,v}({\bf x}_A,{\bf x},\omega)\partial_3G_{\rm h}^{p,q}({\bf x},{\bf x}_B,\omega)\Bigr){\rm d}^2{\bf x},\label{eqrepgenwghgtraburtswgrw}
\end{eqnarray}
with $G_{\rm h}^{p,q}({\bf x}_A,{\bf x}_B,\omega)$ defined in Eq. (\ref{eq1254}).

\subsection{Source and receiver redatuming}\label{sec6.3}

Redatuming is the process of moving sources and/or receivers from the acquisition boundary to positions inside the medium 
\cite{Berkhout82Book, Berryhill84GEO, Wapenaar89Book, Hokstad2000GEO}. Traditionally this is done with Kirchhoff-Helmholtz integrals with Green's functions defined in a 
smooth background medium, thus ignoring multiple scattering. Here we derive a unified redatuming method that accounts for multiple scattering, 
using the single-sided homogeneous Green's matrix representation of section \ref{sec6.2} as the starting point.

\begin{figure}
\vspace{-.4cm}\centerline{\epsfysize=6 cm \epsfbox{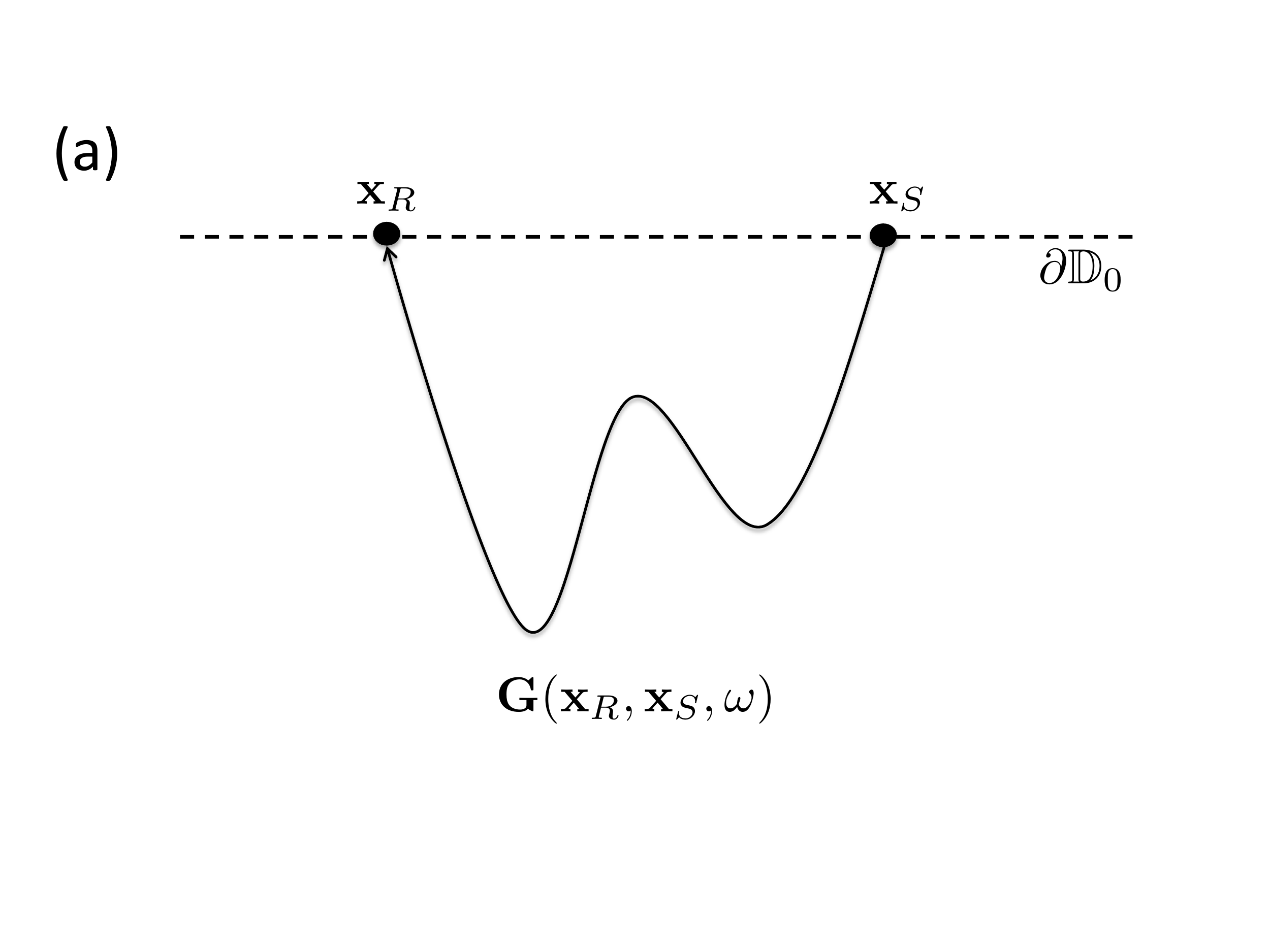}}\vspace{-1.2cm}
\centerline{\epsfysize=6 cm \epsfbox{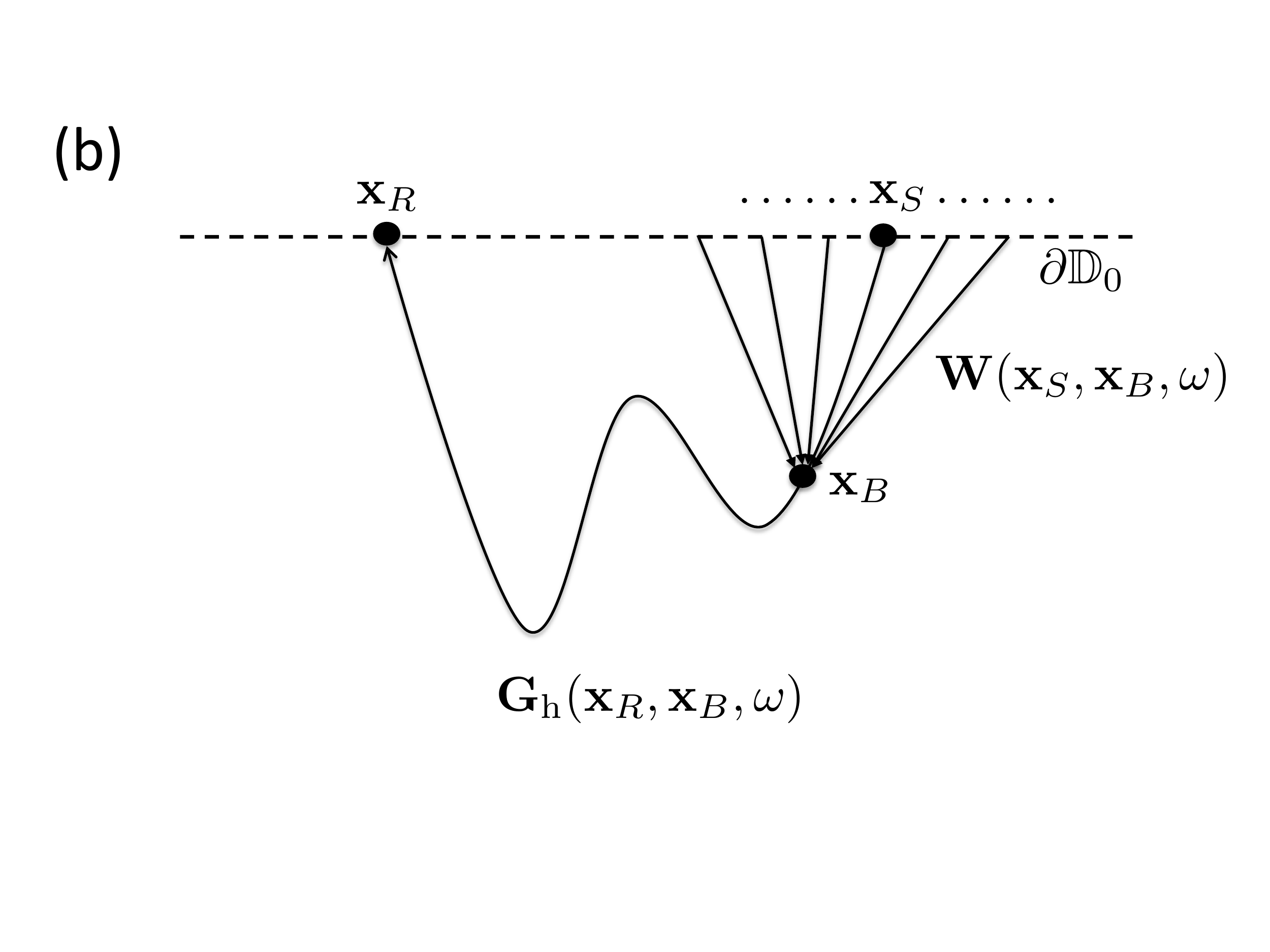}}\vspace{-1.2cm}
\centerline{\epsfysize=6 cm \epsfbox{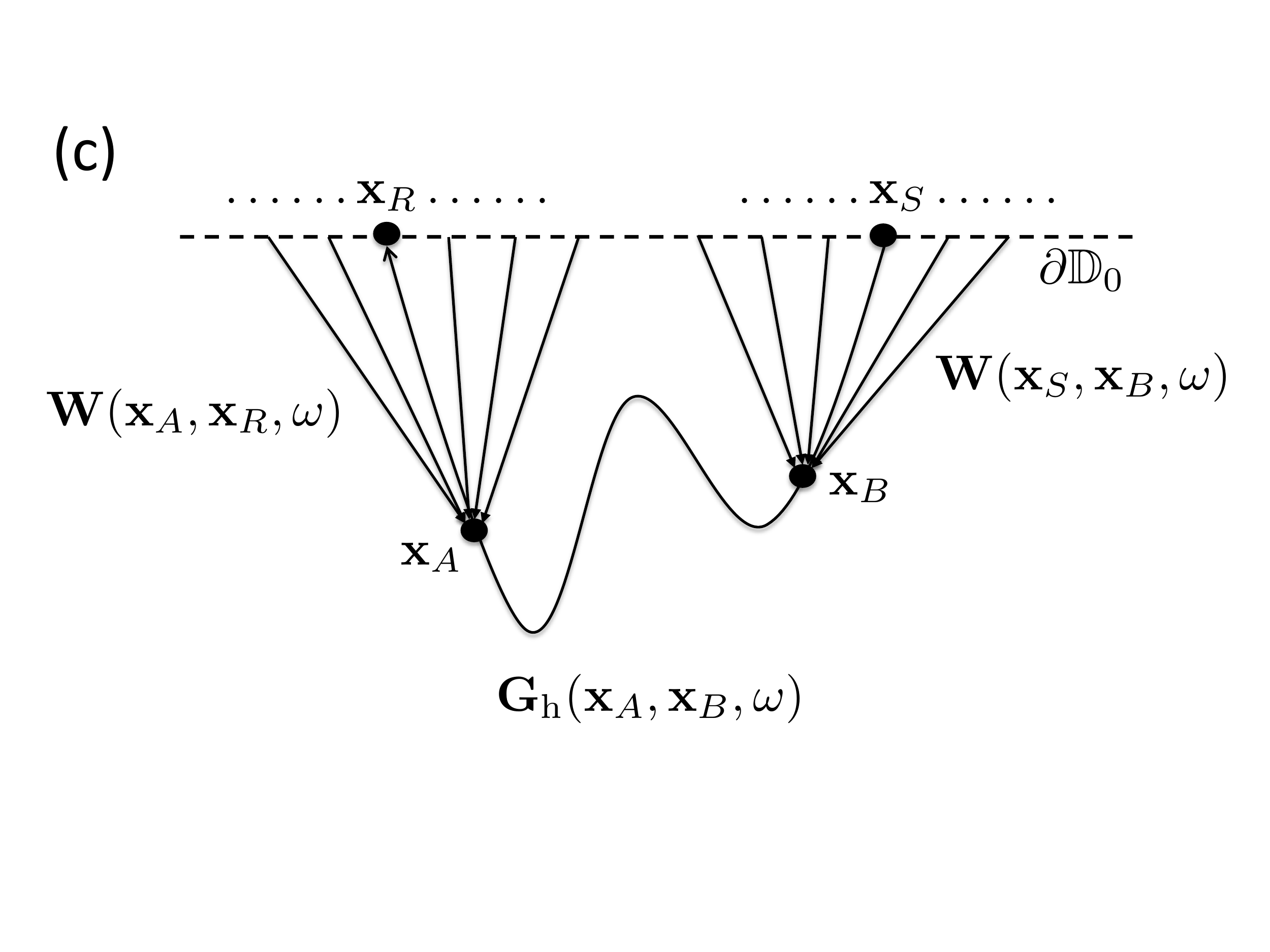}}\vspace{-1.2cm}
\caption{Illustration of source and receiver redatuming.
All responses in this figure are represented by simple rays, but in reality these are multi-component wave fields, including primaries, multiples, converted, refracted and evanescent waves.
(a) The  response ${\bf G}({{\bf x}_R},{{\bf x}_S},\omega)$ at the surface. The homogeneous Green's matrix ${\bf G}_{\rm h}({{\bf x}_R},{{\bf x}_S},\omega)$ is obtained from this response with Eq. (\ref{eq750}). 
(b) Source redatuming. Using Eq. (\ref{eqrepgenwghgtrab}), the homogeneous Green's matrix ${\bf G}_{\rm h}({{\bf x}_R},{\bf x}_B,\omega)$ is obtained for a virtual source at ${\bf x}_B$.
(c) Receiver redatuming. Using Eq. (\ref{eqrepgenwghomagR}), the homogeneous Green's matrix ${\bf G}_{\rm h}({\bf x}_A,{\bf x}_B,\omega)$ is obtained for a virtual receiver at ${\bf x}_A$. 
}\label{Fig3}
\end{figure}

Let ${\bf x}_S$ and ${\bf x}_R$ denote source and receiver coordinates, respectively, at the acquisition boundary ${{\partial\mathbb{D}}}_0$. Then the Green's matrix
${\bf G}({\bf x}_R,{\bf x}_S,\omega)$ represents the medium's  response, measured with sources and receivers at the surface,  see Fig. \ref{Fig3}(a).  Assuming the medium is lossless,
the homogeneous Green's matrix ${\bf G}_{\rm h}({\bf x}_R,{\bf x}_S,\omega)$ is obtained from this Green's matrix via Eq. (\ref{eq750}).

First we discuss a method for source redatuming.
We rename some variables in Eq. (\ref{eqrepgenwghomag}) (${\bf x}_B\to{\bf x}_R$, ${\bf x}_A\to{\bf x}_B$, ${\bf x}\to{\bf x}_S$) and transpose all matrices, which gives
\begin{eqnarray}
{\bf G}_{\rm h}^t({\bf x}_B,{\bf x}_R,\omega)=\int_{{{\partial\mathbb{D}}}_0} {\bf G}_{\rm h}^t({\bf x}_S,{\bf x}_R,\omega){\bf W}^t({\bf x}_B,{\bf x}_S,\omega){\rm d}^2{\bf x}_S.\label{eqrepgenwghgtr}
\end{eqnarray}
Using Eqs. (\ref{eq65ah}) and (\ref{eq65aw}) we obtain
\begin{eqnarray}
{\bf G}_{\rm h}({\bf x}_R,{\bf x}_B,\omega)=\int_{{{\partial\mathbb{D}}}_0} {\bf G}_{\rm h}({\bf x}_R,{\bf x}_S,\omega){\bf W}({\bf x}_S,{\bf x}_B,\omega){\rm d}^2{\bf x}_S.\label{eqrepgenwghgtrab}
\end{eqnarray}
The latter expression describes redatuming of the sources from ${\bf x}_S$ at the acquisition boundary to a virtual-source position ${\bf x}_B$ inside the medium, see 
Fig. \ref{Fig3}(b).

Next, we discuss receiver redatuming. We replace ${\bf x}$ by ${\bf x}_R$ in Eq. (\ref{eqrepgenwghomag}), which gives
\begin{eqnarray}
&&\hspace{-0.9cm}{\bf G}_{\rm h}({\bf x}_A,{\bf x}_B,\omega)=\int_{{{\partial\mathbb{D}}}_0} {\bf W}({\bf x}_A,{\bf x}_R,\omega){\bf G}_{\rm h}({\bf x}_R,{\bf x}_B,\omega){\rm d}^2{\bf x}_R.
\label{eqrepgenwghomagR}
\end{eqnarray}
This expression describes redatuming of the receivers from ${\bf x}_R$ at the acquisition boundary to a virtual-receiver position ${\bf x}_A$ inside the medium, see 
Fig. \ref{Fig3}(c).
The Green's matrix under the integral is the output of the source redatuming method, described by Eq. (\ref{eqrepgenwghgtrab}).

Combining Eq. (\ref{eqrepgenwghgtrab})  with Eq. (\ref{eqrepgenwghomagR}) gives the following expression for combined source and receiver redatuming
\begin{eqnarray}
&&\hspace{-1.5cm}{\bf G}_{\rm h}({\bf x}_A,{\bf x}_B,\omega)=
\int_{{{\partial\mathbb{D}}}_0}\int_{{{\partial\mathbb{D}}}_0} {\bf W}({\bf x}_A,{\bf x}_R,\omega)
{\bf G}_{\rm h}({\bf x}_R,{\bf x}_S,\omega){\bf W}({\bf x}_S,{\bf x}_B,\omega){\rm d}^2{\bf x}_S{\rm d}^2{\bf x}_R.
\label{eqrepgenwghgsr}
\end{eqnarray}
This expression redatums the actual sources and receivers from the acquisition boundary ${{\partial\mathbb{D}}}_0$ to virtual sources and receivers inside the medium.
\rev{Since it takes multiple scattering into account, it generalizes classical source and receiver redatuming \cite{Berkhout82Book, Berryhill84GEO, Wapenaar89Book, Hokstad2000GEO}.
Equation (\ref{eqrepgenwghgsr}) resembles expressions for source-receiver interferometry \cite{Curtis2010PRE, Halliday2012GJI},
but with the integrals along  a closed boundary 
replaced by integrals along the open acquisition boundary ${{\partial\mathbb{D}}}_0$, and with two of the Green's functions
replaced by}
 propagator matrices. When the medium is known, these matrices can be numerically modelled.
Alternatively, the propagator matrices can be expressed in terms of focusing functions which, at least for the acoustic situation, 
can be retrieved via the Marchenko method from the reflection response at the acquisition boundary.
In the next section we introduce relations between the propagator matrix and  focusing functions (section \ref{sec7.2}) 
and derive Marchenko-type representations which relate the focusing functions
to the reflection response at the acquisition boundary (section \ref{sec7.3}).

\section{Representations with focusing functions}\label{sec7}

\subsection{Relation between propagator matrix and focusing functions}\label{sec7.2}

In section \ref{sec2.5} we indicated that there is a relation between the propagator matrix and the Marchenko-type focusing functions. Here we derive this relation for 
the  propagator matrix of the unified matrix-vector wave equation, assuming the medium is lossless.
Our starting point is Eq. (\ref{eqrepgenws}), which is repeated here for convenience
\begin{eqnarray}
{\bf q}({\bf x}_A,\omega)&=&\int_{{{\partial\mathbb{D}}}_0} {\bf W}({\bf x}_A,{\bf x},\omega){\bf q}({\bf x},\omega){\rm d}^2{\bf x},\label{eqrepgenwsag}
\end{eqnarray}
with boundary condition
\begin{eqnarray}
{\bf W}({{\bf x}_A},{\bf x},\omega)|_{x_{3,A}={x_{3,0}}} = {\bf I}\delta({{\bf x}_{{\rm H},A}}-{{\bf x}_{\rm H}}),\label{eq9998dqq}
\end{eqnarray}
for ${\bf x}$ at ${{\partial\mathbb{D}}}_0$. 
From here onward we assume that the half-space above ${{\partial\mathbb{D}}}_0$  (including ${{\partial\mathbb{D}}}_0$) is homogeneous and isotropic;
the medium below ${{\partial\mathbb{D}}}_0$ is arbitrary inhomogeneous, anisotropic and source-free, and ${{\bf x}_A}$ is chosen at or below ${{\partial\mathbb{D}}}_0$ (hence $x_{3,A}\ge x_{3,0}$).

Without loss of generality, we can decompose ${\bf q}({\bf x},\omega)$ in the upper half-space into downgoing and upgoing plane waves.
To this end, we defined the spatial Fourier transform $\tilde u({{\bf s}},x_3,\omega)$ of a space- and frequency-dependent function $u({\bf x},\omega)$ in Eq. (\ref{eq99950b}).
For a function  of two space variables, $u({{\bf x}_A},{\bf x},\omega)$, we define the spatial Fourier transform along the horizontal components of the second space variable as
\begin{eqnarray}
&&\hspace{-.5cm}\tilde u({{\bf x}_A},{{\bf s}},x_3,\omega)=
\int_{{\mathbb{R}}^2}u({{\bf x}_A},{{\bf x}_{\rm H}},x_3,\omega)\exp\{i\omega{{\bf s}}\cdot{{\bf x}_{\rm H}}\}{\rm d}^2{{\bf x}_{\rm H}}.\label{eq999329}
\end{eqnarray}
Using these definitions and Parseval's theorem, we rewrite Eq. (\ref{eqrepgenwsag}) as
\begin{eqnarray}
{\bf q}({{\bf x}_A},\omega)=\frac{\omega^2}{4\pi^2}\int_{{\mathbb{R}}^2} \tilde{\bf W}({{\bf x}_A},{{\bf s}},{x_{3,0}},\omega)\tilde{\bf q}({{\bf s}},{x_{3,0}},\omega){\rm d}^2{{\bf s}},
 \label{eq99910}
\end{eqnarray}
with boundary condition 
\begin{eqnarray}
 \tilde{\bf W}({{\bf x}_A},{{\bf s}},{x_{3,0}},\omega)|_{x_{3,A}={x_{3,0}}}={\bf I}\exp\{i\omega{{\bf s}}\cdot{{\bf x}_{{\rm H},A}}\}.\label{eq99930}
\end{eqnarray}
In the upper half-space we relate the wave\rev{-field} vector $\tilde{\bf q}$ to downgoing and upgoing wave\rev{-field} vectors $\tilde{\bf p}^+$ and $\tilde{\bf p}^-$, respectively, via
\begin{eqnarray}
\tilde{\bf q}=
\begin{pmatrix} \tilde{\bf q}_1 \\ \tilde{\bf q}_2 \end{pmatrix} =
\begin{pmatrix} \tilde{\bf L}_1^+ & \tilde{\bf L}_1^- \\ \tilde{\bf L}_2^+ & \tilde{\bf L}_2^- \end{pmatrix}
\begin{pmatrix} \tilde{\bf p}^+\\\tilde{\bf p}^- \end{pmatrix},\quad\mbox{for}\quad x_3 \le {x_{3,0}}\label{eq305k}
\end{eqnarray}
\cite{Kennett78GJRAS, Fishman84JMP, Ursin83GEO, Wapenaar89Book}.
Next, we renormalize the downgoing and upgoing wave\rev{-field} vectors. To this end, we define the downgoing and upgoing parts of $\tilde{\bf q}_1$ as 
\begin{eqnarray}
\tilde{\bf q}_1^\pm= \tilde{\bf L}_1^\pm\tilde{\bf p}^\pm\label{eq1479}
\end{eqnarray}
and rewrite Eq. (\ref{eq305k}) as
\begin{eqnarray}
\tilde{\bf q}=
\begin{pmatrix} \tilde{\bf q}_1 \\ \tilde{\bf q}_2 \end{pmatrix} =
\underbrace{\begin{pmatrix} {\bf I}& {\bf I} \\ \tilde{\bf D}_1^+ & \tilde{\bf D}_1^- \end{pmatrix}}_{\tilde{\bf D}_1}
\underbrace{\begin{pmatrix} \tilde{\bf q}_1^+\\\tilde{\bf q}_1^- \end{pmatrix}}_{\tilde{\bf b}_1},\quad\mbox{for}\quad x_3 \le {x_{3,0}},\label{eq305kb}
\end{eqnarray}
with
\begin{eqnarray}
\tilde{\bf D}_1^\pm=\tilde{\bf L}_2^\pm(\tilde{\bf L}_1^\pm)^{-1}.\label{eq1481}
\end{eqnarray}
In Appendix \ref{AppB} we give explicit expressions for $\tilde{\bf D}_1^\pm$ for the acoustic, 
electromagnetic and elastodynamic situation.
Substitution of Eq. (\ref{eq305kb}) into Eq. (\ref{eq99910}) gives
\begin{eqnarray}
{\bf q}({{\bf x}_A},\omega)=\frac{\omega^2}{4\pi^2}\int_{{\mathbb{R}}^2} \tilde{\bf Y}_1({{\bf x}_A},{{\bf s}},{x_{3,0}},\omega)\tilde{\bf b}_1({{\bf s}},{x_{3,0}},\omega){\rm d}^2{{\bf s}},
 \label{eq99910y}
\end{eqnarray}
with
\begin{eqnarray}
 \tilde{\bf Y}_1({{\bf x}_A},{{\bf s}},{x_{3,0}},\omega)=\tilde{\bf W}({{\bf x}_A},{{\bf s}},{x_{3,0}},\omega)\tilde{\bf D}_1({{\bf s}}).\label{eq99943y}
 \end{eqnarray}
Let $\tilde{\bf F}_1({{\bf x}_A},{{\bf s}},{x_{3,0}},\omega)$ be the upper-right $N/2\,\times\, N/2$ matrix of block matrix $\tilde{\bf Y}_1({{\bf x}_A},{{\bf s}},{x_{3,0}},\omega)$. 
According to Eqs. (\ref{eq10W}), (\ref{eq99943y}) and the definition of $\tilde{\bf D}_1({{\bf s}})$  in Eq. (\ref{eq305kb}), it is defined as
\begin{eqnarray}
\tilde{\bf F}_1({{\bf x}_A},{{\bf s}},{x_{3,0}},\omega)=(\tilde{\bf W}_{11}+\tilde{\bf W}_{12}\tilde{\bf D}_1^-)({{\bf x}_A},{{\bf s}},{x_{3,0}},\omega).\label{eq8884}
\end{eqnarray}
This is a generalization of the definition of the acoustic focusing function for a horizontally layered medium, defined in Eq. (\ref{eq1135}).
From Eqs. (\ref{eq10W}) and (\ref{eq99930}) it follows that  $\tilde{\bf F}_1({{\bf x}_A},{{\bf s}},{x_{3,0}},\omega)$ obeys the boundary condition
\begin{eqnarray}
 \tilde {\bf F}_1({{\bf x}_A},{{\bf s}},{x_{3,0}},\omega)|_{x_{3,A}={x_{3,0}}}={\bf I}\exp\{i\omega{{\bf s}}\cdot{{\bf x}_{{\rm H},A}}\},\label{eq99930Yaa}
\end{eqnarray}
or, applying an inverse spatial and temporal Fourier transform,
\begin{eqnarray}
 {\bf F}_1({{\bf x}_A},{\bf x},t)|_{x_{3,A}={x_{3,0}}}={\bf I}\delta({{\bf x}_{{\rm H},A}}-{{\bf x}_{\rm H}})\delta(t),\label{eq99930Ybb}
\end{eqnarray}
for ${\bf x}$ at ${{\partial\mathbb{D}}}_0$. Hence, $ {\bf F}_1({{\bf x}_A},{\bf x},t)$ is indeed a focusing function (with ${{\bf x}_A}$ being a variable and ${\bf x}$ being the focal point at ${{\partial\mathbb{D}}}_0$).

To analyse  the upper-left $N/2\,\times\, N/2$ matrix of block matrix $\tilde{\bf Y}_1({{\bf x}_A},{{\bf s}},{x_{3,0}},\omega)$, 
we first establish some symmetry properties.
The symmetry of $\tilde{\bf W}$ follows from the Fourier transform of Eq. (\ref{eq65aws}) for the lossless situation, hence
\begin{eqnarray}\label{eq65awswn}
\tilde{\bf W}({\bf x}_A,{{\bf s}},{x_{3,0}},\omega)={\bf J}\tilde{\bf W}^*({\bf x}_A,-{{\bf s}},{x_{3,0}},\omega){\bf J}.
\end{eqnarray}
For the sub-matrices of $\tilde{\bf W}$ this implies
\begin{eqnarray}\label{eq65awswnsub}
\tilde{\bf W}_{\alpha\beta}({\bf x}_A,{{\bf s}},{x_{3,0}},\omega)={\bf J}_{\alpha\alpha}\tilde{\bf W}_{\alpha\beta}^*({\bf x}_A,-{{\bf s}},{x_{3,0}},\omega){\bf J}_{\beta\beta}
\end{eqnarray}
(no summation convention) with ${\bf J}_{11}=-{\bf J}_{22}={\bf I}$.
The symmetry of $\tilde{\bf D}_1^\pm$ follows from its explicit definitions in Appendix \ref{AppB}. 
In all cases, the following symmetry holds
\begin{eqnarray}
\tilde{\bf D}_1^+({{\bf s}})=
-\tilde{\bf D}_1^-(-{{\bf s}}).\label{eqdsym}
\end{eqnarray}
When we ignore evanescent waves at and above ${{\partial\mathbb{D}}}_0$, then $\tilde{\bf D}_1^\pm({{\bf s}})$ is real-valued, see Appendix \ref{AppB}. 
Hence, given that $\tilde{\bf F}_1({{\bf x}_A},{{\bf s}},{x_{3,0}},\omega)$ is the upper-right $N/2\,\times\, N/2$ matrix of  $\tilde{\bf Y}_1({{\bf x}_A},{{\bf s}},{x_{3,0}},\omega)$,
we derive from Eqs. (\ref{eq10W}), (\ref{eq99943y}), (\ref{eq65awswnsub}) and (\ref{eqdsym})
and the structure of $\tilde{\bf D}_1$ indicated in Eq. (\ref{eq305kb}), that the upper-left $N/2\,\times\, N/2$ matrix 
is equal to  $\tilde{\bf F}_1^*({{\bf x}_A},-{{\bf s}},{x_{3,0}},\omega)$. 
Using this in Eq. (\ref{eq99910y}) we obtain for the upper $N/2\,\times\, 1$ vector of ${\bf q}({{\bf x}_A},\omega)$ 
\begin{eqnarray}
{\bf q}_1({{\bf x}_A},\omega)&=&\frac{\omega^2}{4\pi^2}\int_{{\mathbb{R}}^2} \tilde{\bf F}_1^*({{\bf x}_A},-{{\bf s}},{x_{3,0}},\omega)\tilde{\bf q}_1^+({{\bf s}},{x_{3,0}},\omega){\rm d}^2{{\bf s}}\nonumber\\
&+&\frac{\omega^2}{4\pi^2}\int_{{\mathbb{R}}^2} \tilde{\bf F}_1({{\bf x}_A},{{\bf s}},{x_{3,0}},\omega)\tilde{\bf q}_1^-({{\bf s}},{x_{3,0}},\omega){\rm d}^2{{\bf s}}.
\label{eq99910y2}
\end{eqnarray}
Applying Parseval's theorem again, we obtain
\begin{eqnarray}
{\bf q}_1({{\bf x}_A},\omega)&=&\int_{{{\partial\mathbb{D}}}_0} {\bf F}_1^*({{\bf x}_A},{\bf x},\omega){\bf q}_1^+({\bf x},\omega){\rm d}^2{\bf x} 
+\int_{{{\partial\mathbb{D}}}_0} {\bf F}_1({{\bf x}_A},{\bf x},\omega){\bf q}_1^-({\bf x},\omega){\rm d}^2{\bf x}.
\label{eq99912av}
\end{eqnarray}
This relation has previously been derived via another route for the situations of acoustic waves (${\bf q}_1=p$) and elastodynamic waves (${\bf q}_1={\bf v}$) \cite{Wapenaar2021GJI},
\rev{but without explicitly defining the focusing function ${\bf F}_1({{\bf x}_A},{\bf x},\omega)$. Here we have an explicit expression for the Fourier transform of this focusing function (Eq. (\ref{eq8884})).}
In section \ref{sec7.3} we use Eq. (\ref{eq99912av}) as the basis for deriving Marchenko-type Green's matrix representations. 

Next, we derive a representation similar to Eq. (\ref{eq99912av}) for  ${\bf q}_2({{\bf x}_A},\omega)$.
To this end, we define the downgoing and upgoing parts of $\tilde{\bf q}_2$ as  
\begin{eqnarray}
\tilde{\bf q}_2^\pm= \tilde{\bf L}_2^\pm\tilde{\bf p}^\pm\label{eq1495}
\end{eqnarray}
and rewrite Eq. (\ref{eq305k}),
analogous to Eq. (\ref{eq305kb}), as
\begin{eqnarray}
\tilde{\bf q}=
\begin{pmatrix} \tilde{\bf q}_1 \\ \tilde{\bf q}_2 \end{pmatrix} =
\underbrace{\begin{pmatrix}  \tilde{\bf D}_2^+ & \tilde{\bf D}_2^- \\{\bf I}& {\bf I}  \end{pmatrix}}_{\tilde{\bf D}_2}
\underbrace{\begin{pmatrix} \tilde{\bf q}_2^+\\\tilde{\bf q}_2^- \end{pmatrix}}_{\tilde{\bf b}_2},\quad\mbox{for}\quad x_3 \le {x_{3,0}},\label{eq305kb2}
\end{eqnarray}
with
\begin{eqnarray}
\tilde{\bf D}_2^\pm=\tilde{\bf L}_1^\pm(\tilde{\bf L}_2^\pm)^{-1}=(\tilde{\bf D}_1^\pm)^{-1}.\label{eq1493}
\end{eqnarray}
Substitution of Eq. (\ref{eq305kb2}) into Eq. (\ref{eq99910}) gives
\begin{eqnarray}
{\bf q}({{\bf x}_A},\omega)=\frac{\omega^2}{4\pi^2}\int_{{\mathbb{R}}^2} \tilde{\bf Y}_2({{\bf x}_A},{{\bf s}},{x_{3,0}},\omega)\tilde{\bf b}_2({{\bf s}},{x_{3,0}},\omega){\rm d}^2{{\bf s}},
 \label{eq99910yk}
\end{eqnarray}
with
\begin{eqnarray}
 \tilde{\bf Y}_2({{\bf x}_A},{{\bf s}},{x_{3,0}},\omega)=\tilde{\bf W}({{\bf x}_A},{{\bf s}},{x_{3,0}},\omega)\tilde{\bf D}_2({{\bf s}}).\label{eq99943yk}
 \end{eqnarray}
Let $\tilde{\bf F}_2({{\bf x}_A},{{\bf s}},{x_{3,0}},\omega)$ be the lower-right  $N/2\,\times\, N/2$ matrix of block-matrix $\tilde{\bf Y}_2({{\bf x}_A},{{\bf s}},{x_{3,0}},\omega)$. 
According to Eqs. (\ref{eq10W}), (\ref{eq99943yk}) and the definition of $\tilde{\bf D}_2({{\bf s}})$  in Eq. (\ref{eq305kb2}), it is defined as 
\begin{eqnarray}
\tilde{\bf F}_2({{\bf x}_A},{{\bf s}},{x_{3,0}},\omega)=(\tilde{\bf W}_{21}\tilde{\bf D}_2^-+\tilde{\bf W}_{22})({{\bf x}_A},{{\bf s}},{x_{3,0}},\omega).\label{eq8898}
\end{eqnarray}
$\tilde{\bf F}_2$ is a focusing function, with similar focusing properties as $\tilde{\bf F}_1$, expressed in equations 
(\ref{eq99930Yaa}) and (\ref{eq99930Ybb}).

To analyse  the lower-left $N/2\,\times\, N/2$ matrix of block matrix $\tilde{\bf Y}_2({{\bf x}_A},{{\bf s}},{x_{3,0}},\omega)$, 
we first derive from Eqs. (\ref{eqdsym}) and (\ref{eq1493}) 
\begin{eqnarray}
\tilde{\bf D}_2^+({{\bf s}})=-\tilde{\bf D}_2^-(-{{\bf s}}).\label{eqdsym2}
\end{eqnarray}
When we ignore evanescent waves at and above ${{\partial\mathbb{D}}}_0$, then $\tilde{\bf D}_2^\pm({{\bf s}})$ is real-valued. 
Hence, given that $\tilde{\bf F}_2({{\bf x}_A},{{\bf s}},{x_{3,0}},\omega)$ is the lower-right $N/2\,\times\, N/2$ matrix of  $\tilde{\bf Y}_2({{\bf x}_A},{{\bf s}},{x_{3,0}},\omega)$,
we derive from Eqs. (\ref{eq10W}),  (\ref{eq65awswnsub}), (\ref{eq99943yk}) and (\ref{eqdsym2})
and the structure of $\tilde{\bf D}_2$ indicated in Eq. (\ref{eq305kb2}), that the  lower-left $N/2\,\times\, N/2$ matrix 
is equal to $\tilde{\bf F}_2^*({{\bf x}_A},-{{\bf s}},{x_{3,0}},\omega)$. 
Using this in Eq. (\ref{eq99910yk}) we obtain for the lower $N/2\,\times\, 1$ vector of ${\bf q}({{\bf x}_A},\omega)$
\begin{eqnarray}
{\bf q}_2({{\bf x}_A},\omega)&=&\frac{\omega^2}{4\pi^2}\int_{{\mathbb{R}}^2} \tilde{\bf F}_2^*({{\bf x}_A},-{{\bf s}},{x_{3,0}},\omega)\tilde{\bf q}_2^+({{\bf s}},{x_{3,0}},\omega){\rm d}^2{{\bf s}}\nonumber\\
&+&\frac{\omega^2}{4\pi^2}\int_{{\mathbb{R}}^2} \tilde{\bf F}_2({{\bf x}_A},{{\bf s}},{x_{3,0}},\omega)\tilde{\bf q}_2^-({{\bf s}},{x_{3,0}},\omega){\rm d}^2{{\bf s}}.
\label{eq99910y2k}
\end{eqnarray}
Applying Parseval's theorem again, we obtain
\begin{eqnarray}
{\bf q}_2({{\bf x}_A},\omega)&=&\int_{{{\partial\mathbb{D}}}_0} {\bf F}_2^*({{\bf x}_A},{\bf x},\omega){\bf q}_2^+({\bf x},\omega){\rm d}^2{\bf x} 
+\int_{{{\partial\mathbb{D}}}_0} {\bf F}_2({{\bf x}_A},{\bf x},\omega){\bf q}_2^-({\bf x},\omega){\rm d}^2{\bf x}.
\label{eq99912avk}
\end{eqnarray}

From Eqs. (\ref{eq1479}), (\ref{eq1481}) and (\ref{eq1495}) we obtain
 $\tilde{\bf q}_2^\pm = \tilde{\bf D}_1^\pm\tilde{\bf q}_1^\pm$. Substituting this into Eq. (\ref{eq99910y2k})
we obtain a representation for ${\bf q}_2({{\bf x}_A},\omega)$ in terms of $\tilde{\bf q}_1^+$ and $\tilde{\bf q}_1^-$.
Combining this with Eq. (\ref{eq99910y2}) into a single equation, we obtain Eq. (\ref{eq99910y}), with (using Eq. (\ref{eqdsym}))
\begin{eqnarray}
&&\hspace{-0.7cm}\tilde{\bf Y}_1({{\bf x}_A},{{\bf s}},{x_{3,0}},\omega)=
\begin{pmatrix} \tilde {\bf F}_1^*({{\bf x}_A},-{{\bf s}},{x_{3,0}},\omega) & \tilde {\bf F}_1({{\bf x}_A},{{\bf s}},{x_{3,0}},\omega)\\
-\tilde {\bf F}_2^*({{\bf x}_A},-{{\bf s}},{x_{3,0}},\omega)\tilde{\bf D}_1^-(-{{\bf s}}) & \tilde {\bf F}_2({{\bf x}_A},{{\bf s}},{x_{3,0}},\omega)\tilde{\bf D}_1^-({{\bf s}}) 
\end{pmatrix}.\label{eq934}
\end{eqnarray}
Once $\tilde{\bf Y}_1$ is known, the propagator matrix $\tilde{\bf W}$ follows from inverting Eq. (\ref{eq99943y}), according to
\begin{eqnarray}
 \tilde{\bf W}({{\bf x}_A},{{\bf s}},{x_{3,0}},\omega)=\tilde{\bf Y}_1({{\bf x}_A},{{\bf s}},{x_{3,0}},\omega)\{\tilde{\bf D}_1({{\bf s}})\}^{-1}.\label{eq99943invk}
 \end{eqnarray}
Equations (\ref{eq934}) and (\ref{eq99943invk}), with $\tilde{\bf D}_1({{\bf s}})$ defined in Eq. (\ref{eq305kb}),  express the propagator matrix explicitly in terms of focusing functions.

For the acoustic situation, using Eq. (\ref{eq99961bB}), 
Eqs. (\ref{eq8884}) and (\ref{eq8898})  become
\begin{eqnarray}
&&\hspace{-0.7cm}\tilde F^p({{\bf x}_A},{{\bf s}},{x_{3,0}},\omega)=\Bigl(\tilde W^{p,p}-\frac{s_{3,0}}{\rho_0}\tilde W^{p,v}\Bigr)({{\bf x}_A},{{\bf s}},{x_{3,0}},\omega),
\label{eq8884ac}\\
&&\hspace{-0.7cm}\tilde F^v({{\bf x}_A},{{\bf s}},{x_{3,0}},\omega)=\Bigl(-\frac{\rho_0}{s_{3,0}}\tilde W^{v,p}+\tilde W^{v,v}\Bigr)({{\bf x}_A},{{\bf s}},{x_{3,0}},\omega),
\label{eq8884acb}
\end{eqnarray}
with $s_3$ defined in Eq. (\ref{eq99953prsq}).
Equation (\ref{eq934}) yields for the acoustic situation
\begin{eqnarray}
&&\hspace{-0.7cm}\tilde{\bf Y}_1({{\bf x}_A},{{\bf s}},{x_{3,0}},\omega)=
\begin{pmatrix} \{\tilde F^p({{\bf x}_A},-{{\bf s}},{x_{3,0}},\omega)\}^* & \tilde F^p({{\bf x}_A},{{\bf s}},{x_{3,0}},\omega)\\
\frac{s_{3,0}}{\rho_0}\{\tilde F^v({{\bf x}_A},-{{\bf s}},{x_{3,0}},\omega)\}^* & -\frac{s_{3,0}}{\rho_0}\tilde F^v({{\bf x}_A},{{\bf s}},{x_{3,0}},\omega)
\end{pmatrix}.\label{eq934ac}
\end{eqnarray}
Using this in Eq. (\ref{eq99943invk}), together with
\begin{eqnarray}
\{\tilde{\bf D}_1({{\bf s}})\}^{-1}=\begin{pmatrix}{\frac{1}{2}} & \frac{\rho_0}{2s_{3,0}}\\{\frac{1}{2}} & -\frac{\rho_0}{2s_{3,0}}\end{pmatrix},
\end{eqnarray}
we obtain an explicit expression for $ \tilde{\bf W}({{\bf x}_A},{{\bf s}},{x_{3,0}},\omega)$ in terms of acoustic focusing functions. Transforming this to
the space-frequency domain, 
replacing $s_{3,0}$ by \rev{$\frac{1}{\omega}{\cal H}_1$ (where ${\cal H}_1$ is the square-root of the Helmholtz operator $\omega^2/c_0^2+\partial_\alpha\partial_\alpha$
in the homogeneous upper half-space \cite{Corones75JMAA, Fishman84JMP, Wapenaar89Book})}, yields \cite{Wapenaar2021GEO}
\begin{eqnarray}
&&\hspace{-0.7cm}{\bf W}({{\bf x}_A},{\bf x},\omega)=
\begin{pmatrix}  \Re\{F^p({{\bf x}_A},{\bf x},\omega)\} &\rev{ -i\omega\rho_0{\cal H}_1^{-1}({\bf x},\omega)\Im\{F^p({{\bf x}_A},{\bf x},\omega)}\}\\
\rev{\frac{1}{i\omega\rho_0}{\cal H}_1({\bf x},\omega)\Im\{F^v({{\bf x}_A},{\bf x},\omega)\}} &  \Re\{F^v({{\bf x}_A},{\bf x},\omega)\}
\end{pmatrix},\label{eq934acsf}
\end{eqnarray}
for ${\bf x}$ at ${{\partial\mathbb{D}}}_0$, \rev{where $\Im$ denotes the imaginary part}.

\subsection{Marchenko-type Green's matrix representations with focusing functions}\label{sec7.3}

We use Eqs. (\ref{eq99912av}) and (\ref{eq99912avk}) as the starting point for deriving two Marchenko-type Green's function representations. 
We follow a similar procedure as reference \cite[Appendix B]{Wapenaar2021GJI}, generalized for the different wave phenomena considered in this paper.
We replace the $N/2\,\times\, 1$ vector ${\bf q}_1({\bf x},\omega)$ by a modified version ${\bf \Gamma}_{12}({\bf x},{\bf x}_S,\omega)$ 
of the $N/2\,\times\, N/2$ Green's matrix ${\bf G}_{12}({\bf x},{\bf x}_S,\omega)$.
Here ${\bf G}_{12}({\bf x},{\bf x}_S,\omega)$ stands for the ${\bf q}_1$-type field observed at ${\bf x}$, in response to a unit ${\bf d}_2$-type source at ${\bf x}_S$.
We choose ${\bf x}_S=({\bf x}_{{\rm H},S},x_{3,S})$  in the upper half-space, at a vanishing distance $\epsilon$ above ${{\partial\mathbb{D}}_0}$, hence, $x_{3,S}={x_{3,0}}-\epsilon$.
Our aim is to modify this Green's matrix such that  for ${\bf x}$ at ${{\partial\mathbb{D}}_0}$ (i.e., just below the source) its downgoing part simplifies to
\begin{eqnarray}
{\bf \Gamma}_{12}^+({\bf x},{\bf x}_S,\omega)|_{x_3={x_{3,0}}}&=&{\bf I}\delta({\bf x}_{\rm H}-{\bf x}_{{\rm H},S}).\label{eq4Gag}
\end{eqnarray}
First we derive the properties of the downgoing part of ${\bf G}_{12}({\bf x},{\bf x}_S,\omega)$ for ${\bf x}$ at ${{\partial\mathbb{D}}_0}$.
To this end, consider the inverse of Eq. (\ref{eq305kb}), which reads
\begin{eqnarray}
\begin{pmatrix} \tilde{\bf q}_1^+\\\tilde{\bf q}_1^- \end{pmatrix}=
\underbrace{\begin{pmatrix}-(\tilde{\bf \Delta}_1)^{-1}\tilde{\bf D}_1^-& (\tilde{\bf \Delta}_1)^{-1}\\
(\tilde{\bf \Delta}_1)^{-1}\tilde{\bf D}_1^+ & -(\tilde{\bf \Delta}_1)^{-1} \end{pmatrix}}_{(\tilde{\bf D}_1)^{-1}}
\begin{pmatrix} \tilde{\bf q}_1 \\ \tilde{\bf q}_2 \end{pmatrix},
\label{eq305kbinv}
\end{eqnarray}
for $\, x_3 \le {x_{3,0}}$, with
\begin{eqnarray}
\tilde{\bf \Delta}_1=\tilde{\bf D}_1^+- \tilde{\bf D}_1^-.\label{eq2107}
\end{eqnarray}
The upper-right $N/2\,\times\, N/2$ matrix $(\tilde{\bf \Delta}_1)^{-1}$ transforms $\tilde{\bf q}_2$ into a downgoing field vector $\tilde{\bf q}_1^+$.
In a similar way, this matrix transforms a unit ${\bf d}_2$-type source in a homogeneous half-space into the downgoing part 
of the Green's matrix  $\tilde{\bf G}_{12}$ just below this source, according to
\begin{eqnarray}
&&\hspace{-.7cm}\lim_{x_3\downarrow x_{3,S}}\tilde {\bf G}_{12}^+({\bf s},x_3,{\bf 0},x_{3,S},\omega)=\{\tilde{\bf \Delta}_1({\bf s})\}^{-1}.\label{eq325}
\end{eqnarray}
Explicit expressions for $(\tilde{\bf \Delta}_1)^{-1}$ are given in Appendix \ref{AppB}.
In Eq. (\ref{eq325}) the source is located at $({\bf 0}, x_{3,S})$.
Next, we consider ${\bf G}_{12}({\bf x},{\bf x}_S,\omega)$ for a laterally shifted source position $({\bf x}_{{\rm H},S}, x_{3,S})$.
Applying a spatial Fourier transform along the horizontal source coordinate ${\bf x}_{{\rm H},S}$, using Eq. (\ref{eq999329}) with ${\bf x}_{\rm H}$ replaced by ${\bf x}_{{\rm H},S}$, 
yields  $\tilde{\bf G}_{12}({\bf x},{\bf s},x_{3,S},\omega)$. For the downgoing part just below the source we obtain a phase-shifted version of 
the Green's function of Eq. (\ref{eq325}), according to
\begin{eqnarray}
\lim_{x_3\downarrow x_{3,S}}\tilde{\bf G}_{12}^+({\bf x},{\bf s},x_{3,S},\omega)
=\{\tilde{\bf \Delta}_1({{\bf s}})\}^{-1}\exp\{i\omega{\bf s}\cdot{\bf x}_{\rm H}\}.\label{eq330k}
\end{eqnarray}
This suggests to define the modified Green's matrix as
\begin{eqnarray}
\tilde{\bf \Gamma}_{12}({\bf x},{\bf s},x_{3,S},\omega)=\tilde{\bf G}_{12}({\bf x},{\bf s},x_{3,S},\omega)\tilde{\bf \Delta}_1({\bf s}),\label{eq331}
\end{eqnarray}
such that
\begin{eqnarray}
\lim_{x_{3}\downarrow x_{3,S}}\tilde{\bf \Gamma}_{12}^+ ({\bf x},{\bf s},x_{3,S},\omega)
={\bf I}\exp\{i\omega{\bf s}\cdot{\bf x}_{\rm H}\}.\label{eq544}
\end{eqnarray}
Transforming this back to the space domain yields indeed Eq. (\ref{eq4Gag}).
Next, we define the reflection response ${\bf R}_{12}({\bf x},{\bf x}_S,\omega)$
of the medium below ${{\partial\mathbb{D}}_0}$ 
as the upgoing part of the modified Green's function ${\bf \Gamma}_{12}({\bf x},{\bf x}_S,\omega)$,
for ${\bf x}$ at ${{\partial\mathbb{D}}_0}$, hence
\begin{eqnarray}
{\bf R}_{12}({\bf x},{\bf x}_S,\omega)&=&{\bf \Gamma}_{12}^-({\bf x},{\bf x}_S,\omega).
\label{eq4Rag}
\end{eqnarray}
Substituting  ${\bf q}_1({\bf x}_A,\omega)={\bf \Gamma}_{12}({\bf x}_A,{\bf x}_S,\omega)$ and ${\bf q}_1^\pm({\bf x},\omega)={\bf \Gamma}_{12}^\pm({\bf x},{\bf x}_S,\omega)$ into Eq. (\ref{eq99912av}),
using Eqs. (\ref{eq4Gag}) and (\ref{eq4Rag}), gives
 \begin{eqnarray}
{\bf \Gamma}_{12}({\bf x}_A,{\bf x}_S,\omega)&=&\int_{{{\partial\mathbb{D}}_0}} {\bf F}_1({\bf x}_A,{\bf x},\omega){\bf R}_{12}({\bf x},{\bf x}_S,\omega){\rm d}^2{\bf x}
+{\bf F}_1^*({\bf x}_A,{\bf x}_S,\omega),\label{eq339}
\end{eqnarray}
for $x_{3,A} \ge {x_{3,0}}$. This is the first Marchenko-type representation.
In a similar way, we derive a second Marchenko-type representation, for  a modified version ${\bf \Gamma}_{22}({\bf x},{\bf x}_S,\omega)$ of the Green's matrix ${\bf G}_{22}({\bf x},{\bf x}_S,\omega)$.
We derive the properties of the downgoing part of ${\bf G}_{22}({\bf x},{\bf x}_S,\omega)$ for ${\bf x}$ at ${{\partial\mathbb{D}}}_0$.
Consider the inverse of Eq. (\ref{eq305kb2}), which reads 
\begin{eqnarray}
\begin{pmatrix} \tilde{\bf q}_2^+\\\tilde{\bf q}_2^- \end{pmatrix}=
\underbrace{\begin{pmatrix} (\tilde{\bf \Delta}_2)^{-1} & -(\tilde{\bf \Delta}_2)^{-1}\tilde{\bf D}_2^-\\
 -(\tilde{\bf \Delta}_2)^{-1} & (\tilde{\bf \Delta}_2)^{-1}\tilde{\bf D}_2^+  \end{pmatrix}}_{(\tilde{\bf D}_2)^{-1}}
\begin{pmatrix} \tilde{\bf q}_1 \\ \tilde{\bf q}_2 \end{pmatrix},
\label{eq305kb2inv}
\end{eqnarray}
for $x_3 \le {x_{3,0}}$, with
\begin{eqnarray}
\tilde{\bf \Delta}_2=\tilde{\bf D}_2^+- \tilde{\bf D}_2^-.\label{eq2118}
\end{eqnarray}
The upper-right $N/2\,\times\, N/2$ matrix $-(\tilde{\bf \Delta}_2)^{-1}\tilde{\bf D}_2^-$
transforms $\tilde{\bf q}_2$ into a downgoing field vector $\tilde{\bf q}_2^+$.
In a similar way, this matrix transforms a unit ${\bf d}_2$-type source in a homogeneous half-space into the downgoing part of the Green's matrix
$\tilde{\bf G}_{22}$ just below this source, according to
\begin{eqnarray}
&&\hspace{-1.2cm}\lim_{x_3\downarrow x_{3,S}}\tilde {\bf G}_{22}^+({\bf s},x_3,{\bf 0},x_{3,S},\omega)=-\{\tilde{\bf \Delta}_2({{\bf s}})\}^{-1}\tilde{\bf D}_2^-({\bf s}).\label{eq325q}
\end{eqnarray}
Explicit expressions for $-(\tilde{\bf \Delta}_2)^{-1}\tilde{\bf D}_2^-$ are given in Appendix \ref{AppB}.
Similar steps as below Eq. (\ref{eq325}) lead to
\begin{eqnarray}
&&\hspace{-1.2cm}\lim_{x_3\downarrow x_{3,S}}\tilde {\bf G}_{22}^+({\bf x},{\bf s},x_{3,S},\omega)=
-\{\tilde{\bf \Delta}_2({{\bf s}})\}^{-1}\tilde{\bf D}_2^-({\bf s})\exp\{i\omega{\bf s}\cdot{\bf x}_{\rm H}\}.
\label{eq325qh}
\end{eqnarray}
This suggests to define the modified Green's matrix as
\begin{eqnarray}
&&\hspace{-0.7cm}\tilde{\bf \Gamma}_{22}({\bf x},{\bf s},x_{3,S},\omega)=
-\tilde{\bf G}_{22}({\bf x},{\bf s},x_{3,S},\omega)\{\tilde{\bf D}_2^-({\bf s})\}^{-1}\tilde{\bf \Delta}_2({\bf s}),\label{eq331q}
\end{eqnarray}
such that 
\begin{eqnarray}
{\bf \Gamma}_{22}^+({\bf x},{\bf x}_S,\omega)|_{x_3={x_{3,0}}}&=&{\bf I}\delta({\bf x}_{\rm H}-{\bf x}_{{\rm H},S}).\label{eq4Gagk}
\end{eqnarray}
We define the reflection response ${\bf R}_{22}({\bf x},{\bf x}_S,\omega)$ as
\begin{eqnarray}
{\bf R}_{22}({\bf x},{\bf x}_S,\omega)&=&{\bf \Gamma}_{22}^-({\bf x},{\bf x}_S,\omega)
\label{eq4Ragq}
\end{eqnarray}
for ${\bf x}$ at ${{\partial\mathbb{D}}_0}$. 
Substituting  ${\bf q}_2({\bf x}_A,\omega)={\bf \Gamma}_{22}({\bf x}_A,{\bf x}_S,\omega)$ and ${\bf q}_2^\pm({\bf x},\omega)={\bf \Gamma}_{22}^\pm({\bf x},{\bf x}_S,\omega)$ 
into Eq. (\ref{eq99912avk}), using Eqs. (\ref{eq4Gagk}) and (\ref{eq4Ragq}), gives
 \begin{eqnarray}
{\bf \Gamma}_{22}({\bf x}_A,{\bf x}_S,\omega)&=&\int_{{{\partial\mathbb{D}}_0}} {\bf F}_2({\bf x}_A,{\bf x},\omega){\bf R}_{22}({\bf x},{\bf x}_S,\omega){\rm d}^2{\bf x}
+{\bf F}_2^*({\bf x}_A,{\bf x}_S,\omega),
\label{eq3392}
\end{eqnarray}
for $x_{3,A} \ge {x_{3,0}}$. This is the second Marchenko-type representation.

For the acoustic case, using Eqs. (\ref{eq99961bB}) and (\ref{eq99961bBg}),
Eqs. (\ref{eq331}) and (\ref{eq331q}) become
\begin{eqnarray}
\tilde\Gamma_{12}({\bf x},{\bf s},x_{3,S},\omega)&=&\frac{2s_{3,0}}{\rho_0}\tilde G^{p,q}({\bf x},{\bf s},x_{3,S},\omega),\label{eq331ac}\\
&=&-\frac{2}{i\omega\rho_0}\partial_{3,S}\tilde G^{p,q}({\bf x},{\bf s},x_{3,S},\omega)\nonumber\\
\tilde\Gamma_{22}({\bf x},{\bf s},x_{3,S},\omega)&=&2\tilde G^{v,q}({\bf x},{\bf s},x_{3,S},\omega),\label{eq331qac}
\end{eqnarray}
or, in the space-frequency domain (using Eq. (\ref{eq1125})),
\begin{eqnarray}
\Gamma_{12}({\bf x},{\bf x}_S,\omega)&=&2G^{p,f}({\bf x},{\bf x}_S,\omega),\\
\Gamma_{22}({\bf x},{\bf x}_S,\omega)&=&2G^{v,q}({\bf x},{\bf x}_S,\omega).
\end{eqnarray}
For this situation, $R_{12}({\bf x},{\bf x}_S,\omega)$ and $R_{22}({\bf x},{\bf x}_S,\omega)$ in the representations of Eqs. (\ref{eq339}) and (\ref{eq3392}) are the
upgoing parts of $2G^{p,f}({\bf x},{\bf x}_S,\omega)$ and $2G^{v,q}({\bf x},{\bf x}_S,\omega)$, respectively, for ${\bf x}$ at ${{\partial\mathbb{D}}}_0$. 
Note that, according to Eq. (\ref{eq65a}), $G^{v,q}({\bf x},{\bf x}_S,\omega)=-G^{p,f}({\bf x}_S,{\bf x},\omega)$.

In their general form, the representations of Eqs. (\ref{eq339}) and (\ref{eq3392}) are generalizations of previously derived representations for the 3D Marchenko method for acoustic 
\cite{Wapenaar2013PRL, Slob2014GEO, Neut2015GJI} and elastodynamic wave fields \cite{Wapenaar2014GJI, Costa2014PRE} (but note that
the subscripts 1 and 2 of the focusing functions have a different meaning than in those papers; here they 
refer  to the wave-field components ${\bf q}_1$ and ${\bf q}_2$). 
In most previous work on the 3D Marchenko method, one of the underlying assumptions is that the wave field can
be decomposed into downgoing and upgoing waves in the interior of the medium. Only recently several authors proposed to avoid decomposition inside the medium
\cite{Kiraz2021JASA, Diekmann2021PRR, Wapenaar2021GJI}. The representations discussed in this section expand on this.
\rev{In these representations, decomposition into downgoing and upgoing waves and negligence of evanescent waves occurs only in the upper half-space.}
However, inside the medium no wave-field decomposition takes place. Moreover,
evanescent waves inside the medium (for example in high-velocity layers) are accounted for by the representations of Eqs. (\ref{eq339}) and (\ref{eq3392}).
These representations, transformed to the time domain, form the basis for the development of Marchenko schemes, 
aiming at resolving the  focusing functions ${\bf F}_1$ and ${\bf F}_2$ from 
the reflection responses ${\bf R}_{12}$ and ${\bf R}_{22}$ at the acquisition boundary ${{\partial\mathbb{D}}}_0$. Such schemes have been successfully developed for precritical
acoustic data \cite{Ravasi2016GJI, Staring2018GEO, Jia2018GEO, Brackenhoff2019JGR, Pereira2019SEG, Zhang2020GEO, Staring2020GP}.
For more complex situations, research on retrieving the focusing functions from reflection responses is ongoing (for example \cite{Reinicke2020GEO}). 
Once the focusing functions are found, they can be used to define the propagator matrix via Eqs. (\ref{eq934}) and (\ref{eq99943invk}),
which can subsequently be used in the representations of sections \ref{sec5} and \ref{sec6}.

\section{Conclusions}\label{sec8}

We have discussed different types of wave-field representation in a systematic way. Classical wave-field representations contain Green's functions. 
Starting with a unified matrix-vector wave equation, we have formulated wave-field representations with Green's matrices, analogous to the classical representations.
For example, the classical Kirchhoff-Helmholtz integral follows as a special case of the unified representation with the Green's matrix. 
Another special case is the classical homogeneous Green's function representation.

Using the same matrix-vector formalism, we  formulated wave-field representations with propagator matrices.
Unlike a Green's matrix, a propagator matrix depends only on the medium parameters between the two depth levels for which this matrix is defined.
The representations with the propagator matrices have a similar form as those with the Green's matrices.
An important difference is that the boundary integrals in the representations with the propagator matrices are single-sided. 

We also formulated representations containing a mix of Green's matrices and propagator matrices.
A special case is the single-sided homogeneous Green's function representation, as the counterpart of the classical closed-boundary homogeneous Green's function representation.

We have shown that the propagator matrix is related to Marchenko-type focusing functions. We have used this relation to reformulate the representations with the propagator matrix
into representations with focusing functions. For the acoustic situation, these focusing functions can be retrieved from the single-sided reflection response of the medium by applying the 
Marchenko method. For more complex situations, research on retrieving these focusing functions from the reflection response is ongoing. Once the focusing functions are known,
they can be used to construct the propagator matrix. Subsequently, the propagator matrix can be used in the representations to obtain the wave field inside the medium which, in turn,
can be used for example for imaging or monitoring. 
Unlike earlier imaging methods that use the propagator matrix, this is a data-driven approach (since the propagator matrix is retrieved from the reflection response) and hence
it does not require a detailed model of the medium.

\section*{Acknowledgments}

The author thanks Roel Snieder, Sjoerd de Ridder and Evert Slob for fruitful discussions 
\rev{and two anonymous reviewers for their positive evaluations of the original manuscript and their useful suggestions for further improvements}.
This research is funded by the European Research Council (ERC) under the European Union's Horizon 2020 research and innovation programme (grant agreement No: 742703).

\appendix

\section{Explicit expressions for some matrices in the homogeneous isotropic upper half-space}\label{AppB}

We derive explicit expressions for $\tilde{\bf D}_1^\pm$, $(\tilde{\bf \Delta}_1)^{-1}$ and $-(\tilde{\bf \Delta}_2)^{-1}\tilde{\bf D}_2^-$ 
in the homogeneous isotropic upper half-space $x_3\le x_{3,0}$
for  the acoustic, electromagnetic and elastodynamic situation.

\subsection*{Unified expressions in horizontal slowness domain}

For the lossless homogeneous isotropic upper half-space $x_3\le x_{3,0}$, we transform the unified matrix-vector wave equation (\ref{eq2.1}) to the horizontal slowness domain, 
using the spatial Fourier transform of Eq. (\ref{eq99950b}).
This yields 
\begin{eqnarray}\label{eq2.1BQ}
\partial_3\tilde{\bf q} - \tilde{\bf A}\tilde{\bf q} =\tilde{\bf d},
\end{eqnarray}
with
\begin{eqnarray}\label{Aeq7mvbbprffB}
\tilde{\bf q}=\begin{pmatrix} \tilde{\bf q}_1 \\ \tilde{\bf q}_2 \end{pmatrix},\quad
                \tilde{\bf d}=\begin{pmatrix} \tilde{\bf d}_1 \\ \tilde{\bf d}_2 \end{pmatrix},\quad
\tilde{\bf A}= \begin{pmatrix}\tilde{\bf A}_{11}      & \tilde{\bf A}_{12} \\
               \tilde{\bf A}_{21} & \tilde{\bf A}_{22}    \end{pmatrix}.
\end{eqnarray}
The symmetry property defined by Eq. (\ref{eqsym}) transforms to
\begin{eqnarray}
\{\tilde{\bf A}(-{{\bf s}},\omega)\}^t{\bf N}&=&-{\bf N}\tilde{\bf A}({{\bf s}},\omega).\label{eqsymB}
\end{eqnarray}
We decompose matrix $\tilde{\bf A}$ as follows \cite{Kennett78GJRAS, Ursin83GEO, Wapenaar89Book}
\begin{eqnarray}
\tilde{\bf A}&=&\tilde{\bf L}\tilde{{{\mbox{\boldmath $\Lambda$}}}}\tilde{\bf L}^{-1},\label{eqBB5}
\end{eqnarray}
with
\begin{eqnarray}\label{Aeq7mvbbprffBB}
\tilde{\bf L}= \begin{pmatrix}\tilde{\bf L}_1^+      & \tilde{\bf L}_1^- \\
              \tilde{\bf L}_2^+ & \tilde{\bf L}_2^-    \end{pmatrix},\quad
\tilde{{{\mbox{\boldmath $\Lambda$}}}}= \begin{pmatrix}\tilde{{{\mbox{\boldmath $\Lambda$}}}}^+      &{\bf O}\\
              {\bf O} & \tilde{{{\mbox{\boldmath $\Lambda$}}}}^-   \end{pmatrix}.
\end{eqnarray}
With a proper scaling of the columns of matrix $\tilde{\bf L}$, these matrices obey the following symmetry properties
\begin{eqnarray}
\{\tilde{\bf L}(-{{\bf s}})\}^t{\bf N}&=&-{\bf N}\{\tilde{\bf L}({{\bf s}})\}^{-1},\label{eqsymLB}\\
\{\tilde{{{\mbox{\boldmath $\Lambda$}}}}(-{{\bf s}},\omega)\}^t{\bf N}&=&-{\bf N}\tilde{{{\mbox{\boldmath $\Lambda$}}}}({{\bf s}},\omega).\label{eqsymBLambda}
\end{eqnarray}
Substitution of Eq. (\ref{eqBB5}) into Eq. (\ref{eq2.1BQ}) and premultiplying all terms with $\tilde{\bf L}^{-1}$ yields
\begin{eqnarray}\label{eq2.1B}
\partial_3\tilde{\bf p} -\tilde{{{\mbox{\boldmath $\Lambda$}}}}\tilde{\bf p} =\tilde{\bf L}^{-1}\tilde{\bf d},
\end{eqnarray}
with
\begin{eqnarray}
\tilde{\bf p}=\tilde{\bf L}^{-1}\tilde{\bf q}=\begin{pmatrix} \tilde{\bf p}^+ \\ \tilde{\bf p}^- \end{pmatrix},\label{eqBB4}
\end{eqnarray}
where $\tilde{\bf p}^+$ and $\tilde{\bf p}^-$ are wave\rev{-field} vectors containing downgoing ($+$) and upgoing ($-$) waves, respectively.
Following Eqs. (\ref{eq1481}), (\ref{eq1493}), (\ref{eq2107}) and (\ref{eq2118}) we define
\begin{eqnarray}
&&\hspace{-1.4cm}\tilde{\bf D}_1^\pm({{\bf s}})=\tilde{\bf L}_2^\pm(\tilde{\bf L}_1^\pm)^{-1},\quad \{\tilde{\bf \Delta}_1({{\bf s}})\}^{-1}=(\tilde{\bf D}_1^+- \tilde{\bf D}_1^-)^{-1}, \label{eq1481B}\\
&&\hspace{-1.4cm}\tilde{\bf D}_2^\pm({{\bf s}})=\tilde{\bf L}_1^\pm(\tilde{\bf L}_2^\pm)^{-1},\quad \{\tilde{\bf \Delta}_2({{\bf s}})\}^{-1}=(\tilde{\bf D}_2^+- \tilde{\bf D}_2^-)^{-1}. \label{eq1481Bg}
\end{eqnarray}

\subsection*{Acoustic wave equation}\label{AppB2}

For the acoustic wave equation the $1\times 1$ sub-matrices of $\tilde{\bf A}$ are
\begin{eqnarray}
\tilde{\bf A}_{12}=i\omega \rho,\quad \tilde{\bf A}_{21}= i\omega(\kappa-s_r^2/\rho),
\quad \tilde{\bf A}_{11}=\tilde{\bf A}_{22}=0,
\end{eqnarray}
with radial slowness $s_r$ defined as
\begin{eqnarray}
s_r^2=s_\alpha s_\alpha=s_1^2+s_2^2.\label{eqBBB13}
\end{eqnarray} 
Here $\kappa$ and $\rho$ are the compressibility and mass density of the homogeneous upper half-space.
For convenience, here and in the remainder of this appendix we do not use subscripts 0 to indicate parameters of the upper half-space. 
The $1\times 1$ sub-matrices of $\tilde{\bf L}$ and $\tilde{{{\mbox{\boldmath $\Lambda$}}}}$ are
\begin{eqnarray}
\tilde{\bf L}_1^\pm=\Bigl(\frac{\rho}{2s_3}\Bigr)^{1/2},\quad \tilde{\bf L}_2^\pm= \pm\Bigl(\frac{s_3}{2\rho}\Bigr)^{1/2},\quad
\tilde{{{\mbox{\boldmath $\Lambda$}}}}^\pm=\pm i\omega s_3,\label{eqBB13}
\end{eqnarray}
with the vertical slowness $s_3$ defined as 
\begin{eqnarray}
s_3=\begin{cases}
\sqrt{1/c^2-s_r^2}, &\mbox{for } s_r^2\le 1/c^2,\\
i\sqrt{s_r^2-1/c^2}, &\mbox{for } s_r^2> 1/c^2,
\end{cases}\label{eq99953prsq}
\end{eqnarray}
with propagation velocity $c$ defined as $c=1/\sqrt{\kappa\rho}$. The two expressions in Eq. (\ref{eq99953prsq}) represent propagating and evanescent waves, respectively.
Upon substitution of Eq. (\ref{eqBB13}) into Eqs. (\ref{eq1481B}) and (\ref{eq1481Bg}) we obtain
\begin{eqnarray}
&&\hspace{-0.7cm}\tilde {\bf D}_1^\pm({{\bf s}}) = \pm\frac{s_3}{\rho},\quad
\{\tilde{\bf \Delta}_1({{\bf s}})\}^{-1}=\frac{\rho}{2s_3},\label{eq99961bB}\\
&&\hspace{-0.7cm}\tilde {\bf D}_2^\pm({{\bf s}}) = \pm\frac{\rho}{s_3},\quad
-\{\tilde{\bf \Delta}_2({{\bf s}})\}^{-1}\tilde{\bf D}_2^-({{\bf s}})=\frac{1}{2}.\label{eq99961bBg}
\end{eqnarray}

\subsection*{Electromagnetic wave equation}

For the electromagnetic wave equation the $2\times 2$ sub-matrices of $\tilde{\bf A}$ are
\begin{eqnarray}
&&\hspace{-0.7cm}\tilde{\bf A}_{12}=i\omega \begin{pmatrix} 
\mu-\frac{s_1^2}{\varepsilon}&-\frac{s_1s_2}{\varepsilon}\\
-\frac{s_1s_2}{\varepsilon}&\mu-\frac{s_2^2}{\varepsilon}
\end{pmatrix},\,
\tilde{\bf A}_{21}=i\omega \begin{pmatrix} 
\varepsilon-\frac{s_2^2}{\mu}&\frac{s_1s_2}{\mu}\\
\frac{s_1s_2}{\mu}&\varepsilon-\frac{s_1^2}{\mu}
\end{pmatrix},\\
&&\hspace{-0.7cm}\tilde{\bf A}_{11}=\tilde{\bf A}_{22}={\bf O}.
\end{eqnarray}
Here $\varepsilon$ and $\mu$ are the permittivity and permeability of the upper half-space.
The $2\times 2$ sub-matrices of $\tilde{\bf L}$ and $\tilde{{{\mbox{\boldmath $\Lambda$}}}}$ are
\begin{eqnarray}
\tilde{\bf L}_1^\pm&=&\frac{i}{\sqrt{2}s_r}\begin{pmatrix}
s_1\bigl(\frac{s_3}{\varepsilon}\bigr)^{1/2} & -s_2\bigl(\frac{\mu}{s_3}\bigr)^{1/2}\\
s_2\bigl(\frac{s_3}{\varepsilon}\bigr)^{1/2} & s_1\bigl(\frac{\mu}{s_3}\bigr)^{1/2}
\end{pmatrix},\label{eqBBB23}\\
\tilde{\bf L}_2^\pm&=&\pm\frac{i}{\sqrt{2}s_r}\begin{pmatrix}
s_1\bigl(\frac{\varepsilon}{s_3}\bigr)^{1/2} & -s_2\bigl(\frac{s_3}{\mu}\bigr)^{1/2}\\
s_2\bigl(\frac{\varepsilon}{s_3}\bigr)^{1/2} & s_1\bigl(\frac{s_3}{\mu}\bigr)^{1/2}
\end{pmatrix},\label{eqBBB24}\\
\tilde{{{\mbox{\boldmath $\Lambda$}}}}^\pm&=&
\pm i\omega\begin{pmatrix}s_3 & 0  \\ 0 & s_3 
\end{pmatrix},
\end{eqnarray}
with $s_r$ and $s_3$ defined in Eqs. (\ref{eqBBB13}) and (\ref{eq99953prsq}), respectively, but this time with propagation velocity $c$ defined as $c=1/\sqrt{\varepsilon\mu}$. 
Upon substitution of Eqs. (\ref{eqBBB23}) and (\ref{eqBBB24}) into Eqs. (\ref{eq1481B}) and (\ref{eq1481Bg}) we obtain
\begin{eqnarray}
\tilde{\bf D}_1^\pm({{\bf s}})&=&\pm\frac{1}{\mu s_3}\begin{pmatrix}
c^{-2}-s_2^2 & s_1s_2\\s_1s_2&c^{-2}-s_1^2
\end{pmatrix},\\
\{\tilde{\bf \Delta}_1({{\bf s}})\}^{-1}&=&\frac{1}{2\varepsilon s_3}\begin{pmatrix}
c^{-2}-s_1^2 & -s_1s_2\\-s_1s_2&c^{-2}-s_2^2
\end{pmatrix},\\
\tilde{\bf D}_2^\pm({{\bf s}})&=&\pm\frac{1}{\varepsilon s_3}\begin{pmatrix}
c^{-2}-s_1^2 & -s_1s_2\\-s_1s_2&c^{-2}-s_2^2
\end{pmatrix},\\
-(\tilde{\bf \Delta}_2)^{-1}\tilde{\bf D}_2^-&=&\begin{pmatrix} \frac{1}{2} & 0 \\ 0 & \frac{1}{2} \end{pmatrix}.
\end{eqnarray}

\subsection*{Elastodynamic wave equation}

For the elastodynamic wave equation the $3\times 3$ sub-matrices of $\tilde{\bf A}$ are
\begin{eqnarray}
&&\hspace{-1cm}\tilde{\bf A}_{11}=i\omega \begin{pmatrix} 
0&0&-s_1\\0&0&-s_2\\-\frac{\lambda}{\lambda+2\mu}s_1&-\frac{\lambda}{\lambda+2\mu}s_2&0
\end{pmatrix},\\
&&\hspace{-1cm}\tilde{\bf A}_{12}=i\omega \begin{pmatrix} 
\frac{1}{\mu}&0&0\\0&\frac{1}{\mu}&0\\0&0&\frac{1}{\lambda+2\mu}
\end{pmatrix},
\end{eqnarray}
\begin{eqnarray}
&&\hspace{-1cm}\tilde{\bf A}_{21}=i\omega \begin{pmatrix} 
\rho-\nu_1s_1^2-\mu s_2^2&\,-(\nu_2+\mu)s_1s_2&\,0\\
-(\nu_2+\mu)s_1s_2 &\, \rho-\mu s_1^2-\nu_1s_2^2 &\, 0\\
0&\,0&\,\rho
\end{pmatrix},\\
&&\hspace{-1cm} \tilde{\bf A}_{22}=(\tilde{\bf A}_{11})^t,
\end{eqnarray}
with
\begin{eqnarray}
\nu_1=4\mu\Bigl(\frac{\lambda+\mu}{\lambda+2\mu}\Bigr),\,\nu_2=2\mu\Bigl(\frac{\lambda}{\lambda+2\mu}\Bigr),
\end{eqnarray}
where $\lambda$ and $\mu$ are the Lam\'e parameters  and $\rho$ the mass density of the upper half-space.
The $3\times 3$ sub-matrices of $\tilde{\bf L}$ and $\tilde{{{\mbox{\boldmath $\Lambda$}}}}$ are
%
\begin{eqnarray}
&&\hspace{-.7cm}\tilde{\bf L}_1^\pm=\frac{1}{(2\rho)^{1/2}}\begin{pmatrix}\pm \frac{s_1}{(s_3^P)^{1/2}} & \mp\frac{s_1(s_3^S)^{1/2}}{s_r} & \mp\frac{s_2}{c_S s_r (s_3^S)^{1/2}}\\
\pm \frac{s_2}{(s_3^P)^{1/2}} & \mp\frac{s_2(s_3^S)^{1/2}}{s_r} & \pm\frac{s_1}{c_S s_r (s_3^S)^{1/2}}\\
 (s_3^P)^{1/2} & \frac{s_r}{(s_3^S)^{1/2}}&0
\end{pmatrix},\label{eqBBB35}\\
&&\hspace{-.7cm}\tilde{\bf L}_2^\pm=\Bigl(\frac{\rho}{2}\Bigr)^{1/2}c_S^2\begin{pmatrix}  2s_1(s_3^P)^{1/2} & -\frac{s_1(c_S^{-2}-2s_r^2)}{s_r(s_3^S)^{1/2}} & -\frac{s_2(s_3^S)^{1/2}}{c_Ss_r}\\
  2s_2(s_3^P)^{1/2} & -\frac{s_2(c_S^{-2}-2s_r^2)}{s_r(s_3^S)^{1/2}} & \frac{s_1(s_3^S)^{1/2}}{c_Ss_r}\\
 \pm\frac{(c_S^{-2}-2s_r^2)}{(s_3^P)^{1/2}} & \pm 2s_r (s_3^S)^{1/2} & 0
\end{pmatrix},\label{eqBBB36}\\
&&\hspace{-.7cm}\tilde{{{\mbox{\boldmath $\Lambda$}}}}^\pm=
\pm i\omega\begin{pmatrix}s_3^P & 0 & 0 \\ 0 & s_3^S & 0 \\ 0 & 0 & s_3^S
\end{pmatrix},
\end{eqnarray}
with $s_r$ defined in Eq. (\ref{eqBBB13}), and the vertical slownesses $s_3^P$ and $s_3^S$ defined as
\begin{eqnarray}
s_3^{P,S}=\begin{cases}
\sqrt{1/c_{P,S}^2-s_r^2}, &\mbox{for } s_r^2\le 1/c_{P,S}^2,\\
i\sqrt{s_r^2-1/c_{P,S}^2}, &\mbox{for } s_r^2> 1/c_{P,S}^2.
\end{cases}\label{eq324}
\end{eqnarray}
Here $c_P$ and $c_S$ are the $P$- and $S$-wave velocities of the upper half-space,
defined as $c_P=\sqrt{(\lambda+2\mu)/\rho}$ and $c_S=\sqrt{\mu/\rho}$, respectively.
Upon substitution of Eqs. (\ref{eqBBB35}) and (\ref{eqBBB36}) into Eqs. (\ref{eq1481B}) and (\ref{eq1481Bg}) we obtain
\begin{eqnarray}
&&\hspace{-.7cm}\tilde{\bf D}_1^\pm({\bf s})=\frac{\rho c_S^2}{s_3^P s_3^S +s_r^2}
\begin{pmatrix} \pm((c_S^{-2}-s_2^2)s_3^P + s_2^2s_3^S) & \pm s_1s_2(s_3^P -s_3^S) & -s_1(c_S^{-2}-2S)\\
 \pm s_1s_2(s_3^P -s_3^S) & \pm((c_S^{-2}-s_1^2)s_3^P + s_1^2s_3^S)&  -s_2(c_S^{-2}-2S)\\
 s_1(c_S^{-2}-2S) &s_2(c_S^{-2}-2S) &\pm s_3^S c_S^{-2}
\end{pmatrix},\nonumber\\
&&\\
&&\hspace{-.7cm}\{\tilde{\bf \Delta}_1({{\bf s}})\}^{-1}=
\frac{1}{2\rho}\begin{pmatrix}
\frac{s_1^2}{s_3^P }+\Bigl(\frac{1}{c_S^2}-s_1^2\Bigr)\frac{1}{s_3^S} & \Bigl(\frac{1}{s_3^P }-\frac{1}{s_3^S}\Bigr)s_1s_2 & 0\\
\Bigl(\frac{1}{s_3^P }-\frac{1}{s_3^S}\Bigr)s_1s_2 & \frac{s_2^2}{s_3^P }+\Bigl(\frac{1}{c_S^2}-s_2^2\Bigr)\frac{1}{s_3^S} & 0\\
0 & 0 & s_3^P +\frac{s_r^2}{s_3^S}
\end{pmatrix},
\label{eq325ee}\\
&&\hspace{-0.7cm}-\{\tilde{\bf \Delta}_2({{\bf s}})\}^{-1}\tilde{\bf D}_2^-({\bf s})=
\begin{pmatrix} \frac{1}{2} & 0 & \frac{s_1\bigl(S c_S^2-\frac{1}{2}\bigr)}{s_3^S}\\ 
0 & \frac{1}{2} &  \frac{s_2\bigl(S c_S^2-\frac{1}{2}\bigr)}{s_3^S} \\
 - \frac{s_1\bigl(S c_S^2-\frac{1}{2}\bigr)}{s_3^P} & - \frac{s_2\bigl(S c_S^2-\frac{1}{2}\bigr)}{s_3^P} & \frac{1}{2} 
 \end{pmatrix}.
\end{eqnarray}
with $S=s_3^P s_3^S+s_r^2$.

\end{spacing}

\newpage
\centerline{{\Large References}}



\begin{thebibliography}{10}
\newcommand{\enquote}[1]{``#1''}
\expandafter\ifx\csname url\endcsname\relax
  \def\url#1{\texttt{#1}}\fi
\expandafter\ifx\csname urlprefix\endcsname\relax\def\urlprefix{URL }\fi
\providecommand{\bibinfo}[2]{#2}
\providecommand{\noopsort}[1]{}
\providecommand{\switchargs}[2]{#2#1}

\bibitem{Green1828Book}
\bibinfo{author}{G.~Green},  {\bibinfo{title}{\it An essay on the application
  of mathematical analysis to the theories of electricity and magnetism}}
  (\bibinfo{publisher}{Originally published as book in Nottingham, 1828.
  Reprinted in three parts in Journal f\"ur die reine und angewandte Mathematik
  Vol. 39, 1 (1850) p. 73–89, Vol. 44, 4 (1852) p. 356–74, and Vol. 47, 3
  (1854) p. 161–221.}) (\bibinfo{year}{1828}).

\bibitem{Challis2003PhysicsToday}
\bibinfo{author}{L.~Challis} and \bibinfo{author}{F.~Sheard},
  \enquote{\bibinfo{title}{The {G}reen of {G}reen functions}},
  \bibinfo{journal}{Physics Today} \textbf{\bibinfo{volume}{56}},
  \bibinfo{pages}{41--46} (\bibinfo{year}{2003}).

\bibitem{Born65Book}
\bibinfo{author}{M.~Born} and \bibinfo{author}{E.~Wolf},
  {\bibinfo{title}{\it Principles of optics}} (\bibinfo{publisher}{Pergamon
  {P}ress, {L}ondon}) (\bibinfo{year}{1965}), Chap. 8.

\bibitem{Rayleigh78Book}
\bibinfo{author}{J.~W.~S. Rayleigh}, {\bibinfo{title}{\it The theory of sound.
  {V}olume {II}}} (\bibinfo{publisher}{Dover {P}ublications, {I}nc. ({R}eprint
  1945)}) (\bibinfo{year}{1878}), Chap. 14.

\bibitem{Bleistein84Book}
\bibinfo{author}{N.~Bleistein},  {\bibinfo{title}{\it Mathematical methods for
  wave phenomena}} (\bibinfo{publisher}{Academic {P}ress, {I}nc., {O}rlando})
  (\bibinfo{year}{1984}), Chap. 6.

\bibitem{Knopoff56JASA}
\bibinfo{author}{L.~Knopoff}, \enquote{\bibinfo{title}{Diffraction of elastic
  waves}}, \bibinfo{journal}{J. Acoust. Soc.  Am.}
  \textbf{\bibinfo{volume}{28}}, \bibinfo{pages}{217--229}
  (\bibinfo{year}{1956}).

\bibitem{Hoop58PHD}
\bibinfo{author}{A.~T. de~{H}oop}, \enquote{\bibinfo{title}{Representation
  theorems for the displacement in an elastic solid and their applications to
  elastodynamic diffraction theory}}, Ph.D. thesis, \bibinfo{school}{Delft
  {U}niversity of {T}echnology, {D}elft} (\bibinfo{year}{1958}).

\bibitem{Gangi70JGR}
\bibinfo{author}{A.~F. Gangi}, \enquote{\bibinfo{title}{A derivation of the
  seismic representation theorem using seismic reciprocity}},
  \bibinfo{journal}{J. Geophys. Res.}
  \textbf{\bibinfo{volume}{75}}, \bibinfo{pages}{2088--2095}
  (\bibinfo{year}{1970}).

\bibitem{Pao76JASA}
\bibinfo{author}{Y.~H. Pao} and \bibinfo{author}{V.~Varatharajulu},
  \enquote{\bibinfo{title}{Huygens' principle, radiation conditions, and
  integral formulations for the scattering of elastic waves}},
  \bibinfo{journal}{J. Acoust. Soc.  Am.}
  \textbf{\bibinfo{volume}{59}}, \bibinfo{pages}{1361--1371}
  (\bibinfo{year}{1976}).

\bibitem{Kong86Book}
\bibinfo{author}{J.~A. Kong}, {\bibinfo{title}{\it Electromagnetic wave
  theory}} (\bibinfo{publisher}{Wiley Interscience}) (\bibinfo{year}{1986}), Chap. 5.

\bibitem{Altman91Book}
\bibinfo{author}{C.~Altman} and \bibinfo{author}{K.~Suchy},
  {\bibinfo{title}{\it Reciprocity, spatial mapping and time reversal in
  electromagnetics}} (\bibinfo{publisher}{Kluwer, Dordrecht})
  (\bibinfo{year}{1991}), Chap 3.

\bibitem{Hoop95Book}
\bibinfo{author}{A.~T. de~Hoop}, {\bibinfo{title}{\it Handbook of radiation
  and scattering of waves}} (\bibinfo{publisher}{Academic Press, London})
  (\bibinfo{year}{1995}), Chaps. 7, 15 and 28.

\bibitem{Hilterman70GEO}
\bibinfo{author}{F.~J. Hilterman}, \enquote{\bibinfo{title}{Three-dimensional
  seismic modeling}}, \bibinfo{journal}{Geophysics}
  \textbf{\bibinfo{volume}{35}}, \bibinfo{pages}{1020--1037}
  (\bibinfo{year}{1970}).

\bibitem{Frazer85GJRAS}
\bibinfo{author}{L.~N. Frazer} and \bibinfo{author}{M.~K. Sen},
  \enquote{\bibinfo{title}{Kirchhoff-{H}elmholtz reflection seismograms in a
  laterally inhomogeneous multi-layered elastic medium, {I}, {T}heory}},
  \bibinfo{journal}{Geophys. J. R. Astr. Soc.}
  \textbf{\bibinfo{volume}{80}}, \bibinfo{pages}{121--147}
  (\bibinfo{year}{1985}).

\bibitem{Mansuripur2021NP}
\bibinfo{author}{M.~Mansuripur}, \enquote{\bibinfo{title}{A tutorial on the
  classical theories of electromagnetic scattering and diffraction}},
  \bibinfo{journal}{Nanophotonics} \textbf{\bibinfo{volume}{10}},
  \bibinfo{pages}{315--342} (\bibinfo{year}{2021}).

\bibitem{Porter82JOSA}
\bibinfo{author}{R.~P. Porter} and \bibinfo{author}{A.~J. Devaney},
  \enquote{\bibinfo{title}{Holography and the inverse source problem}},
  \bibinfo{journal}{J. Opt. Soc. Am.}
  \textbf{\bibinfo{volume}{72}}, \bibinfo{pages}{327--330}
  (\bibinfo{year}{1982}).

\bibitem{Devaney82UI}
\bibinfo{author}{A.~J. Devaney}, \enquote{\bibinfo{title}{A filtered
  backpropagation algorithm for diffraction tomography}},
  \bibinfo{journal}{Ultrasonic {I}maging} \textbf{\bibinfo{volume}{4}},
  \bibinfo{pages}{336--350} (\bibinfo{year}{1982}).

\bibitem{Bojarski83JASA}
\bibinfo{author}{N.~N. Bojarski}, \enquote{\bibinfo{title}{Generalized reaction
  principles and reciprocity theorems for the wave equations, and the
  relationship between the time-advanced and time-retarded fields}},
  \bibinfo{journal}{J. Acoust. Soc.  Am.}
  \textbf{\bibinfo{volume}{74}}, \bibinfo{pages}{281--285}
  (\bibinfo{year}{1983}).

\bibitem{Oristaglio89IP}
\bibinfo{author}{M.~L. Oristaglio}, \enquote{\bibinfo{title}{An inverse
  scattering formula that uses all the data}}, \bibinfo{journal}{Inverse
  Probl.} \textbf{\bibinfo{volume}{5}}, \bibinfo{pages}{1097--1105}
  (\bibinfo{year}{1989}).

\bibitem{Porter70JOSA}
\bibinfo{author}{R.~P. Porter}, \enquote{\bibinfo{title}{Diffraction-limited,
  scalar image formation with holograms of arbitrary shape}},
  \bibinfo{journal}{J. Opt. Soc. Am.}
  \textbf{\bibinfo{volume}{60}}, \bibinfo{pages}{1051--1059}
  (\bibinfo{year}{1970}).

\bibitem{Schneider78GEO}
\bibinfo{author}{W.~A. Schneider}, \enquote{\bibinfo{title}{Integral
  formulation for migration in two and three dimensions}},
  \bibinfo{journal}{Geophysics} \textbf{\bibinfo{volume}{43}},
  \bibinfo{pages}{49--76} (\bibinfo{year}{1978}).

\bibitem{Berkhout82Book}
\bibinfo{author}{A.~J. Berkhout}, {\bibinfo{title}{\it Seismic {M}igration.
  {I}maging of acoustic energy by wave field extrapolation. {A}. {T}heoretical
  aspects}} (\bibinfo{publisher}{Elsevier}) (\bibinfo{year}{1982}), Chap. 7.

\bibitem{Maynard85JASA}
\bibinfo{author}{J.~D. Maynard}, \bibinfo{author}{E.~G. Williams}, and
  \bibinfo{author}{Y.~Lee}, \enquote{\bibinfo{title}{Nearfield acoustic
  holography: {I}. {T}heory of generalized holography and the development of
  {NAH}}}, \bibinfo{journal}{J. Acoust. Soc.  Am.}
  \textbf{\bibinfo{volume}{78}}, \bibinfo{pages}{1395--1413}
  (\bibinfo{year}{1985}).

\bibitem{Esmersoy88GEO}
\bibinfo{author}{C.~Esmersoy} and \bibinfo{author}{M.~Oristaglio},
  \enquote{\bibinfo{title}{Reverse-time wave-field extrapolation, imaging, and
  inversion}}, \bibinfo{journal}{Geophysics} \textbf{\bibinfo{volume}{53}},
  \bibinfo{pages}{920--931} (\bibinfo{year}{1988}).

\bibitem{Lindsey2004AJSS}
\bibinfo{author}{C.~Lindsey} and \bibinfo{author}{D.~C. Braun},
  \enquote{\bibinfo{title}{Principles of seismic holography for diagnostics of
  the shallow subphotosphere}}, \bibinfo{journal}{The Astrophysical Journal
  Supplement Series} \textbf{\bibinfo{volume}{155}}, \bibinfo{pages}{209--225}
  (\bibinfo{year}{2004}).

\bibitem{Fink2001IP}
\bibinfo{author}{M.~Fink} and \bibinfo{author}{C.~Prada},
  \enquote{\bibinfo{title}{Acoustic time-reversal mirrors}},
  \bibinfo{journal}{Inverse Problems} \textbf{\bibinfo{volume}{17}},
  \bibinfo{pages}{R1--R38} (\bibinfo{year}{2001}).

\bibitem{Derode2003JASA}
\bibinfo{author}{A.~Derode}, \bibinfo{author}{E.~Larose},
  \bibinfo{author}{M.~Tanter}, \bibinfo{author}{J.~de~{R}osny},
  \bibinfo{author}{A.~Tourin}, \bibinfo{author}{M.~Campillo}, and
  \bibinfo{author}{M.~Fink}, \enquote{\bibinfo{title}{Recovering the {G}reen's
  function from field-field correlations in an open scattering medium ({L})}},
  \bibinfo{journal}{J. Acoust. Soc.  Am.}
  \textbf{\bibinfo{volume}{113}}, \bibinfo{pages}{2973--2976}
  (\bibinfo{year}{2003}).

\bibitem{Wapenaar2003GEO}
\bibinfo{author}{K.~Wapenaar}, \enquote{\bibinfo{title}{Synthesis of an
  inhomogeneous medium from its acoustic transmission response}},
  \bibinfo{journal}{Geophysics} \textbf{\bibinfo{volume}{68}},
  \bibinfo{pages}{1756--1759} (\bibinfo{year}{2003}).

\bibitem{Weaver2004JASA}
\bibinfo{author}{R.~L. Weaver} and \bibinfo{author}{O.~I. Lobkis},
  \enquote{\bibinfo{title}{Diffuse fields in open systems and the emergence of
  the {G}reen's function ({L})}}, \bibinfo{journal}{J. Acoust. Soc. Am.} \textbf{\bibinfo{volume}{116}},
  \bibinfo{pages}{2731--2734} (\bibinfo{year}{2004}).

\bibitem{Thomson50JAP}
\bibinfo{author}{W.~T. Thomson}, \enquote{\bibinfo{title}{Transmission of
  elastic waves through a stratified solid medium}}, \bibinfo{journal}{J.  Appl. Phys.} \textbf{\bibinfo{volume}{21}}, \bibinfo{pages}{89--93}
  (\bibinfo{year}{1950}).

\bibitem{Haskell53BSSA}
\bibinfo{author}{N.~A. Haskell}, \enquote{\bibinfo{title}{The dispersion of
  surface waves on multilayered media}}, \bibinfo{journal}{Bull. Seismol. Soc. Am.} \textbf{\bibinfo{volume}{43}},
  \bibinfo{pages}{17--34} (\bibinfo{year}{1953}).

\bibitem{Gilbert66GEO}
\bibinfo{author}{F.~Gilbert} and \bibinfo{author}{G.~E. Backus},
  \enquote{\bibinfo{title}{Propagator matrices in elastic wave and vibration
  problems}}, \bibinfo{journal}{Geophysics} \textbf{\bibinfo{volume}{31}},
  \bibinfo{pages}{326--332} (\bibinfo{year}{1966}).

\bibitem{Kennett72BS}
\bibinfo{author}{B.~L.~N. Kennett}, \enquote{\bibinfo{title}{The connection
  between elastodynamic representation theorems and propagator matrices}},
  \bibinfo{journal}{Bull. Seismol. Soc. Am.}
  \textbf{\bibinfo{volume}{62}}, \bibinfo{pages}{973--983}
  (\bibinfo{year}{1972}).

\bibitem{Kennett72GJRAS}
\bibinfo{author}{B.~L.~N. Kennett}, \enquote{\bibinfo{title}{Seismic waves in
  laterally inhomogeneous media}}, \bibinfo{journal}{Geophys. J. R. Astr. Soc.} \textbf{\bibinfo{volume}{27}},
  \bibinfo{pages}{301--325} (\bibinfo{year}{1972}).

\bibitem{Woodhouse74GJR}
\bibinfo{author}{J.~H. Woodhouse}, \enquote{\bibinfo{title}{Surface waves in a
  laterally varying layered structure}}, \bibinfo{journal}{Geophys. J. R. Astr. Soc.} \textbf{\bibinfo{volume}{37}},
  \bibinfo{pages}{461--490} (\bibinfo{year}{1974}).

\bibitem{Haines88GJI}
\bibinfo{author}{A.~J. Haines}, \enquote{\bibinfo{title}{Multi-source,
  multi-receiver synthetic seismograms for laterally heterogeneous media using
  {F}-{K} domain propagators}}, \bibinfo{journal}{Geophys. J. Int.} \textbf{\bibinfo{volume}{95}}, \bibinfo{pages}{237--260}
  (\bibinfo{year}{1988}).

\bibitem{Kennett90GJI}
\bibinfo{author}{B.~L.~N. Kennett}, \bibinfo{author}{K.~Koketsu}, and
  \bibinfo{author}{A.~J. Haines}, \enquote{\bibinfo{title}{Propagation
  invariants, reflection and transmission in anisotropic, laterally
  heterogeneous media}}, \bibinfo{journal}{Geophys. J. Int.}
  \textbf{\bibinfo{volume}{103}}, \bibinfo{pages}{95--101}
  (\bibinfo{year}{1990}).

\bibitem{Koketsu91GJI}
\bibinfo{author}{K.~Koketsu}, \bibinfo{author}{B.~L.~N. Kennett}, and
  \bibinfo{author}{H.~Takenaka}, \enquote{\bibinfo{title}{2-{D} reflectivity
  method and synthetic seismograms for irregularly layered structures - {II}.
  {I}nvariant embedding approach}}, \bibinfo{journal}{Geophys. J. Int.} \textbf{\bibinfo{volume}{105}}, \bibinfo{pages}{119--130}
  (\bibinfo{year}{1991}).

\bibitem{Takenaka93WM}
\bibinfo{author}{H.~Takenaka}, \bibinfo{author}{B.~L.~N. Kennett}, and
  \bibinfo{author}{K.~Koketsu}, \enquote{\bibinfo{title}{The integral operator
  representation of propagation invariants for elastic waves in irregularly
  layered media}}, \bibinfo{journal}{Wave Motion}
  \textbf{\bibinfo{volume}{17}}, \bibinfo{pages}{299--317}
  (\bibinfo{year}{1993}).

\bibitem{Wapenaar87GEO}
\bibinfo{author}{C.~P.~A. Wapenaar}, \bibinfo{author}{N.~A. Kinneging}, and
  \bibinfo{author}{A.~J. Berkhout}, \enquote{\bibinfo{title}{Principle of
  prestack migration based on the full elastic two-way wave equation}},
  \bibinfo{journal}{Geophysics} \textbf{\bibinfo{volume}{52}},
  \bibinfo{pages}{151--173} (\bibinfo{year}{1987}).

\bibitem{Rose2001PRA}
\bibinfo{author}{J.~H. Rose}, \enquote{\bibinfo{title}{ ``{S}ingle-sided''
  focusing of the time-dependent {S}chr\"{o}dinger equation}},
  \bibinfo{journal}{Phys. Rev. A} \textbf{\bibinfo{volume}{65}},
  \bibinfo{pages}{012707} (\bibinfo{year}{2001}).

\bibitem{Rose2002IP}
\bibinfo{author}{J.~H. Rose}, \enquote{\bibinfo{title}{ `{S}ingle-sided'
  autofocusing of sound in layered materials}}, \bibinfo{journal}{Inverse
  Probl.} \textbf{\bibinfo{volume}{18}}, \bibinfo{pages}{1923--1934}
  (\bibinfo{year}{2002}).

\bibitem{Broggini2012EJP}
\bibinfo{author}{F.~Broggini} and \bibinfo{author}{R.~Snieder},
  \enquote{\bibinfo{title}{Connection of scattering principles: a visual and
  mathematical tour}}, \bibinfo{journal}{Eur. J. Phys.}
  \textbf{\bibinfo{volume}{33}}, \bibinfo{pages}{593--613}
  (\bibinfo{year}{2012}).

\bibitem{Wapenaar2014JASA}
\bibinfo{author}{K.~Wapenaar}, \bibinfo{author}{J.~Thorbecke},
  \bibinfo{author}{J.~van~der Neut}, \bibinfo{author}{F.~Broggini},
  \bibinfo{author}{E.~Slob}, and \bibinfo{author}{R.~Snieder},
  \enquote{\bibinfo{title}{Green's function retrieval from reflection data, in absence of a receiver at the virtual source position}},
  \bibinfo{journal}{J. Acoust. Soc. Am.}
  \textbf{\bibinfo{volume}{135}}, \bibinfo{pages}{2847--2861}
  (\bibinfo{year}{2014}).

\bibitem{Ravasi2016GJI}
\bibinfo{author}{M.~Ravasi}, \bibinfo{author}{I.~Vasconcelos},
  \bibinfo{author}{A.~Kritski}, \bibinfo{author}{A.~Curtis},
  \bibinfo{author}{C.~A. da~Costa~Filho}, and \bibinfo{author}{G.~A. Meles},
  \enquote{\bibinfo{title}{Target-oriented {M}archenko imaging of a {N}orth
  {S}ea field}}, \bibinfo{journal}{Geophys. J. Int.}
  \textbf{\bibinfo{volume}{205}}, \bibinfo{pages}{99--104}
  (\bibinfo{year}{2016}).

\bibitem{Staring2018GEO}
\bibinfo{author}{M.~Staring}, \bibinfo{author}{R.~Pereira},
  \bibinfo{author}{H.~Douma}, \bibinfo{author}{J.~van~der Neut}, and
  \bibinfo{author}{K.~Wapenaar}, \enquote{\bibinfo{title}{Source-receiver
  {M}archenko redatuming on field data using an adaptive double-focusing
  method}}, \bibinfo{journal}{Geophysics} \textbf{\bibinfo{volume}{83}},
  \bibinfo{pages}{S579--S590} (\bibinfo{year}{2018}).

\bibitem{Jia2018GEO}
\bibinfo{author}{X.~Jia}, \bibinfo{author}{A.~Guitton}, and
  \bibinfo{author}{R.~Snieder}, \enquote{\bibinfo{title}{A practical
  implementation of subsalt {M}archenko imaging with a {G}ulf of {M}exico data
  set}}, \bibinfo{journal}{Geophysics} \textbf{\bibinfo{volume}{83}},
  \bibinfo{pages}{S409--S419} (\bibinfo{year}{2018}).

\bibitem{Neut2017JASA}
\bibinfo{author}{J.~Van~der Neut}, \bibinfo{author}{J.~L. Johnson},
  \bibinfo{author}{K.~van Wijk}, \bibinfo{author}{S.~Singh},
  \bibinfo{author}{E.~Slob}, and \bibinfo{author}{K.~Wapenaar},
  \enquote{\bibinfo{title}{A {M}archenko equation for acoustic inverse source
  problems}}, \bibinfo{journal}{J. Acoust. Soc.  Am.}
  \textbf{\bibinfo{volume}{141}}, \bibinfo{pages}{4332--4346}
  (\bibinfo{year}{2017}).

\bibitem{Wapenaar96JASA}
\bibinfo{author}{C.~P.~A. Wapenaar}, \enquote{\bibinfo{title}{Reciprocity
  theorems for two-way and one-way wave vectors: a comparison}},
  \bibinfo{journal}{J. Acoust. Soc.  Am.}
  \textbf{\bibinfo{volume}{100}}, \bibinfo{pages}{3508--3518}
  (\bibinfo{year}{1996}).

\bibitem{Haines96JMP}
\bibinfo{author}{A.~J. Haines} and \bibinfo{author}{M.~V. de~{H}oop},
  \enquote{\bibinfo{title}{An invariant imbedding analysis of general wave
  scattering problems}}, \bibinfo{journal}{J.\ Math.\ Phys.}
  \textbf{\bibinfo{volume}{37}}, \bibinfo{pages}{3854--3881}
  (\bibinfo{year}{1996}).

\bibitem{Wapenaar2019GJI}
\bibinfo{author}{K.~Wapenaar}, \enquote{\bibinfo{title}{Unified matrix-vector
  wave equation, reciprocity and representations}},
  \bibinfo{journal}{Geophys. J. Int.}
  \textbf{\bibinfo{volume}{216}}, \bibinfo{pages}{560--583}
  (\bibinfo{year}{2019}).

\bibitem{Hoop88JASA}
\bibinfo{author}{A.~T. de~{H}oop}, \enquote{\bibinfo{title}{Time-domain
  reciprocity theorems for acoustic wave fields in fluids with relaxation}},
  \bibinfo{journal}{J. Acoust. Soc.  Am.}
  \textbf{\bibinfo{volume}{84}}, \bibinfo{pages}{1877--1882}
  (\bibinfo{year}{1988}).

\bibitem{Schoenberg83JASA}
\bibinfo{author}{M.~Schoenberg} and \bibinfo{author}{P.~N. Sen},
  \enquote{\bibinfo{title}{Properties of a periodically stratified acoustic
  half-space and its relation to a {B}iot fluid}}, \bibinfo{journal}{J. Acoust. Soc. Am.} \textbf{\bibinfo{volume}{73}},
  \bibinfo{pages}{61--67} (\bibinfo{year}{1983}).

\bibitem{Corones75JMAA}
\bibinfo{author}{J.~P. Corones}, \enquote{\bibinfo{title}{Bremmer series that
  correct parabolic approximations}}, \bibinfo{journal}{J.\ Math.\ Anal.\
  Appl.} \textbf{\bibinfo{volume}{50}}, \bibinfo{pages}{361--372}
  (\bibinfo{year}{1975}).

\bibitem{Ursin83GEO}
\bibinfo{author}{B.~Ursin}, \enquote{\bibinfo{title}{Review of elastic and
  electromagnetic wave propagation in horizontally layered media}},
  \bibinfo{journal}{Geophysics} \textbf{\bibinfo{volume}{48}},
  \bibinfo{pages}{1063--1081} (\bibinfo{year}{1983}).

\bibitem{Fishman84JMP}
\bibinfo{author}{L.~Fishman} and \bibinfo{author}{J.~J. McCoy},
  \enquote{\bibinfo{title}{Derivation and application of extended parabolic
  wave theories. {I}. {T}he factorized {H}elmholtz equation}},
  \bibinfo{journal}{J.\ Math.\ Phys.} \textbf{\bibinfo{volume}{25}},
  \bibinfo{pages}{285--296} (\bibinfo{year}{1984}).

\bibitem{Wapenaar89Book}
\bibinfo{author}{C.~P.~A. Wapenaar} and \bibinfo{author}{A.~J. Berkhout},
  {\bibinfo{title}{\it Elastic wave field extrapolation}}
  (\bibinfo{publisher}{Elsevier, Amsterdam}) (\bibinfo{year}{1989}), Chaps 3, 4, 11 and 12.

\bibitem{Hoop96JMP}
\bibinfo{author}{M.~V. de~{H}oop}, \enquote{\bibinfo{title}{Generalization of
  the {B}remmer coupling series}}, \bibinfo{journal}{J.\ Math.\ Phys.}
  \textbf{\bibinfo{volume}{37}}, \bibinfo{pages}{3246--3282}
  (\bibinfo{year}{1996}).

\bibitem{Stoffa89Book}
\bibinfo{author}{P.~L. Stoffa}, {\bibinfo{title}{\it Tau-p - {A} plane wave
  approach to the analysis of seismic data}} (\bibinfo{publisher}{Kluwer
  Academic Publishers}, \bibinfo{address}{Dordrecht}) (\bibinfo{year}{1989}), Chap. 2.

\bibitem{Wapenaar2021GEO}
\bibinfo{author}{K.~Wapenaar} and \bibinfo{author}{S.~de~Ridder},
  \enquote{\bibinfo{title}{On the relation between the propagator matrix and
  the {M}archenko focusing function}}, \bibinfo{journal}{Geophysics  \textbf{\bibinfo{volume}{87}}(2), \bibinfo{pages}{A7-A11} } 
   (\bibinfo{year}{2022}).

\bibitem{Lorentz1895KNAW}
\bibinfo{author}{H.~A. Lorentz}, \enquote{\bibinfo{title}{The theorem of
  {P}oynting concerning the energy in the electromagnetic field and two general
  propositions concerning the propagation of light}},
  \bibinfo{journal}{Verslagen der Afdeeling Natuurkunde van de Koninklijke
  Akademie van Wetenschappen} \textbf{\bibinfo{volume}{4}},
  \bibinfo{pages}{176--187} (\bibinfo{year}{1895}).

\bibitem{Knopoff59GEO}
\bibinfo{author}{L.~Knopoff} and \bibinfo{author}{A.~F. Gangi},
  \enquote{\bibinfo{title}{Seismic reciprocity}}, \bibinfo{journal}{Geophysics}
  \textbf{\bibinfo{volume}{24}}, \bibinfo{pages}{681--691}
  (\bibinfo{year}{1959}).

\bibitem{Hoop66ASR}
\bibinfo{author}{A.~T. de~Hoop}, \enquote{\bibinfo{title}{An elastodynamic
  reciprocity theorem for linear, viscoelastic media}},
  \bibinfo{journal}{Applied Scientific Research} \textbf{\bibinfo{volume}{16}},
  \bibinfo{pages}{39--45} (\bibinfo{year}{1966}).

\bibitem{Flekkoy99PRE}
\bibinfo{author}{E.~G. Flekk{\o}y} and \bibinfo{author}{S.~R. Pride},
  \enquote{\bibinfo{title}{Reciprocity and cross coupling of two-phase flow in
  porous media from {O}nsager theory}}, \bibinfo{journal}{Phys. Rev. E}
  \textbf{\bibinfo{volume}{60}}, \bibinfo{pages}{4130--4137}
  (\bibinfo{year}{1999}).

\bibitem{Auld79WM}
\bibinfo{author}{B.~A. Auld}, \enquote{\bibinfo{title}{General
  electromechanical reciprocity relations applied to the calculation of elastic
  wave scattering coefficients}}, \bibinfo{journal}{Wave Motion}
  \textbf{\bibinfo{volume}{1}}, \bibinfo{pages}{3--10} (\bibinfo{year}{1979}).

\bibitem{Achenbach2003Book}
\bibinfo{author}{J.~D. Achenbach}, {\bibinfo{title}{\it Reciprocity in
  elastodynamics}} (\bibinfo{publisher}{Cambridge University Press, Cambridge})
  (\bibinfo{year}{2003}), Chap. 14.

\bibitem{Pride96JASA}
\bibinfo{author}{S.~R. Pride} and \bibinfo{author}{M.~W. Haartsen},
  \enquote{\bibinfo{title}{Electroseismic wave properties}},
  \bibinfo{journal}{J. Acoust. Soc.  Am.}
  \textbf{\bibinfo{volume}{100}}, \bibinfo{pages}{1301--1315}
  (\bibinfo{year}{1996}).

\bibitem{Slob2007GRL}
\bibinfo{author}{E.~Slob} and \bibinfo{author}{K.~Wapenaar},
  \enquote{\bibinfo{title}{Electromagnetic {G}reen's functions retrieval by
  cross-correlation and cross-convolution in media with losses}},
  \bibinfo{journal}{Geophys. Res. Lett.}
  \textbf{\bibinfo{volume}{34}}, \bibinfo{pages}{L05307}
  (\bibinfo{year}{2007}).

\bibitem{Wapenaar2010JASA}
\bibinfo{author}{K.~Wapenaar} and \bibinfo{author}{J.~van~der Neut},
  \enquote{\bibinfo{title}{A representation for {G}reen's function retrieval by
  multidimensional deconvolution}}, \bibinfo{journal}{J. Acoust. Soc. Am.} \textbf{\bibinfo{volume}{128}},
  \bibinfo{pages}{EL366--EL371} (\bibinfo{year}{2010}).

\bibitem{Wapenaar2006PRL}
\bibinfo{author}{K.~Wapenaar}, \bibinfo{author}{E.~Slob}, and
  \bibinfo{author}{R.~Snieder}, \enquote{\bibinfo{title}{Unified {G}reen's
  function retrieval by cross correlation}}, \bibinfo{journal}{Phys. Rev.
  Lett.} \textbf{\bibinfo{volume}{97}}, \bibinfo{pages}{234301}
  (\bibinfo{year}{2006}).

\bibitem{Snieder2007PRE}
\bibinfo{author}{R.~Snieder}, \bibinfo{author}{K.~Wapenaar}, and
  \bibinfo{author}{U.~Wegler}, \enquote{\bibinfo{title}{Unified {G}reen's
  function retrieval by cross-correlation; connection with energy principles}},
  \bibinfo{journal}{Phys. Rev. E} \textbf{\bibinfo{volume}{75}},
  \bibinfo{pages}{036103} (\bibinfo{year}{2007}).

\bibitem{Slob2014GEO}
\bibinfo{author}{E.~Slob}, \bibinfo{author}{K.~Wapenaar},
  \bibinfo{author}{F.~Broggini}, and \bibinfo{author}{R.~Snieder},
  \enquote{\bibinfo{title}{Seismic reflector imaging using internal multiples
  with {M}archenko-type equations}}, \bibinfo{journal}{Geophysics}
  \textbf{\bibinfo{volume}{79}}, \bibinfo{pages}{S63--S76}
  (\bibinfo{year}{2014}).

\bibitem{Berryhill84GEO}
\bibinfo{author}{J.~R. Berryhill}, \enquote{\bibinfo{title}{Wave-equation
  datuming before stack}}, \bibinfo{journal}{Geophysics}
  \textbf{\bibinfo{volume}{49}}, \bibinfo{pages}{2064--2066}
  (\bibinfo{year}{1984}).

\bibitem{Hokstad2000GEO}
\bibinfo{author}{K.~Hokstad}, \enquote{\bibinfo{title}{Multicomponent
  {K}irchhoff migration}}, \bibinfo{journal}{Geophysics}
  \textbf{\bibinfo{volume}{65}}, \bibinfo{pages}{861--873}
  (\bibinfo{year}{2000}).

\rev{
\bibitem{Curtis2010PRE}
\bibinfo{author}{A.~Curtis} and \bibinfo{author}{D.~Halliday},
  \enquote{\bibinfo{title}{Source-receiver wavefield interferometry}},
  \bibinfo{journal}{Phys. Rev. E} \textbf{\bibinfo{volume}{81}},
  \bibinfo{pages}{046601} (\bibinfo{year}{2010}).
}
\rev{
\bibitem{Halliday2012GJI}
\bibinfo{author}{D.~Halliday}, \bibinfo{author}{A.~Curtis}, and
  \bibinfo{author}{K.~Wapenaar}, \enquote{\bibinfo{title}{Generalized {PP + PS
  = SS} from seismic interferometry}}, \bibinfo{journal}{Geophys. J.
  Int.} \textbf{\bibinfo{volume}{189}}, \bibinfo{pages}{1015--1024}
  (\bibinfo{year}{2012}).
}
\bibitem{Kennett78GJRAS}
\bibinfo{author}{B.~L.~N. Kennett}, \bibinfo{author}{N.~J. Kerry}, and
  \bibinfo{author}{J.~H. Woodhouse}, \enquote{\bibinfo{title}{Symmetries in the
  reflection and transmission of elastic waves}}, \bibinfo{journal}{Geophys. J. R. Astr. Soc.} \textbf{\bibinfo{volume}{52}},
  \bibinfo{pages}{215--229} (\bibinfo{year}{1978}).

\bibitem{Wapenaar2021GJI}
\bibinfo{author}{K.~Wapenaar}, \bibinfo{author}{R.~Snieder},
  \bibinfo{author}{S.~de~Ridder}, and \bibinfo{author}{E.~Slob},
  \enquote{\bibinfo{title}{Green’s function representations for {M}archenko
  imaging without up/down decomposition}}, \bibinfo{journal}{Geophys. J. Int.} \textbf{\bibinfo{volume}{227}},
  \bibinfo{pages}{184--203} (\bibinfo{year}{2021}).

\bibitem{Wapenaar2013PRL}
\bibinfo{author}{K.~Wapenaar}, \bibinfo{author}{F.~Broggini},
  \bibinfo{author}{E.~Slob}, and \bibinfo{author}{R.~Snieder},
  \enquote{\bibinfo{title}{Three-dimensional single-sided {M}archenko inverse
  scattering, data-driven focusing, {G}reen's function retrieval, and their
  mutual relations}}, \bibinfo{journal}{Phys. Rev. Lett.}
  \textbf{\bibinfo{volume}{110}}, \bibinfo{pages}{084301}
  (\bibinfo{year}{2013}).

\bibitem{Neut2015GJI}
\bibinfo{author}{J.~Van~der Neut}, \bibinfo{author}{I.~Vasconcelos}, and
  \bibinfo{author}{K.~Wapenaar}, \enquote{\bibinfo{title}{On {G}reen's function
  retrieval by iterative substitution of the coupled {M}archenko equations}},
  \bibinfo{journal}{Geophys. J. Int.}
  \textbf{\bibinfo{volume}{203}}, \bibinfo{pages}{792--813}
  (\bibinfo{year}{2015}).

\bibitem{Wapenaar2014GJI}
\bibinfo{author}{K.~Wapenaar} and \bibinfo{author}{E.~Slob},
  \enquote{\bibinfo{title}{On the {M}archenko equation for multicomponent
  single-sided reflection data}}, \bibinfo{journal}{Geophys. J. Int.} \textbf{\bibinfo{volume}{199}}, \bibinfo{pages}{1367--1371}
  (\bibinfo{year}{2014}).

\bibitem{Costa2014PRE}
\bibinfo{author}{C.~A. da~Costa~Filho}, \bibinfo{author}{M.~Ravasi},
  \bibinfo{author}{A.~Curtis}, and \bibinfo{author}{G.~A. Meles},
  \enquote{\bibinfo{title}{Elastodynamic {G}reen's function retrieval through
  single-sided {M}archenko inverse scattering}}, \bibinfo{journal}{Phys. Rev. E} \textbf{\bibinfo{volume}{90}}, \bibinfo{pages}{063201}
  (\bibinfo{year}{2014}).

\bibitem{Kiraz2021JASA}
\bibinfo{author}{M.~S.~R. Kiraz}, \bibinfo{author}{R.~Snieder}, and
  \bibinfo{author}{K.~Wapenaar}, \enquote{\bibinfo{title}{Focusing waves in an
  unknown medium without wavefield decomposition}}, \bibinfo{journal}{JASA
  Express Letters} \textbf{\bibinfo{volume}{1}}, \bibinfo{pages}{055602}
  (\bibinfo{year}{2021}).

\bibitem{Diekmann2021PRR}
\bibinfo{author}{L.~Diekmann} and \bibinfo{author}{I.~Vasconcelos},
  \enquote{\bibinfo{title}{Focusing and {G}reen's function retrieval in
  three-dimensional inverse scattering revisited: {A} single-sided {M}archenko
  integral for the full wave field}}, \bibinfo{journal}{Phys. Rev.
  Res.} \textbf{\bibinfo{volume}{3}}, \bibinfo{pages}{013206}
  (\bibinfo{year}{2021}).

\bibitem{Brackenhoff2019JGR}
\bibinfo{author}{J.~Brackenhoff}, \bibinfo{author}{J.~Thorbecke}, and
  \bibinfo{author}{K.~Wapenaar}, \enquote{\bibinfo{title}{Virtual sources and
  receivers in the real {E}arth: {C}onsiderations for practical applications}},
  \bibinfo{journal}{J. Geophys. Res.}
  \textbf{\bibinfo{volume}{124}}, \bibinfo{pages}{11,802--11,821}
  (\bibinfo{year}{2019}).

\bibitem{Pereira2019SEG}
\bibinfo{author}{R.~Pereira}, \bibinfo{author}{M.~Ramzy},
  \bibinfo{author}{P.~Griscenco}, \bibinfo{author}{B.~Huard},
  \bibinfo{author}{H.~Huang}, \bibinfo{author}{L.~Cypriano}, and
  \bibinfo{author}{A.~Khalil}, \enquote{\bibinfo{title}{Internal multiple
  attenuation for {OBN} data with overburden/target separation}}, in
  {\bibinfo{booktitle}{\it SEG, Expanded Abstracts}},
  \bibinfo{pages}{4520--4524} (\bibinfo{year}{2019}).

\bibitem{Zhang2020GEO}
\bibinfo{author}{L.~Zhang} and \bibinfo{author}{E.~Slob},
  \enquote{\bibinfo{title}{A field data example of {M}archenko multiple
  elimination}}, \bibinfo{journal}{Geophysics} \textbf{\bibinfo{volume}{85}},
  \bibinfo{pages}{S65--S70} (\bibinfo{year}{2020}).

\bibitem{Staring2020GP}
\bibinfo{author}{M.~Staring} and \bibinfo{author}{K.~Wapenaar},
  \enquote{\bibinfo{title}{Three-dimensional {M}archenko internal multiple
  attenuation on narrow azimuth streamer data of the {S}antos {B}asin,
  {B}razil}}, \bibinfo{journal}{Geophys. Prosp.}
  \textbf{\bibinfo{volume}{68}}, \bibinfo{pages}{1864--1877}
  (\bibinfo{year}{2020}).

\bibitem{Reinicke2020GEO}
\bibinfo{author}{C.~Reinicke}, \bibinfo{author}{M.~Dukalski}, and
  \bibinfo{author}{K.~Wapenaar}, \enquote{\bibinfo{title}{Comparison of
  monotonicity challenges encountered by the inverse scattering series and the
  {M}archenko demultiple method for elastic waves}},
  \bibinfo{journal}{Geophysics} \textbf{\bibinfo{volume}{85}},
  \bibinfo{pages}{Q11--Q26} (\bibinfo{year}{2020}).

\end{thebibliography}
\end{document}